\documentclass[
  journal=pasa,
  manuscript=research-paper, %% or "review"
  year=2026,
  volume=YY,
]{cup-journal}

\usepackage{amsmath}
\usepackage{amssymb,microtype,siunitx,booktabs}
\usepackage{threeparttable}
\usepackage{hyperref}
\usepackage{cleveref}
\usepackage{tabularx} 
\usepackage{graphicx}
\usepackage{orcidlink}
\usepackage{subcaption}
\usepackage[table]{xcolor}
\usepackage[table]{xcolor}
\usepackage{seqsplit}
\definecolor{brown}{RGB}{218, 165, 32} % A professional "Goldenrod" tone
\definecolor{darkblue}{RGB}{235, 230, 250}
\definecolor{darkgreen}{RGB}{235, 230, 250}
\definecolor{darkpurple}{RGB}{235, 230, 250}

\def\lya{Ly$\alpha$}

\def\oii{[O~{\sc ii}]}
\def\oiii{O~{\sc iii}]}

\def\sii{[S~{\sc ii}]}

\def\Oiii{[O~{\sc iii}]}

\def\hi{H~{\sc i}}
\def\hii{H~{\sc ii}}

\def\ciii{C~{\sc iii}]}
\def\civ{C~{\sc iv}}

\def\heii{He~{\sc ii}}

\def\siiii{Si~{\sc iii}]}

\def\niii{N~{\sc iii}]}

\def\nv{N~{\sc v}}
\def\oiv{O~{\sc iv}}

\usepackage{hyperref}
\hypersetup{
    colorlinks=True,
    linkcolor=blue,
    urlcolor=blue,
    citecolor=blue
    }

\title{The MAGPI Survey: Investigating Properties of \heii\ $\lambda1640$\,\AA\ Emitters at $z\sim 2-4$}

\author{I. U. Aalia
\textsuperscript{\orcidlink{0009-0005-0489-6232}}}
\affiliation{School of Mathematical and Physical Sciences, Macquarie University, NSW 2109, Australia}
\alsoaffiliation{Astrophysics and Space Technologies Research Centre, Macquarie University, Sydney, NSW 2109, Australia}
\email[Aalia Imam Uzma]{aaliaimam.uzma@hdr.mq.edu.au}
\author{T. Zafar
\textsuperscript{\orcidlink{0000-0003-3935-7018}}}
\affiliation{School of Mathematical and Physical Sciences, Macquarie University, NSW 2109, Australia}
\alsoaffiliation{Astrophysics and Space Technologies Research Centre, Macquarie University, Sydney, NSW 2109, Australia}

\author{T. Mukherjee
\textsuperscript{\orcidlink{0009-0004-7639-869X}}}
\affiliation{School of Mathematical and Physical Sciences, Macquarie University, NSW 2109, Australia}
\alsoaffiliation{Astrophysics and Space Technologies Research Centre, Macquarie University, Sydney, NSW 2109, Australia}

\author{T. Nanayakkara
\textsuperscript{\orcidlink{0000-0003-2804-0648}}}
\affiliation{Centre for Astrophysics and Supercomputing, Swinburne University of Technology, PO Box 218, Hawthorn 3122, VIC, Australia}

\author{A. Hopkins
\textsuperscript{\orcidlink{0000-0002-6097-2747}}}
\affiliation{School of Mathematical and Physical Sciences, Macquarie University, NSW 2109, Australia}
\alsoaffiliation{Astrophysics and Space Technologies Research Centre, Macquarie University, Sydney, NSW 2109, Australia}

\author{E. Wisnioski
\textsuperscript{\orcidlink{0000-0003-1657-7878}}}
\affiliation{Research School of Astronomy and Astrophysics, Australian National University, Canberra, ACT 2611, Australia}
\alsoaffiliation{ARC Centre of Excellence for All Sky Astrophysics in 3 Dimensions (ASTRO 3D)}

\author{S. M. Croom
\textsuperscript{\orcidlink{0000-0003-2880-9197}}}
\affiliation{Sydney Institute for Astronomy, School of Physics, University of Sydney, NSW 2006, Australia}
\alsoaffiliation{ARC Centre of Excellence for All Sky Astrophysics in 3 Dimensions (ASTRO 3D)}

\author{C. Foster
\textsuperscript{\orcidlink{0000-0003-0247-1204}}}
\affiliation{School of Physics, University of New South Wales, Sydney, NSW2052, Australia}
\alsoaffiliation{ARC Centre of Excellence for All Sky Astrophysics in 3 Dimensions (ASTRO 3D)}

\author{C. D. P. Lagos
\textsuperscript{\orcidlink{0000-0003-3021-8564}}}
\affiliation{International Centre for Radio Astronomy Research (ICRAR), The University of Western Australia, Crawley, WA 6009, Australia}
\alsoaffiliation{ARC Centre of Excellence for All Sky Astrophysics in 3 Dimensions (ASTRO 3D)}

\author{J. T. Mendel
\textsuperscript{\orcidlink{0000-0002-6327-9147}}}
\affiliation{Research School of Astronomy and Astrophysics, Australian National University, Canberra, ACT 2611, Australia}
\alsoaffiliation{ARC Centre of Excellence for All Sky Astrophysics in 3 Dimensions (ASTRO 3D)}

\author{K. E. Harborne
\textsuperscript{\orcidlink{0000-0002-2043-7985}}}
\affiliation{Institute for Computational Cosmology, Durham University, South Road, Durham DH1 3LE, UK}
\alsoaffiliation{Centre for Extragalactic Astronomy, Durham University, South Road, Durham DH1 3LE, UK}
\alsoaffiliation{Department of Physics, Durham University, South Road, Durham DH1 3LE, UK}

\author{L. M. Valenzuela
\textsuperscript{\orcidlink{0000-0002-7972-9675}}}
\affiliation{Universitäts-Sternwarte, Fakultät für Physik, Ludwig-Maximilians-Universität München, Scheinerstr. 1, 81679 München, Germany}

\author{J. Prathap
\textsuperscript{\orcidlink{0009-0004-0251-2672}}}
\affiliation{School of Mathematical and Physical Sciences, Macquarie University, NSW 2109, Australia}
\alsoaffiliation{Australia Telescope National Facility, CSIRO, Space and Astronomy, PO Box 1130, Bentley, WA 6102, Australia)}
\alsoaffiliation{Astrophysics and Space Technologies Research Centre, Macquarie University, Sydney, NSW 2109, Australia}

\author{P. Sharda
\textsuperscript{\orcidlink{0000-0003-3347-7094}}}
\affiliation{Faculty of Science Sterrewacht Observatory Research Observational cosmology}

\author{X. Lyu
\textsuperscript{\orcidlink{0009-0005-6537-9597}}}
\affiliation{School of Mathematical and Physical Sciences, Macquarie University, NSW 2109, Australia}
\alsoaffiliation{Australia Telescope National Facility, CSIRO, Space and Astronomy, P.O. Box 76, Epping, NSW 1710, Australia}

\author{A. Mailvaganam
\textsuperscript{\orcidlink{0009-0003-1221-1630}}}
\affiliation{School of Mathematical and Physical Sciences, Macquarie University, NSW 2109, Australia}
\alsoaffiliation{ARC Centre of Excellence for All Sky Astrophysics in 3 Dimensions (ASTRO 3D)}
\alsoaffiliation{Astrophysics and Space Technologies Research Centre, Macquarie University, Sydney, NSW 2109, Australia}

\author{S. Carlson
\textsuperscript{\orcidlink{0009-0005-2766-2850}}}
\affiliation{School of Mathematical and Physical Sciences, Macquarie University, NSW 2109, Australia}
\alsoaffiliation{Astrophysics and Space Technologies Research Centre, Macquarie University, Sydney, NSW 2109, Australia}

\author{A. Raidani
\textsuperscript{\orcidlink{0000-0001-5980-0174}}}
\affiliation{School of Mathematical and Physical Sciences, Macquarie University, NSW 2109, Australia}

\author{A. J. Battisti
\textsuperscript{\orcidlink{0000-0003-4569-2285}}}
\affiliation{International Centre for Radio Astronomy Research (ICRAR), University of Western Australia, M468, 35 Stirling Highway, Crawley, WA 6009, Australia}
\alsoaffiliation{Research School of Astronomy and Astrophysics, Australian National University, Cotter Road, Weston Creek, ACT 2611, Australia}
\alsoaffiliation{ARC Centre of Excellence for All Sky Astrophysics in 3 Dimensions (ASTRO 3D), Australia}

\author{D. P. Triani
\textsuperscript{\orcidlink{0000-0002-4752-128X}}}
\affiliation{Institute for Theory and Computation, Harvard-Smithsonian Center for Astrophysics Cambridge, MA 02138, USA}
\alsoaffiliation{Center for Astrophysics \text{\textbar} Harvard \& Smithsonian, 60 Garden Street, Cambridge, MA 02138, USA}

\author{I. Breda
\textsuperscript{\orcidlink{0000-0002-3764-5780}}}
\affiliation{Instituto de Astrofísica e Ciências do Espaço-Centro de Astrofísica
da Universidade do Porto (IA-CAUP), Rua das Estrelas, 4150-762
Porto, Portugal}

\author{G. van de Ven
\textsuperscript{\orcidlink{0000-0003-4546-7731}}}
\affiliation{Department of Astrophysics, University of Vienna, T\"urkenschanzstra{\ss}e 17, 1180 Vienna, Austria}

\author{K. Grasha
\textsuperscript{\orcidlink{0000-0002-3247-5321}}}
\affiliation{Research School of Astronomy and Astrophysics, Australian National University, Canberra, ACT 2611, Australia}

\received {dd Mmm YYYY}
\revised  {dd Mmm YYYY}
\accepted {dd Mmm YYYY}
\published{22 September 202X}

 %% First letter not capped
\begin{document}

\begin{abstract}
We present a spectroscopic sample of \heii\ $\lambda1640$ \AA\ emitting galaxies detected within 56 fields of the Middle Ages Galaxy Properties with Integral Field Spectroscopy (MAGPI) survey over 2 $\lesssim z \lesssim$ 4. Leveraging VLT/MUSE integral field spectroscopy, we performed spectral inspection combined with spatial analysis of the data cubes to verify \heii\ detections, resulting in 31 \heii\ emitters. After excluding three broad-line AGN, the final sample comprises 28 \heii\ emitters with a median M$_\mathrm{UV} \approx$ -21.2. Comparison of rest-frame UV emission line diagnostics with photoionisation models shows that the line ratios are consistent with sub-solar metallicity star-forming systems with moderately high ionisation parameters, with no compelling evidence of dominant AGN activity. Interestingly, $\sim$ 77\% of the \heii\ emitters exhibit double-peaked \lya\ profiles, prompting us to explore their connection with \heii. Our results indicate correlations between \heii\ and \lya\ properties, suggesting that hard ionising radiation fields in metal-poor star-forming systems are linked to the neutral gas conditions governing \lya\ escape. Finally, we identify four sources showing \lya\ and \heii\ emission without any UV metal lines, making them potential candidates for chemically primitive systems with very low metallicities. 
\end{abstract}

\section{INTRODUCTION}
\label{sec:int}

%\subsection{A subsection}
One of the major challenges in astronomy is identifying the nature of the primary drivers behind cosmic reionisation. It has become widely accepted that reionisation was primarily driven by low-mass star-forming galaxies \citep{2024MNRAS.527.6139S}, with a potentially small contribution from AGN \citep{dsilva2025dissectingreionisationcosmicstar}, provided that a sufficient fraction of Lyman-continuum photons ($\lambda < 912$\,\AA) were produced and allowed to escape into the intergalactic medium (IGM), leading to the ionisation of neutral hydrogen by $z \sim 6$ \citep{2010Natur.468...49R,2015ApJ...802L..19R,2015ApJ...803...34B}. Recent studies also suggest that reionisation likely ended late by $z \sim 5.3 - 5.5$ \citep{2015MNRAS.447.3402B, 2019ApJ...882..178K, 2021ApJ...917L..37C, 2022MNRAS.514...55B}.

%A key uncertainty, however, is how much of this ionising radiation escaped from galaxies into the intergalactic medium (IGM). Some studies suggest that the escape fraction, $f_\mathrm{es}$c, is expected to be small for typical EoR (Epoch of reionisation) systems, as the gas column density that fuels star formation also efficiently absorbs LyC photons \citep[e.g.][]{2013MNRAS.431.2826F, Xu_2016}. Thus, whether LyC photons escape depends on the potential of stellar feedback to create channels through the interstellar medium \citep[ISM;][]{2014ApJ...788..121K}. Recent simulations have shown that a modest $f_\mathrm{es}$c of $\sim 10-20\%$, achieved from a large population of low-mass, LyC leaking galaxies could be sufficient to re-ionize the Universe without considering any contributions from other ionising sources such as active galactic nuclei \citep[AGN;][]{Finkelstein_2012,2015ApJ...810...71F,2013ApJ...768...71R,2015ApJ...802L..19R}. Moreover, \citet{2019ApJ...879...36F} report that reionisation is achievable with a $f_\mathrm{es}$c < 5\%, provided favourable ionising photon production and a minor contribution from AGNs. Whereas, some observational inferences at $4.5 < z < 9.0$ also point to very low values, with median escape fractions of only $\lesssim$ 1-3\% \citep[see][]{2025arXiv250508870P}.

Another proposed theory involves the contribution of the first generation of metal-free (Population\,III, hereafter Pop\,III) stars, formed from primordial gas and expected to be more massive and extremely high in temperature than later-generation stars \citep[see][and references therein]{2004ApJ...605..579Y,2010ApJ...724..341H, 2023ARA&A..61...65K}. These extreme properties enable them to produce exceptionally hard ionising spectra \citep{2000ApJ...528L..65T,2001ApJ...550L...1T,2001PhR...349..125B,2002A&A...382...28S,2012ApJ...745...50W}. Furthermore, observational studies have shown that high-redshift galaxies exhibit systematically lower gas-phase and stellar metallicities, higher star formation rates, and younger stellar populations compared to their low-redshift counterparts \citep[e.g.][]{1996MNRAS.283.1388M, 2010ApJ...719.1168E,2013ApJ...776L..27H,2014ApJ...795..165S,2014MNRAS.440.2300C,2019MNRAS.487.2038C,2017NatAs...1E..52A, Sanders_2018}, consistent with conditions capable of producing hard radiation fields \citep{2014ApJ...795..165S, 2016ApJ...826..159S} and contributing to the onset of reionisation \citep{2011ARA&A..49..373B}. 

To investigate the presence of extremely massive and hot stars in early galaxies, one of the most informative tracers is the \heii\ $\lambda1640$\,\AA\ emission line, produced by the recombination of doubly ionised helium (He$^{2+}$).
%, especially the ultraviolet (UV) $\lambda1640$\,\AA\ and the optical $\lambda4686$\,\AA\ transitions. 
\heii\ emission requires photons with energies exceeding 54.4\,eV or ($\lambda \leq$ 228\,\AA),\ therefore, it is highly sensitive to hard ionising radiation. Furthermore, its intensity, spatial distribution, and kinematic properties provide valuable diagnostics of the underlying ionising sources. Given these extreme energy requirements, the most commonly invoked source of \heii-ionising photons is an active galactic nucleus (AGN). However, both observations and theoretical models suggest that star-forming galaxies lacking AGN signatures can also generate such hard radiation, particularly at $z \gtrsim 2$ \citep{2012MNRAS.421.1043S, 2022ApJS..262...37J}. In this context, narrow \heii\ lines are typically associated with nebular emission \citep[e.g.][]{2010ApJ...719.1168E}, whereas broader profiles indicate contributions from strong stellar winds of massive stars, or stripped atmospheres of Wolf-Rayet (WR) stars \citep[e.g.][]{2003ApJ...588...65S}. 
%Notably, nebular \heii\ lines are observed more often in high-redshift galaxies than the local ones \citep[see][]{2011A&A...526A.128K, 2013A&A...556A..68C, 2015ApJ...801L..28K,2018MNRAS.480.1081K}, emphasizing the importance of studying the \heii\ emitters to gain insight into the nature of the ionising sources, physical conditions of the ISM, and the stellar populations driving the early stages of cosmic evolution.
%The most expected source of \heii-ionising photons is an active galactic nucleus (AGN). 
%This accretion process results in the release of a tremendous amount of energy capable of ionising He$^{+}$. 

A plausible explanation for the observed \heii\ emission line in high-redshift galaxies is the contribution from Pop\,III stars. They are speculated to begin forming during cosmic dawn \citep[$z \sim$ 20 - 30;][]{2023ARA&A..61...65K}, however, cosmological simulations \citep{2020MNRAS.497.2839L,2022ApJ...935..174S,2023MNRAS.522.3809V} and semi-analytical models \citep{Visbal_2020} suggest that pristine gas reservoirs hosting Pop\,III stars could have lasted until $z \sim $ 6-10. The absence of metals in primordial gas, combined with the presence of a Lyman-Werner background during the epoch of reionisation, favours the formation of extremely massive Pop\,III stars \citep{1902RSPTA.199....1J,2004ARA&A..42...79B,2025arXiv251111314V} with a top-heavy initial mass function \citep[IMF;][]{2002A&A...382...28S}, capable of producing copious \heii-ionising photons, although their direct observational confirmation remains elusive.

Furthermore, the simultaneous presence of \heii\ and \lya\ $\lambda1215.67$\,\AA\ emission has long been recognised as a potential signature of both Pop\,III star formation and cooling radiation from the cold accretion of pristine gas \citep{2001ApJ...550L...1T, 2003A&A...397..527S, 2010A&A...523A..64R, 2013A&A...556A..68C}. This association between dual \heii+\lya\ emission and Pop\,III activity has also been highlighted by others \citep[e.g.][]{2001ApJ...550L...1T, 2003A&A...397..527S, Johnson_2009, 2010A&A...523A..64R}. An indirect way to differentiate the \heii+\lya\ emission from Pop\,III, Pop\,I, or cooling radiation, is to compare their intrinsic \heii/\lya\ luminosity ratios \citep{2003A&A...397..527S, 2006ApJ...646L..33Y}. Although this ratio is not observationally reliable, as \lya\ escape fraction depends strongly on geometry and dynamical state of the system, or the presence of dust that significantly deviates from the intrinsic \heii/\lya\ ratio \citep[see][]{2013A&A...556A..68C,2022MNRAS.517....1S}. In addition, \citet{2013A&A...556A..68C} highlight that WR stars, AGN, and supernovae-driven winds commonly show the dual \heii+\lya\ emission lines.

In addition to Pop\,III stars, WR stars are also considered a source of \heii\ emission. These evolved, massive stars have lost most of their hydrogen envelopes through powerful winds, producing broad emission lines from helium, carbon, oxygen (WC/WO), or nitrogen (WN; \citealt{1996MNRAS.281..163S, 1998MNRAS.296..367C}). 

%A hallmark of classical WR stars is the presence of blended “WR bumps,” notably the blue ($\sim4650$\,\AA) and red ($\sim5808$\,\AA) features produced by fast-moving winds. However, similar spectral signatures can arise from a broader “WR phenomenon,” including fast expanding hot plasma, white dwarf mergers, or low-mass stars exposing hot cores during the planetary nebula phase \citep[e.g.,][]{1981SSRv...28..227V, 2007ApJ...654.1068M, 2011A&A...535A..56G, 2019Natur.569..684G}. Consequently, the detection of WR bumps alone does not uniquely identify classical WR stars as the ionising source \citep{2024AzAJ...19b..73M}.
While helium-burning WR stars are hot enough to emit \heii-ionising photons, their optically thick winds trap this radiation \citep{2022arXiv221105424S}. With decreasing metallicity, these winds weaken \citep{2005A&A...442..587V}, allowing hard photons to escape before losing their characteristics as WR stars \citep{2023A&A...670A..83S}. In the low-$Z$ regime, WN/WO transition stars, with very high temperatures and weaker winds produce significant \heii-ionising flux. This makes them potential drivers of the strong \heii\ emission observed in high-redshift galaxies, despite remaining largely obscured \citep{2025A&A...703L..11G}.

Beyond classical WR stars, several alternative stellar and non-stellar mechanisms have been proposed for nebular \heii\ emission. Binary evolution provides a pathway in which stripped stars formed through mass transfer or envelope removal can sustain long-lived helium-core burning phases and are predicted to produce more \heii-ionising photons than single massive stars \citep{1967ZA.....65..251K,1967AcA....17..355P,2002ApJ...573..283P,2008AIPC..990..230D,2019A&A...629A.134G}. At low-metallicity, binary populatioxn synthesis models further indicate that rotation-driven quasi-homogeneous evolution (QHE) can significantly enhance the hard ionising output, making this particularly relevant at high redshift \citep{2012MNRAS.419..479E}. High-mass X-ray binaries (HMXBs) have also been proposed as a potential source of He$^{+}$-ionising photons, with some studies suggesting they can reproduce the observed \heii\ strengths and their dependence on metallicity in low-metallicity galaxies \citep{2019A&A...622L..10S}, while others find that typical HMXB populations are insufficient to power the observed emission, indicating they are unlikely to be the dominant source \citep[e.g.][]{2020MNRAS.494..941S}. Finally, fast radiative shocks driven by stellar feedback or AGN activity can generate extreme-UV and soft X-ray photons capable of ionising He$^{+}$ \citep[e.g.][]{1996ApJS..102..161D,2005ApJS..161..240T,Allen_2008,2016ApJ...833..136J}, but their overall contribution appears sub-dominant relative to stellar sources and remains poorly constrained, particularly as a function of metallicity \citep{2012MNRAS.427.1229I,2019A&A...624A..89N}.

Extensive studies have been conducted to investigate the genesis of \heii\ emission by observing the UV/optical characteristics of both high- \citep[e.g.][]{2013A&A...556A..68C, 2016ApJ...826..159S, 2018ApJ...859..164B, 2019A&A...624A..89N, 2020A&A...636A..47S, 2025A&A...702A...3G} and low-redshift \citep[e.g.][]{2017MNRAS.472.2608S} galaxies using ground- and space-based telescopes. Young, low-$Z$, highly star-forming systems provide the most favourable environments for identifying the mechanisms powering this emission. At high redshift ($z \gtrsim 1.3$), \heii\ and other UV lines are redshifted into the optical, making them accessible to ground-based optical spectrographs such as the Multi Unit Spectroscopic Explorer \citep[MUSE;][]{2010SPIE.7735E..08B}. With its wide field
of view (FoV)  and advanced Integral Field Spectroscopy (IFS), MUSE can spatially resolve the peaks of \lya,\ and can also map the spatial distribution of \heii,\ and other rest-frame UV lines (e.g. \ciii, \oiii, \siiii, \civ, \nv)\ providing comprehensive information regarding the physical, chemical conditions and gas kinematics. A recent MUSE-based study has found $\sim$\,50\% of \heii+\lya\ sample to be exhibiting double-peaked \lya\ profiles \citep{2025A&A...702A...3G}. 

This work presents a sample of \heii\ $\lambda1640$\,\AA\ emitters at $z \sim 2-4$ identified in MUSE observations as part of the Middle Ages Galaxy Properties with Integral Field
Spectroscopy \citep[MAGPI;][]{2021PASA...38...31F} Survey.  We use rest-frame UV line ratio diagnostics to distinguish stellar from AGN-driven ionisation. A large fraction of our \heii\ emitters also exhibit double-peaked \lya\ profiles \citep{2026PASA...43...21M}, motivating us to explore the presence of \heii\ emission and its link with the mechanisms influencing the \lya\ spectral features.
%Therefore, our study explores the connection between \heii\ emission and the emergence of double-peaked \lya\ line.

This paper is organised as follows: \S \ref{data} describes the sample selection and emission-line fitting procedure of the sample. \S  \ref{results} presents the main results, line ratio diagnostics, their comparison with photoionisation models, and the \lya-\heii\ connection. \S \ref{sec:discussion} provides a discussion of the results, followed by \S \ref{conclusion}, which summarises our findings and conclusions. Throughout this paper, we assume cosmological parameters of $H_0=70\,$km\,s$^{-1}$\,Mpc$^{-1}$, $\Omega_M=0.3$,
$\Omega_\Lambda=0.7$ and $\Omega_{\rm{k}} = 0$.

\section{DATA and METHODOLOGY}
\label{data}
\subsection{The MAGPI Survey}

This work utilised IFS data from the MAGPI survey, a Large Program on MUSE on the European
Southern Observatory’s (ESO) Very Large Telescope (VLT), targeting 56 fields within the Galaxy and Mass Assembly \citep[GAMA;][]{2011MNRAS.413..971D} G12, G15, and G23 regions. All 56 MAGPI fields are used in this analysis.
%MAGPI survey uses . MUSE operates in the visible wavelength range with simultaneous spectral coverage from 4750-9350 \AA.\ It has 24 identical integral field units (IFUs) working together to capture a wide field of view. This instrument features a Narrow Field Mode (NFM) with a field of view of $7.5'' \times 7.5''$ sampled at $0.025''$ /pixel and 
The MUSE data  are taken using a Wide Field Mode (WFM) with a FoV covering $1' \times 1'$ with a spatial sampling of $0.2''$ per pixel.
The survey employs the nominal mode (data are taken with the blue cut-off filter in place), providing a wavelength coverage ranging from 4700-9350 \AA, with a dispersion of 1.25 \AA. In addition, it has a median spectral resolution of R $\sim3000$ which is adequate for precise measurements of emission and absorption lines, gas kinematics, and redshifts. MUSE observations are carried out with VLT's ground layer adaptive optics (GLAO) system to correct for atmospheric turbulence, further enhancing the spatial resolution. The MUSE data reduction procedure is described in detail in \citet{2021PASA...38...31F} and Mendel et al. (in prep). The MAGPI emission-line products and their associated measurements are described in \citep{Battisti2026MAGPI}. The detection and identification techniques of faint, distant background sources within the 3D data cubes are presented in \citet{2026PASA...43...21M}. 

\subsection{Sample Selection}

We selected galaxies with redshifts $1.87 \lesssim z \lesssim 4.70$, ensuring that
the \heii\ $\lambda 1640$ emission line falls within the observed spectral range of the MUSE. This criterion resulted in a subsample of 624 galaxies. We note that the $5800-5970$ \AA\ interval is affected by the MUSE notch filter, which can obscure potential detections of \heii\ for sources whose redshifted line falls within this range. To process our sample, we developed a pipeline using the MUSE Python Data Analysis Framework \citep[\texttt{MPDAF\footnote{\url{https://github.com/mpdf/mpdf}}};][]{2016ascl.soft11003B,2017arXiv171003554P}. %an astronomical data analysis tool designed especially to handle MUSE-like IFU data.
The spectral analysis pipeline consists of the following steps: 

\begin{itemize}
\item For each source, we utilised 1D spectra extracted using circular apertures with radii of $0.6''$, $0.8''$, and $1.0''$. We then converted the wavelengths from air to vacuum using the 
\texttt{pyasl.airtovac2} module from the \texttt{PyAstronomy} package.
%\item The pixel indices were then converted into a wavelength  array. The resulting air wavelengths were then converted to vacuum wavelengths 

\item We modeled the continuum with a 5$^\mathrm{th}$ order polynomial using \texttt{poly\_spec}, and the fitted continuum was subtracted from the observed flux density to obtain the continuum-subtracted spectra.

\item We used the [\ciii\ $\lambda1907$\,\AA\ emission line to determine the systemic redshift, as it provides reliable spectroscopic
redshift confirmation at high redshift \citep{2018MNRAS.478L..60V}, and converted the spectra to the rest-frame.

\item However, for sources where \ciii\ was absent or undetected (e.g., due to the Na notch filter gap), we used other UV emission lines to estimate the systemic redshift and confirm the presence of \heii.

\item For each ultraviolet line of interest (e.g., \heii\ $\lambda1640$, \ciii\ $\lambda\lambda1907,09$, \oiii\ $\lambda\lambda1661,66$, and \siiii\ $\lambda\lambda1883,92$)\ we adopted the rest-frame wavelength as the systemic line centre and converted the wavelength axis into velocity space relative to this reference.

\item We examined the spectra within a velocity range of $\pm2000$ km\,s$^{-1}$ centred on the systemic line position, adequately covering the possible velocity offsets on both the red and blue sides of the expected line position. 
\end{itemize}

We visually inspected the velocity plots of all 624 galaxies to assess the data quality, including sky-line contamination and any residual calibration issues. For potential %sources with robust or uncertain 
\heii\ candidates, we looked into the corresponding data cubes and created narrow-band (NB) images to validate the detection of \heii, as well as other UV emission lines (see Fig.~\ref{fig:NBs_spec}). This process made sure that all the detected features were both spatially and spectrally consistent with the sources. 
%Moreover, a discussion of the sources where only \heii\ and \lya\ are detected, without any other metal lines, is provided in Section XX.

%\subsection{Overview of the MUSE Instrument}

%MAGPI survey uses the \textbf{Multi Unit Spectroscopic Explorer} \citep[MUSE;][]{2010SPIE.7735E..08B} instrument on the European Southern Observatory’s (ESO) Very Large Telescope (VLT). MUSE operates in the visible wavelength range with simultaneous spectral coverage from 4750-9350 \AA.\ It has 24 identical integral field units (IFUs) working together to capture a wide field of view. This instrument features a Narrow Field Mode (NFM) with a field of view of $7.5'' \times 7.5''$ sampled at $0.025''$ /pixel and a Wide Field Mode (WFM) with a FoV covering $1' \times 1'$ with a spatial sampling of $0.2''$ per pixel. In addition, it has a medium spectral resolution of R $\sim3000$\ which is adequate for precise measurements of emission and absorption lines, gas kinematics, and redshifts. The MAGPI survey uses the MUSE instrument with VLT's adaptive optics (AO) system to correct for atmospheric turbulence, further enhancing the spatial resolution.  MUSE stores data in a form of a \textit{minicube}— a 3D data cube with two spatial axes and the third axis corresponding to wavelength, providing a complete spectrum for each pixel.

\subsection{Aperture Selection}
\begin{figure*}[ht!]
    \centering
\includegraphics[width=1.02\linewidth]
{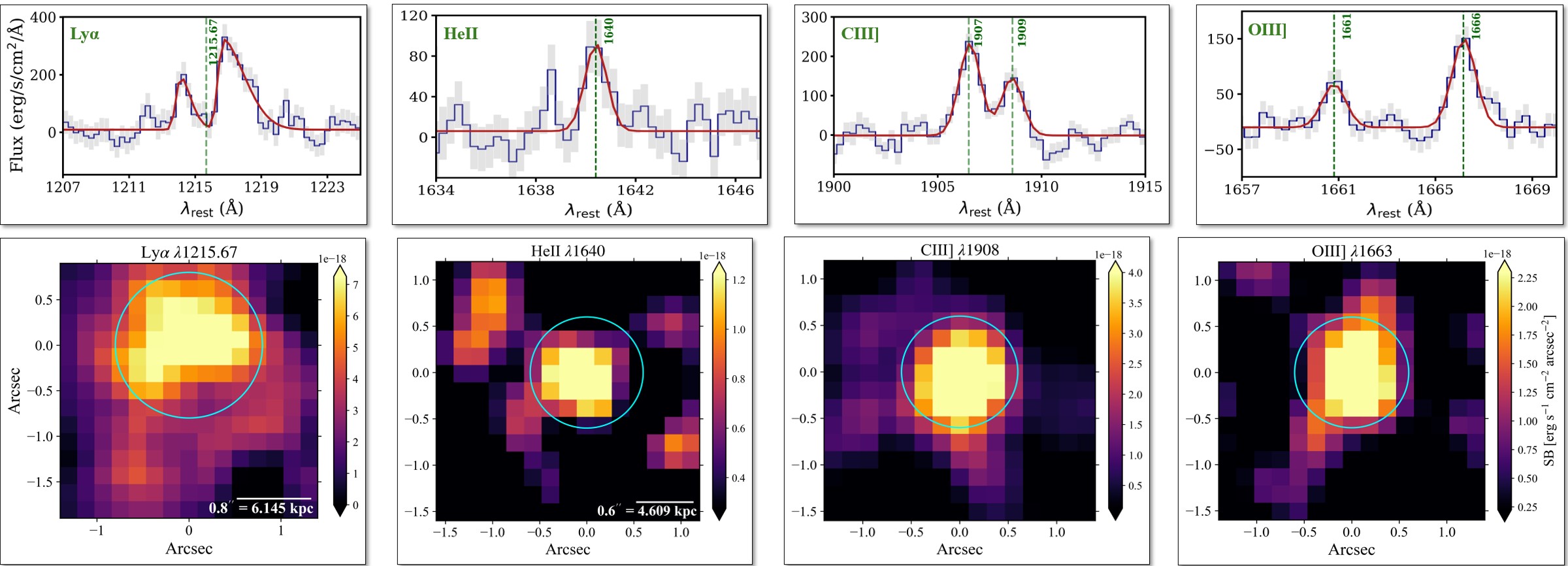}
    \caption{Top row: continuum-subtracted spectra around the rest-frame UV emission lines: \lya\ $\lambda1215.67$, \heii\ $\lambda1640$, \ciii\ $\lambda1908$, and \oiii\ $\lambda1663$ for MAGPI ID: 1533218212 at $z=3.0275$. The observed spectra are shown in blue, with the best-fit Gaussian models overlaid in red, and the grey shaded regions represent the $1\sigma$ flux uncertainties. The \lya\ spectrum and measurements are adopted from \citet{2026PASA...43...21M}, while the \heii, \ciii, and \oiii\ lines are fitted in this work. Bottom row: corresponding continuum-subtracted NB images for each emission line, with the selected circular apertures overlaid in cyan. The aperture radius is $0.6''$ for the \heii, \ciii, and \oiii, and $0.8''$ for the \lya\ image.}
\label{fig:NBs_spec}
\end{figure*}

Prior to the emission-line fitting, we visually inspected the MUSE data cubes of each source using \texttt{QFitsView\footnote{\url{https://www.mpe.mpg.de/~ott/QFitsView/}}} \citep{article} to assess the spatial extent of the emission and determine an appropriate aperture size. We constructed the corresponding continuum-subtracted NB images to verify that the detected emission-line features are spatially coincident with the target galaxy rather than arising from nearby companions or background noise, and to select an appropriate aperture. The aperture size is important for obtaining reliable emission-line fluxes, as an aperture that is too small may exclude part of the source emission, whereas a larger aperture may introduce additional background noise or contamination from nearby sources. Based on the observed spatial extent of the emission, we selected circular apertures individually for each source, with radii ranging from $0.6''$ to $1.0''$ (3--5 pixels at the MUSE spatial sampling of $0.2''/\mathrm{pixel}$). Across the redshift range of our sample, these angular radii correspond to physical radii of approximately $4.3$--$7.9$\,kpc. The selected extraction apertures are overlaid on the NB images in Fig.~\ref{fig:NBs_spec}, and the same apertures were used to extract the corresponding 1D spectra for subsequent emission-line fitting.

\subsection{Emission Line Fitting and \heii\ Emitters Identification}

We developed a custom-built line fitting tool to fit and analyse the observed UV emission lines, and calculate the line properties. The continuum of each spectrum was modelled using a 5$^\mathrm{th}$ order  polynomial, and the resulting continuum-subtracted spectra were used for line fitting. However, in instances where automatic continuum fitting produced an inaccurate estimate of local continuum level, we adopted a manual approach. In this method, we defined the local continuum by fitting a high-order polynomial using \texttt{np.polyfit} to selected line-free spectral regions surrounding the emission feature, providing a more reliable and locally constrained continuum estimate. 

Gaussian profiles were fitted to the continuum-subtracted spectra using \texttt{curve\_fit} via $\chi^{2}$ minimisation, where the uncertainties in the flux densities  were incorporated through the sigma parameter, and \texttt{absolute\_sigma=True} was used to preserve the absolute scaling of the observational uncertainties in the estimated parameter covariance matrix. For \heii\ along with emission doublets (e.g. \oiii\ $\lambda\lambda 1661,66$ and \siiii\ $\lambda\lambda 1883,92$), a single Gaussian profile was fitted within a narrow wavelength window centred around the expected line position, determined from the rest-frame wavelength and the systemic redshift of the source. Conversely, for the closely spaced \ciii\ $\lambda\lambda1907,09$ doublet, we employed a double Gaussian profile to provide a more accurate representation. However, for one source (MAGPI1521166247), where the \ciii\ doublet appears blended, a single Gaussian profile was adopted. The total flux for all isolated emission features 
(including \heii\ and the resolved components of \oiii\ and \siiii) was computed analytically from the best-fit single-Gaussian model. For the \ciii\ $\lambda\lambda 1907,09 $ doublet, the flux of each component was calculated from the corresponding Gaussian profile. 

The flux uncertainties for all emission lines (both singlet and doublets) were estimated by numerically integrating the variance spectrum within the ±5$\sigma$ of the fitted line using Simpson’s rule. Moreover, the full width at half maximum (FWHM) uncertainty was obtained by propagating the 1$\sigma$ uncertainty of the Gaussian width parameter derived from the covariance matrix of the fit, using the standard linear relation between the Gaussian dispersion and FWHM.

Observed-frame equivalent widths (EWs) were obtained by dividing the total flux by the local continuum flux density at the line centre and subsequently converted to the rest-frame EW using the systemic redshift. Uncertainties in EW were estimated via bootstrap
resampling method, generating multiple instances of the
observed spectrum by varying the flux within its uncertainties and then repeating the Gaussian fitting for each
instance. The standard deviation of the resulting EW distribution was adopted as the EW 1$\sigma$ uncertainty. In cases where the local continuum flux density at the line centre fell below the 1$\sigma$ noise level, the continuum value was substituted with its 1$\sigma$ error, and the resulting EW was reported as a lower limit.  %This practice ensured that spectra with undetected continua did not yield falsely high EW estimations.

Following the emission line measurements described above, we identified 31 sources (see \ref{appendix}), with a signal-to-noise ratio  $(S/N)>2$ \footnote{The $S/N$ represents the total integrated line flux divided by the propagated uncertainty.} for the \heii\ emission line, which we refer to as the parent sample. In this sample, three galaxies exhibit broad-line AGN features \citep[e.g.][]{2019A&A...624A..89N, 2020A&A...636A..47S} that we excluded from further analysis (see Fig.~\ref{fig:QS_Combined}).
%The remaining sources were further classified under gold and silver categories. 
%The remaining sample comprises 24 \heii\ emitters identified via a signal-to-noise threshold of  S/N$_\mathrm{He\,II} \geq 2$. 
Within the remaining 28 \heii\ emitters, 13 sources exhibit \heii\ with $S/N \geq3$ and show at least one more UV emission line (\oiii,\ \siiii,\ or both) in addition to \heii\ and \ciii\ which we use for the line ratio diagnostics discussed in \S~\ref{line ratio}. The measured UV emission line properties for the parent sample are summarised in Tables~\ref{tab:magpi_properties} and \ref{tab:line_fluxes}.

\section{RESULTS}
\label{results}
\subsection{UV Emission-Line Diagnostics}

%UV emission lines originate primarily from highly ionised gas in the ISM and CGM of galaxies, as well as from the nebular regions surrounding young massive stars or AGNs. These lines are produced when energetic photons ionise atoms or ions, and the subsequent recombination and collisional excitation processes give rise to radiative transitions that show distinct spectral features in the UV regime (typically between 1200-3000 \AA). Consequently, UV spectroscopy provides powerful constraints on the physical conditions of warm, highly ionised gas in galaxies \citep[e.g.][]{2019ApJ...878L...3B,2022ApJ...930..105S}. 

Rest-frame UV emission-line ratios provide critical diagnostics for the ionising radiation field and help break degeneracies present in classical optical diagnostics \citep[see][]{2016MNRAS.456.3354F,2016ApJ...833..136J,2018A&A...612A..94N,2024ApJ...962...95M}. These ratios probe key nebular properties, including electron density, gas temperature, and ionisation state, often extending to regimes above the critical densities of many strong optical lines \citep{2025ApJ...982...14H}. In addition, metal UV lines offer insights into the chemical enrichment and abundance patterns of early galaxies. This enables the study of star formation activity, radiation hardness, and possible AGN contributions across cosmic time \citep{2014MNRAS.445.3200S, 2016MNRAS.462.1757G,2016ARA&A..54..761S, 2016MNRAS.456.3354F}.

\begin{table*}[ht!]
\centering
\caption{He \textsc{ii} emission line properties of the MAGPI sample}
\setlength{\cmidrulewidth}{0.5pt}  
\label{tab:magpi_properties}
\begin{tabularx}{\textwidth}{l l l *{6}{>{\centering\arraybackslash}X}}
\toprule
 & & & & \multicolumn{4}{c}{He \textsc{ii} $\lambda1640$ \AA} & \\
\cmidrule(lr){5-8}
MAGPI ID & RA (hh:mm:ss) & Dec (dd:mm:ss) & $z$ & Flux & Error & FWHM (\AA) & FWHM$_\mathrm{error}$ (\AA) & M$_{\mathrm{UV,1500}}$ \\
\midrule
% Group 1: Bold Brown IDs (Subsample)
\textbf{\textcolor{brown}{2304252339}} & 22:46:33.35 & -31:17:50.97 & 2.1474 & 265 & 66 & 4.76 & 1.17 & -20.85 \\
\textbf{\textcolor{brown}{1521074276}} & 14:35:13.66 & -01:21:09.07 & 2.2990 & 154 & 38 & 6.06 & 1.02 & -22.82 \\
\textbf{\textcolor{brown}{1511184190}} & 14:26:17.30 & +01:43:49.48 & 2.6563 & 68 & 23 & 2.28 & 0.85 & -21.98 \\
\textbf{\textcolor{brown}{1208071299}} & 12:19:54.88 & -02:28:31.31 & 2.7174 & 224 & 67 & 4.98 & 1.44 & -22.95 \\
\textbf{\textcolor{brown}{1521166247}} & 14:35:12.43 & -01:21:14.87 & 2.8722 & 172 & 43 & 6.52 & 1.36 & -22.28 \\
\textbf{\textcolor{brown}{1201313233}} & 11:38:49.13 & -01:55:34.44 & 2.8923 & 105 & 35 & 3.14 & 1.02 & -19.64 \\
\textbf{\textcolor{brown}{1502159225}} & 14:10:32.35 & +02:33:45.09 & 2.9255 & 236 & 62 & 6.20 & 1.57 & -21.35 \\
\textbf{\textcolor{brown}{1204206169}} & 11:42:38.57 & -00:47:45.25 & 2.9400 & 160 & 40 & 6.93 & 1.67 & -20.15 \\
\textbf{\textcolor{brown}{1533218212}} & 14:51:52.27 & -00:08:27.94 & 3.0275 & 120 & 24 & 4.45 & 0.87 & -21.12 \\
\textbf{\textcolor{brown}{2307213114}} & 23:01:56.07 & -34:30:07.65 & 3.1737 & 159 & 53 & 4.39 & 1.24 & -20.21 \\
\textbf{\textcolor{brown}{1510087249}} & 14:25:08.16 & +02:07:01.13 & 3.1737 & 126 & 41 & 3.10 & 0.81 & —  \\
\textbf{\textcolor{brown}{1201250296}} & 11:38:49.97 & -01:55:21.84 & 3.2374 & 173 & 43 & 2.85 & 0.57 & -20.63 \\
\textbf{\textcolor{brown}{1512115241}} & 14:26:21.81 & -01:42:53.76 & 3.3547 & 101 & 20 & 5.92 & 0.92 & -20.92 \\
\midrule\hline
% Group 2: Full Sample
1207114256 & 12:08:01.07 & -02:28:49.23 & 1.8940 & 203 & 74 & 3.17 & 0.97 & -21.78 \\
1207213340 & 12:07:59.75 & -02:28:32.43 & 2.0739 & 290 & 73 & 13.11 & 3.14 & -21.67 \\
1529187287 & 14:41:28.20 & -00:05:40.26 & 2.1236 & 218 & 48 & 3.14 & 0.71 & -21.84 \\
1516220125 & 14:31:53.69 & +01:17:58.68 & 2.6870 & 128 & 40 & 5.44 & 1.68 & -21.93 \\
1502159347 & 14:10:32.35 & +02:34:09.49 & 2.7065 & 134 & 60 & 5.4 & 2.31 & -22.33 \\
1533222072 & 14:51:52.22 & -00:08:55.94 & 2.9059 & 65 & 28 & 4.03 & 1.35 & -20.07 \\
1522300299 & 14:36:14.88 & +00:48:21.66 & 2.9880 & 129 & 43 & 4.32 & 1.24 & — \\
1526272179 & 14:39:17.36 & +02:15:07.35 & 3.0285 & 108 & 32 & 4.39 & 1.14 & — \\
1512140153 & 14:26:21.63 & -01:43:14.56 & 3.2905 & 85 & 30 & 3.28 & 1.05 & — \\
1529060089 & 14:41:29.89 & -00:06:19.86 & 3.3342 & 106 & 42 & 4.75 & 1.84 & -20.51 \\
2306325152 & 23:00:13.70 & -34:28:12.91 & 3.3490 & 134 & 34 & 5.51 & 1.40 & -21.44 \\
1520248314 & 14:34:21.74 & -00:25:37.58 & 3.3590 & 70 & 23 & 2.91 & 0.97 & — \\
1209141341 & 12:22:24.57 & -01:22:21.53 & 3.5370 & 202 & 54 & 3.04 & 0.69 & — \\
1505089347 & 14:18:39.37 & -01:42:34.77 & 3.7955 & 389 & 124 & 7.03 & 1.65 & -20.11 \\
1511096052 & 14:26:18.48 & +01:43:21.88 & 3.8060 & 208 & 54 & 2.90 & 0.74 & -19.90 \\
\midrule\hline
% Group 3: AGN Section
\multicolumn{9}{l}{\textit{AGN}} \\
\hline 1533349182  & 14:51:50.53 & -00:08:33.94 & 2.1812 & — & — & — & — & -21.93  \\
1528079296 & 14:40:57.33 & 14:40:57.33 & 2.2122 & — & — & — & — & -24.29  \\
1530127219 & 14:48:35.46 & +02:56:33.24 & 2.3792 & — & — & — & — & -25.34 \\
\bottomrule\hline\hline
\end{tabularx}

\smallskip
\parbox{\textwidth}{\small \textbf{Notes:} All line fluxes and their corresponding errors are in units of $10^{-20}\,\mathrm{erg\, cm^{-2}s^{-1}}$. MAGPI IDs highlighted in gold are 13 \heii\ sources with \ciii\ along with at least one additional emission line (\oiii, \siiii, or both). The last section contains identified quasar/AGN-like sources. FWHM (\AA)\ and its associated uncertainty ($\mathrm{FWHM_{error}}$) represent observed-frame measurements. The final column lists the absolute UV magnitude (M$_\mathrm{UV}$) measured at 1500 \AA.}

\end{table*}

\begin{table*}
\centering
\caption{UV Emission line measurements of \heii\ emitters}
\label{tab:line_fluxes}
\setlength{\arrayrulewidth}{0.8pt}   
\setlength{\heavyrulewidth}{1.0pt}   
\setlength{\lightrulewidth}{0.8pt}    
\setlength{\cmidrulewidth}{0.8pt}     

\setlength{\tabcolsep}{11pt} 
\renewcommand{\arraystretch}{1.3} 

\begin{tabular}{c cc cc cc cc cc cc}
\hline\hline % Double line at top
MAGPI ID & 
\multicolumn{2}{c}{[C\,{\sc iii}]1907} & 
\multicolumn{2}{c}{C\,{\sc iii}]1909} & 
\multicolumn{2}{c}{O\,{\sc iii}]1661} & 
\multicolumn{2}{c}{O\,{\sc iii}]1666} & 
\multicolumn{2}{c}{Si\,{\sc iii}]1883} & 
\multicolumn{2}{c}{Si\,{\sc iii}]1892} \\

\cmidrule(lr){2-3} \cmidrule(lr){4-5} \cmidrule(lr){6-7} \cmidrule(lr){8-9} \cmidrule(lr){10-11} \cmidrule(lr){12-13}

 & Flux & Error & Flux & Error & Flux & Error & Flux & Error & Flux & Error & Flux & Error \\
\hline

2304252339 & 298 & 52 & 248 & 66 & 258 & 88 & 402 & 103 & — & — & — & — \\
1521074276 & 200 & 53 & 162 & 134 & — & — & 117 & — & 51 & 17 & 89 & 22 \\
1511184190 & 89 & 20 & 79 & 31 & — & — & 103 & 30 & — & — & — & — \\
1208071299 & 331 & 55 & 142 & 38 & — & — & 193 & 57 & 134 & 38 & 115 & 50 \\
1521166247 & 881 & 61 & — & — & 100 & 30 & 271 & 40 & — & — & — & — \\
1201313233 & 82 & 25 & 240 & 51 & — & — & — & — & 219 & — & — & — \\
1502159225 & 191 & 49 & 75 & 37 & — & — & — & — & 114 & 43 & — & — \\
1204206169 & 134 & 27 & 50 & 17 & 72 & 7.3 & — & — & 150 & — & — & — \\
1533218212 & 294 & 23 & 172 & 25 & 78 & 18 & 173 & 21 & — & — & 78 & — \\
2307213114 & 98 & 23 & 93 & 29 & — & — & 75 & — & — & — & — & — \\
1510087249 & 436 &110 & — & — & 113 & 42 & — & — & — & — & — & — \\
1201250296 & 230 & 36 & 220 & 60 & — & — & 154 & 46 & — & — & — & — \\
1512115241 & — & — & 209 & 78 & 44 & 15 & 150 & 39 & — & — & 68 & 19 \\

\hline\hline % Double line at bottom
\end{tabular}

\smallskip
\parbox{\textwidth}{\small \textbf{Note:} All line fluxes with corresponding uncertainties are in units of $10^{-20}\,\mathrm{erg\, cm^{-2}s^{-1}}$. For sources where the line detection is below the $S/N < 2.5$ threshold, flux values are reported as $3\sigma$ upper limits. In these instances, the uncertainty column is indicated by a dash (—).}
\end{table*}

\subsection{Gas-Phase Oxygen Abundance from Rest-Frame UV Diagnostics}

Gas-phase oxygen abundance (12 + log\ (O/H)), is a primary indicator of the chemical enrichment of the ISM. While the ``direct'' T$_\mathrm{e}$ method, utilizing the auroral \Oiii\ $\lambda4363$ \AA\ line, is the standard for accuracy, this feature is often too faint for detection at high redshift \citep{2019ApJ...887..168G}. Given the MUSE spectral coverage, rest-frame optical lines are unavailable for our high-$z$ sample. Nevertheless, nebular emission-line models predict that UV line fluxes are closely linked to the gas-phase metallicity of galaxies \citep{2017ApJ...840...44B, 2018ApJ...863...14B}. We, therefore, estimated gas-phase metallicities using rest-UV lines, adopting the calibration from \citet{2020ApJ...893....1B}. This calibration assumes solar abundances from \citet{2010Ap&SS.328..179G} and depletion factors from \citet{2013ApJS..208...10D}. The diagnostic combines \heii, \oiii, and \ciii\ emission lines, and provides an estimate of gas-phase oxygen abundance, with typical statistical uncertainties of $0.08$ dex. This calibration is a simplified functional representation of the full \cite{2018ApJ...863...14B} model grid, with abundances tightly correlated with direct interpolation, exhibiting a methodological scatter of $0.03$ dex. The third-order 2D polynomial fit is expressed as,

\begin{equation}
\begin{aligned}
12 + \log\ (\mathrm{O/H}) =\;& 6.88 - 1.13x - 0.46x^2 - 0.61y + 0.02y^2 \\
& - 0.32xy + 0.03xy^2 - 0.03x^3 - 0.04y^3 - \\
0.21x^2y,
\end{aligned}
\end{equation}
where x = log$_\mathrm{10}$ (\oiii\ $\lambda1666$/\ciii\ $\lambda1908$) and y = log$_\mathrm{10}$ (\heii\ $\lambda1640$/\ciii\ $\lambda1908$).

Among the 13 galaxies in our sample, 6 show $S/N\geq3$ across all the emission lines used for estimating gas-phase oxygen abundances with the \citet{2020ApJ...893....1B} calibration. We obtained oxygen abundances in the range 7.20 $\lesssim$ 12 + log\ (O/H) $\lesssim$ 7.71, with a typical (median) uncertainty of $\sim0.12$ dex. Compared to the metallicity range reported by \citet{2025A&A...702A...3G} for similar redshift systems (12 + log\ (O/H) $\sim$ 7.3 - 7.7), our sample extends slightly toward lower metallicities. Notably, one of our sources exhibits 12 + log\ (O/H) $\sim$ 7.20, placing it in close proximity to the extremely metal-poor local blue compact dwarf (BCD), I Zwicky 18 \citep{1972ApJ...173...25S}. IZw 18 is a well-studied high-redshift local analogue, also known to host \heii\ emission and has an oxygen abundance of $\sim$ 7.18 \citep[e.g.][]{1999ApJ...527..757I,2004ApJ...606..213T, 2016MNRAS.459.2992K}. A similar local analogue is the SBS 0335-052E, which possesses a metallicity of $\sim$ 7.2 - 7.3 \citep{2004ApJ...606..213T, 2005ApJ...632..210I,2006A&A...454..119P, 2009A&A...503...61I,2018MNRAS.480.1081K}. These comparable abundances suggest that our sources are potentially associated with chemically primitive environments, where such young, metal-poor star-forming systems are expected to produce the high-energy photons required to power He$^\mathrm{+}$ ionisation.

\subsection{Photoionisation Models by \citet{2016MNRAS.462.1757G}}
\label{line ratio}

To further constrain the physical and ionisation conditions of our sources, we compared the observed line ratios with the photoionisation models of \citet{2016MNRAS.462.1757G} (see Fig.~\ref{fig1:full}). These
models combine the \citet{2003MNRAS.344.1000B} stellar population synthesis (SPS) with the photoionisation code \texttt{CLOUDY} \citep[c13.03;][]{2013RMxAA..49..137F}, following the
framework of \citet{2001MNRAS.323..887C}, to self-consistently model the response of gas and dust to radiation. The resulting grids simulate emission from ensembles of \hii\
regions and diffused ionised gas, incorporating the effects of element
abundances, dust depletion, and ionisation structure. These models are well-suited for probing the ISM conditions and chemical enrichment of high-redshift ($2 \leq z \leq 9$) galaxies \citep{2016MNRAS.456.3354F, 2016MNRAS.462.1757G, Stark_2016, 2019A&A...624A..89N, Atek_2024}.

From the \citet{2016MNRAS.462.1757G} grid, we adopted three metallicity tracks, ($Z=0.0001,0.001$, $0.008$), where $Z_\odot\approx 0.01524$ \citep{Bressan_2012}.
For each track, we considered three carbon-to-oxygen ratios (C/O = 0.14, 0.20, 0.38, where C/O$_\odot =0.44$), suitable for low-$Z$ galaxies at $z\sim3$ \citep[see][]{2017NatAs...1E..52A,2017MNRAS.466.4403N,2023MNRAS.518.3985S}. The hydrogen density was fixed at n$_\mathrm{H}$ = 100\,cm$^{-3}$, close to the typical value ($\sim$ 250\,cm$^{-3}$), observed in high-redshift star-forming galaxies \citep[e.g.,][]{2018ApJ...858...99S}, and we set a dust-to-metal mass ratio of   $\xi_\mathrm{d}$ = 0.5. To account for the contribution of very massive, hot stars, we assumed an upper-mass cutoff of 300 M$\odot$, motivated by observations of very massive stars exceeding the classical $\sim150$ M$\odot$ upper mass limit \citep[e.g.][]{Crowther_2010,Schneider_2018}. The ionisation parameter (log U$_\mathrm{s}$) was allowed to vary freely along each model track.

For the line ratio diagnostics, we focused on the \heii, \ciii, \oiii, and \siiii\ features, selected based on their line strengths and wavelength coverage across the redshift range of our sample. The UV emission lines used are \heii\ $\lambda1640$, \ciii\ $\lambda1908$ = ([\ciii\ $\lambda1907$ + \ciii\ $\lambda1909$), \oiii\ $\lambda1666$, and \siiii\ $\lambda1888$ = (\siiii\ $\lambda1883$ + \siiii\ $\lambda1892$). Due to MUSE wavelength coverage, only a subset of our \heii\ emitters has all the other UV emission lines available. For each diagnostic diagram, sources are treated as detections only if all involved emission lines have $S/N \geq 2.5$, otherwise ratios involving any line below this threshold are treated as limits. In such cases, the noise spectrum constrained the error level, and we used the 3$\sigma$ error level as our upper limit to the flux.

\subsubsection{\textbf{\ciii /\heii\ vs [\oiii /\heii}}

%In this section, we focus on the diagnostic that use the line ratios \oiii /\heii\ versus \ciii /\heii\ (Figure \ref{fig1:c/he vs o/he}). 

\begin{figure*}[t]
    \centering
    \begin{subfigure}[b]{0.33\textwidth}
        \centering
        \includegraphics[width=\textwidth]{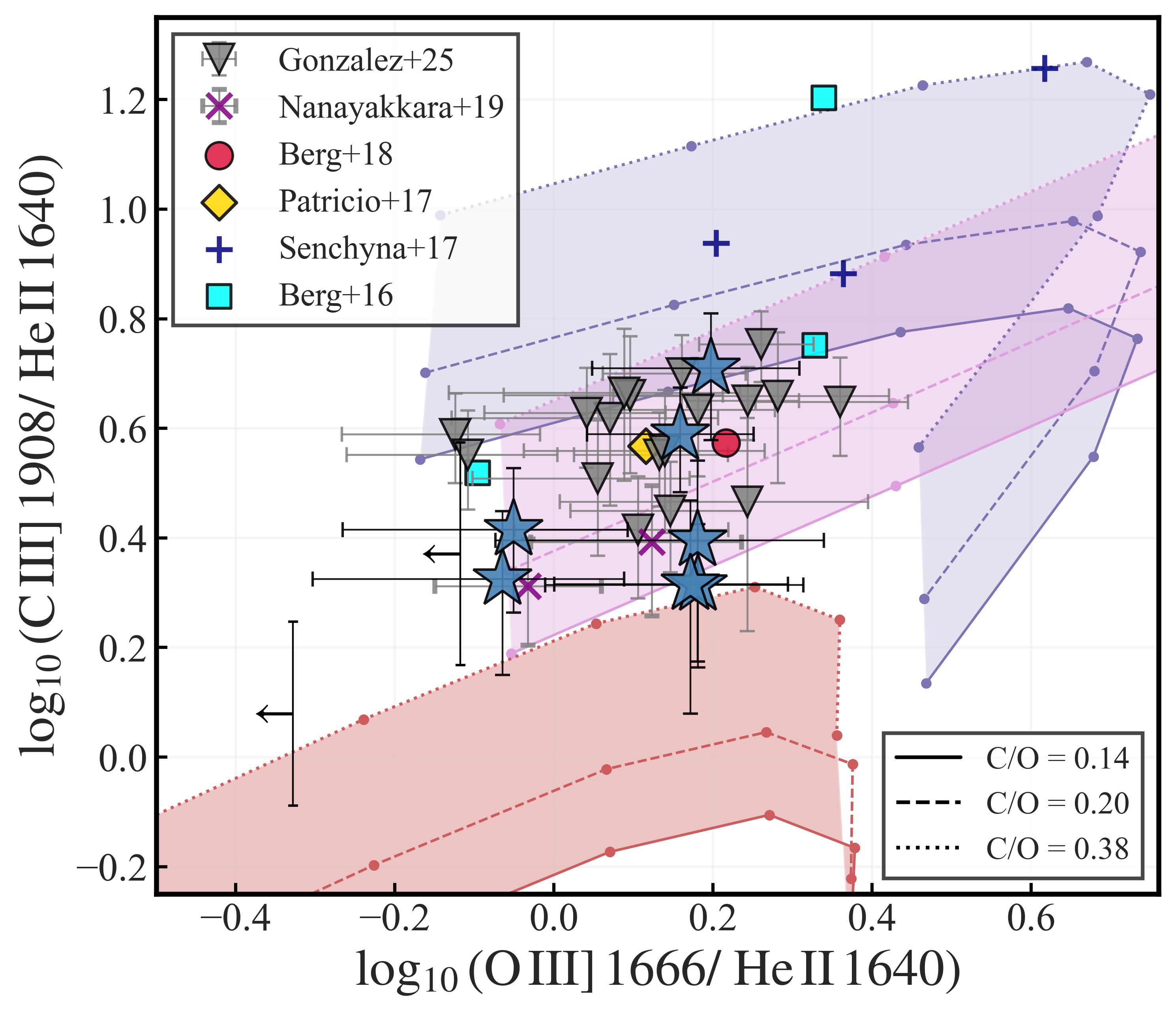}
        \subcaption{} % optional short caption
        \label{fig1:a}
    \end{subfigure}
    \hfill
    \begin{subfigure}[b]{0.33\textwidth}
        \centering
        \includegraphics[width=\textwidth]{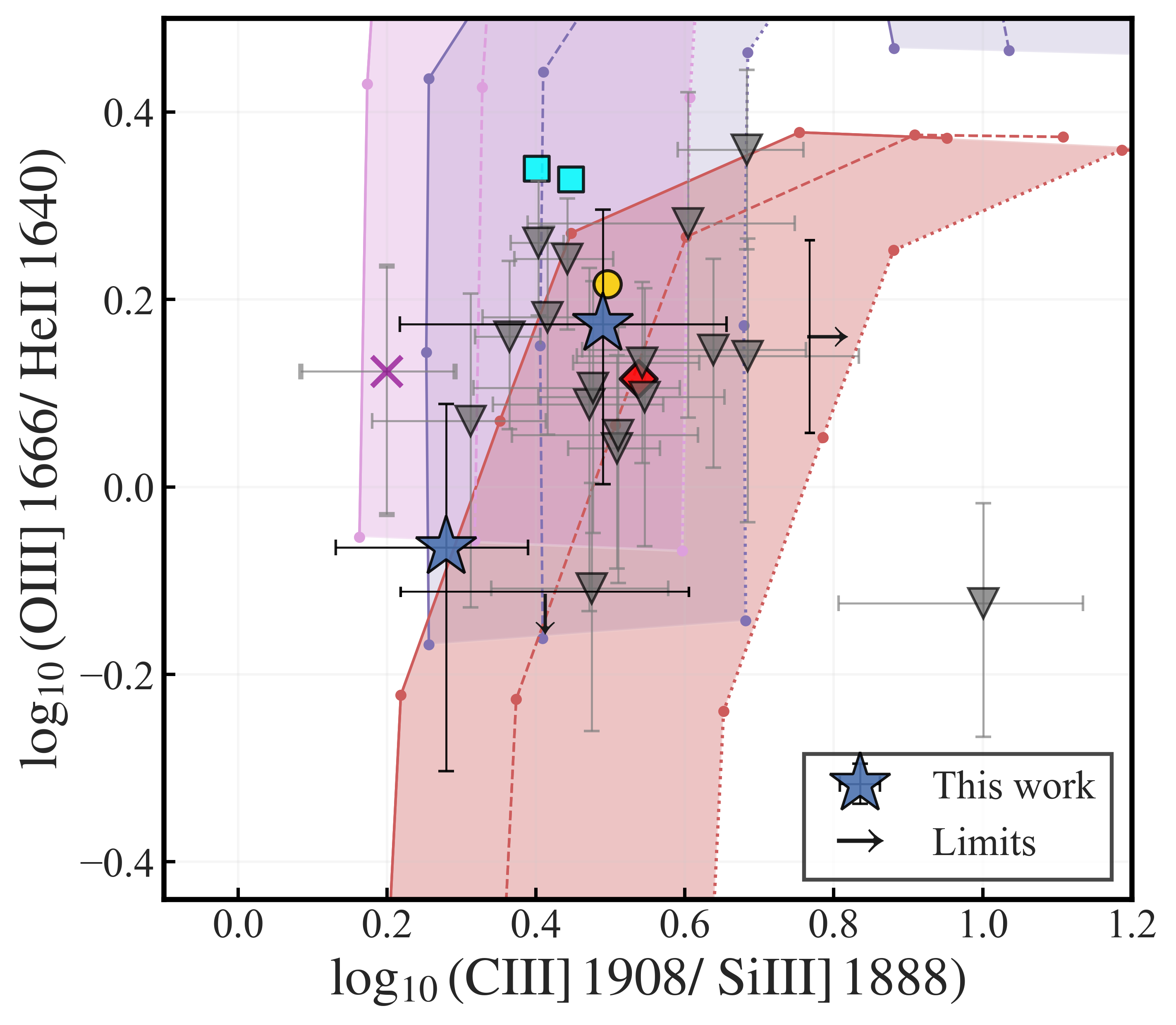}
        \subcaption{}
        \label{fig1:b}
    \end{subfigure}
    \hfill
    \begin{subfigure}[b]{0.33\textwidth}
        \centering
        \includegraphics[width=\textwidth]{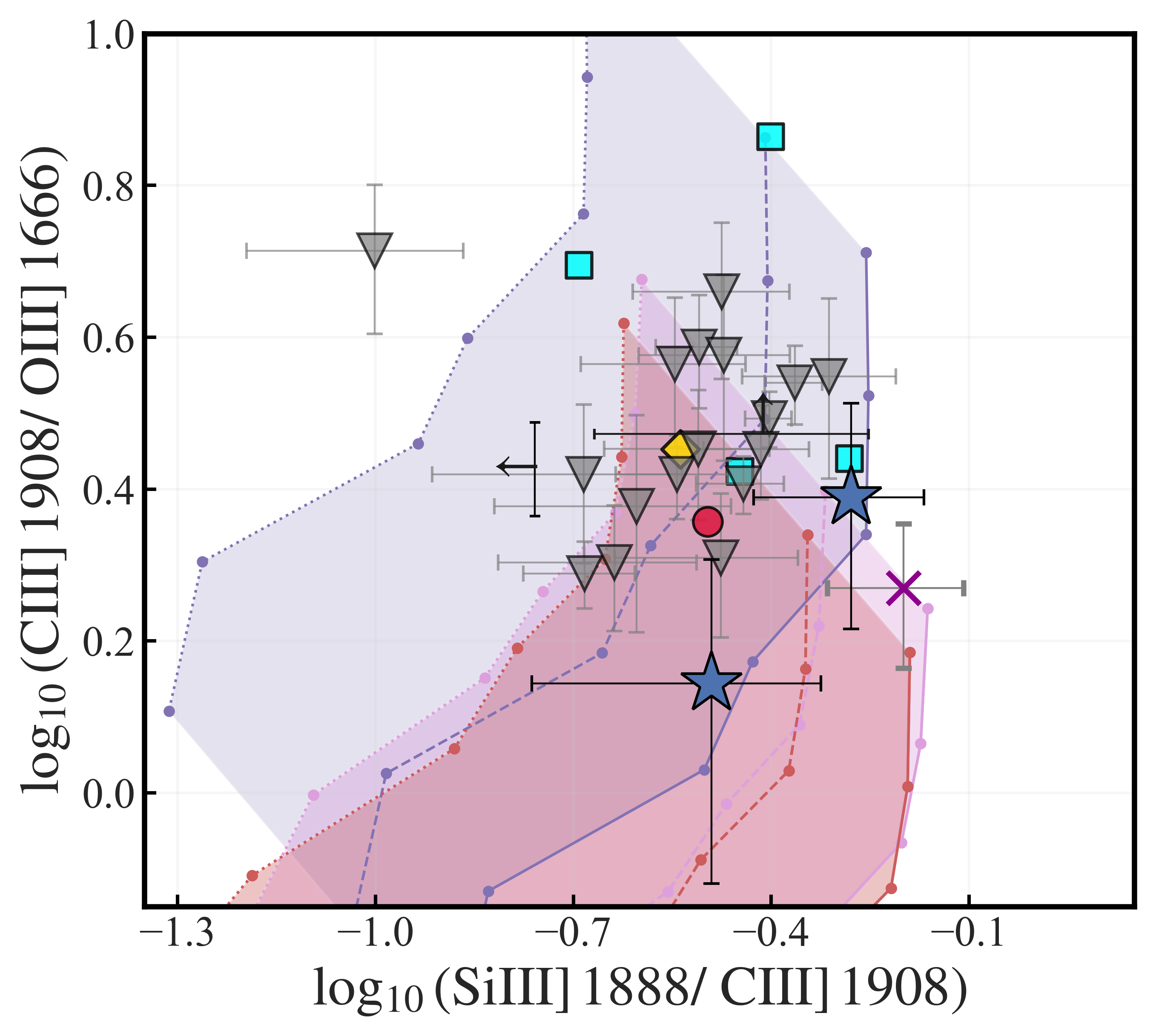}
        \subcaption{}
        \label{fig1:c}
    \end{subfigure}
    \caption{Rest-frame UV emission line ratios of the \heii\ emitters compared with photoionisation models from \citet{2016MNRAS.462.1757G}. Red, pink, and purple tracks correspond to $Z=0.0001,0.001$, and $0.008$ respectively. For each metallicity, the solid, dashed, and dotted curves represent the three adopted C/O ratios ($0.14 - 0.38$), and the region between the minimum and maximum C/O tracks is
shaded by the same colour. The plotted models assume $\xi_\mathrm{d}=0.5$ and m$_\mathrm{up}=300M_\odot$. Points along each track denote ionisation parameters ranging from log U$_\mathrm{s}$ = $-4$ to $-1$ in steps of $0.5$ dex. Ratios with $S/N \geq 2.5$ for relevant emission lines are shown as stars, whereas ratios involving any line with $S/N<2.5$ are represented as upper or lower limits (arrows). \textbf{Left panel:} \ciii /\heii\ vs \oiii/\heii\ diagnostic. \textbf{Middle panel:} \oiii/\heii\ vs \ciii/\siiii\ diagnostic. \textbf{Right panel:} \ciii/\oiii\ vs \siiii/\ciii\ diagnostic. Where available, comparison samples from \citep[$z\approx 0$;][] {2016ApJ...827..126B, 2017MNRAS.472.2608S}, \citep[$z\approx 2.2$;][] {2018ApJ...859..164B}, \citep[$z\approx 3.5$;][]{10.1093/mnras/stv2859}, and \citep[$z\sim 2-4$;][]{2025A&A...702A...3G, 2019A&A...624A..89N} are included.}
    \label{fig1:full}
\end{figure*}

The \ciii /\heii\ versus \oiii /\heii\ diagnostic (Fig.~\ref{fig1:a}; left panel) provides a probe of the ionising radiation field and the chemical enrichment in galaxies. The \oiii /\heii\ ratio is primarily sensitive to the ionisation parameter and the hardness of the radiation field, while \ciii /\heii\ reflects both metallicity and spectral hardness \citep{2018A&A...612A..94N}. Along each model track, the sequence of points corresponds to increasing ionisation parameter, typically progressing slightly diagonally from log U$_\mathrm{s} =-0.4$ (left) to -0.1 (right) in steps of 0.5 dex. For two sources, the \oiii\ emission line falls below the adopted $S/N$ threshold, resulting in upper limits on the \oiii/\heii\ ratio. Our sources are well reproduced by the \citet{2016MNRAS.462.1757G} models, with the majority of the sample lying along the $Z=0.001$ track and consistent with sub-solar abundance ratios of C/O $\sim 0.14 - 0.34$. Notably, two sources lie in close proximity to the lowest-$Z$ track ($Z=0.0001$) at log U$_\mathrm{s}$ $\approx-2$, indicating very metal-poor conditions. Compared to literature samples, our sources show significant overlap with high-redshift populations, such as the $z \sim 2-4$ MUSE \textit{Hubble} Ultra Deep Field sources \citep{2025A&A...702A...3G} as well as galaxies from \citet{2019A&A...624A..89N}, along with low-mass lensed systems at $z\approx 2.2$ \citep{2018ApJ...859..164B} and $z\approx 3.5$ \citep{10.1093/mnras/stv2859}. In contrast, local galaxies ($z \approx 0$) from \cite{2016ApJ...827..126B} and \cite{2017MNRAS.472.2608S} occupy a distinct region in the diagnostic characterised by higher \ciii/\heii\ ratios at a fixed \oiii/\heii\ ratio.

\subsubsection{\textbf{\oiii/\heii\ vs \ciii/\siiii}}

Fig.~\ref{fig1:b} shows \oiii/\heii\ versus \ciii/\siiii\ diagnostic. For two sources, one of the required emission lines falls below the adopted $S/N$ threshold. In one case, \siiii\ detection results in a lower limit on \ciii/\siiii\ while in the other, a weak \oiii\ line yields an upper limit. Along the model grid, log U$_\mathrm{s}$ increases from -4.0 at the bottom to -1.0 toward the upper right. Our sources, including limits, fall on the lowest-$Z$ track ($\approx$ 1/150th\ $Z_{\odot}$) with log U$_\mathrm{s} \sim -2.0$ to $-2.5$. Conversely, the $Z=0.001$ model track can also reproduce our line ratios with log U$_\mathrm{s}$ $\sim-3.0$ to -3.5 and carbon-to-oxygen ratios of C/O $\sim 0.19-0.25$. This is broadly consistent with the parameters found in our previous diagnostic (see Fig.~\ref{fig1:a}). However, this apparent preference is not unique, as the model tracks exhibit substantial overlap in this parameter space. 
%As in the other UV diagnostic plot, this also exhibits a metallicity-ionisation parameter degeneracy where a single line ratio can be observed with different ($Z$, log U$_\mathrm{s}$) combinations.

\subsubsection{\textbf{\ciii/\oiii\ vs \siiii/\ciii}}

In the \ciii/\oiii\ versus \siiii/\ciii\ diagnostic (Fig.~\ref{fig1:c}; right panel), the \siiii/\ciii\ ratio traces the Si/C abundance, while \ciii/\oiii\ is essentially responsive to the C/O ratio, $Z$, and the ionisation parameter. With other parameters fixed, the model tracks move diagonally from the lower right to the upper left, marking the increase in both $Z$ and the C/O ratio. Along each track, the log U$_\mathrm{s}$ varies from -4.0 (upper right) to -1.0 (lower left). Since all of these are metal lines, this plot jointly probes the chemical composition and the ionisation conditions of the gas. As \siiii\ is the weakest line, the observed \siiii/\ciii\ line ratio is treated as the upper limit. In addition, one source has a lower limit on \ciii/\oiii\ as \oiii\ did not meet the $S/N$ threshold. 

\subsection{Rest-frame UV Magnitude}
\label{magnitude}

\begin{figure}[ht!]
    \centering
\includegraphics[width=0.99\linewidth]
{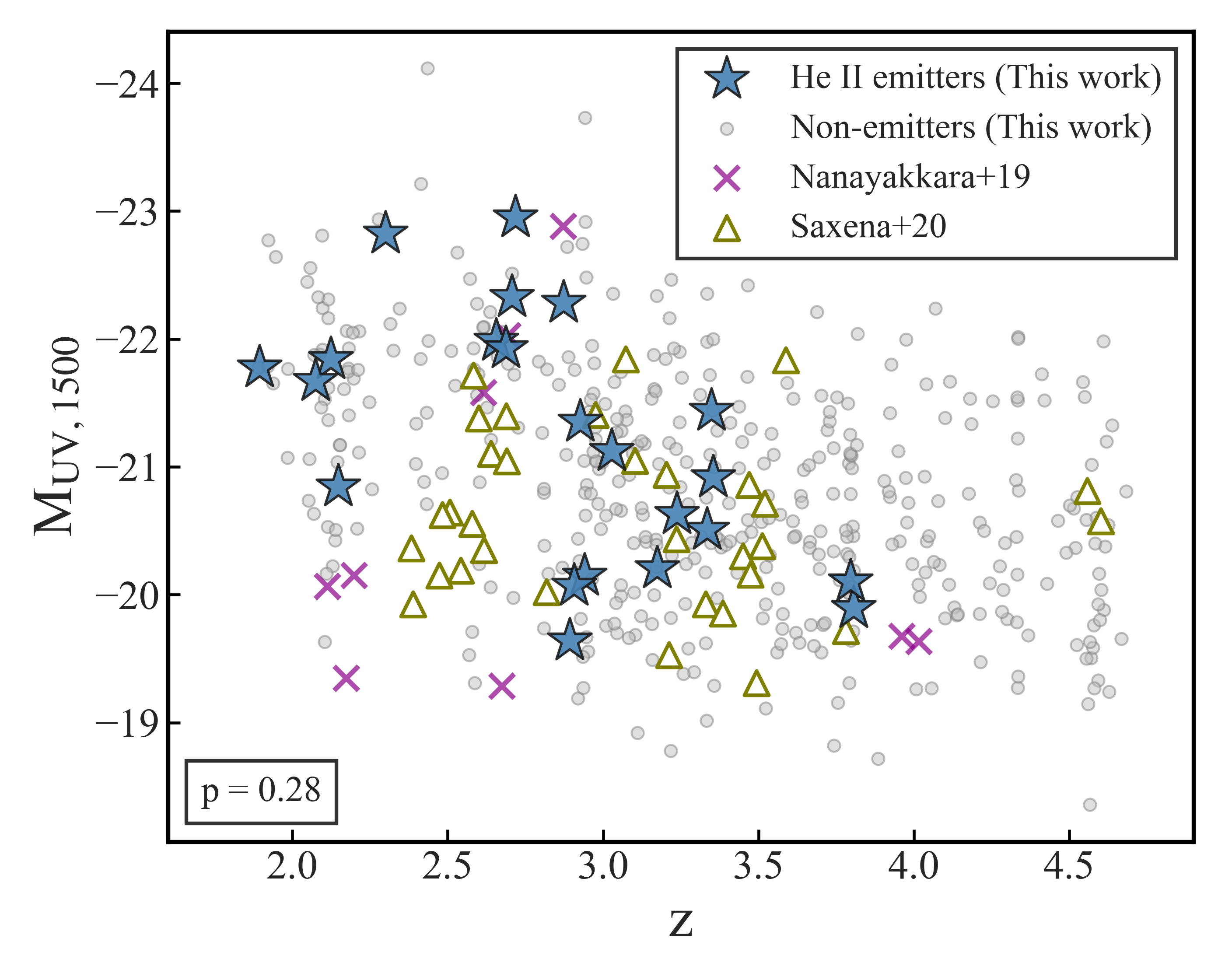}
    \caption{M$_{\mathrm{UV},1500}$ as a function of redshift for our \heii\ emitters (blue stars) and non–\heii\ emitters (grey dots), compared with literature \heii\ samples from \citet[magenta crosses]{2019A&A...624A..89N} and \citet[green triangles]{2020MNRAS.496.3796S}.}
    \label{fig:MUV-z}
\end{figure}

Figure~\ref{fig:MUV-z} illustrates the redshift evolution of the rest-frame UV magnitude (M$_\mathrm{UV}$) at 1500 \AA\ for galaxies with and without detected \heii\ emission. For comparison, we also include \heii\ emitters from \citet{2019A&A...624A..89N} and \citet{2020A&A...636A..47S}, spanning a redshift range similar to our sources. The M$_\mathrm{UV,1500}$ of our \heii\ emitters ranges from $-22.95$ to $-19.64$. Due to the lack of multi-band photometric coverage for our full sample, we derive these values using an approximate transformation formula based on the observed MUSE $i$-band magnitudes and source redshifts under a flat $\Lambda$CDM cosmology.
%These values were computed using the observed apparent MUSE i-band magnitudes and redshifts of the sources under a flat $\Lambda$CDM cosmology. 
We note that not all sources are shown in Fig.~\ref{fig:MUV-z}, as i-band magnitudes are unavailable for a subset of the sample. We find no significant differences between the \heii\ detected and non-detected galaxies in terms of their M$_\mathrm{UV}$, indicating that \heii\ emitters span a wide range of UV luminosities. To statistically assess this, we performed a two-sample Kolmogorov-Smirnov (K-S) test to compare the distributions of M$_\mathrm{UV}$ between the \heii\ detected and non-detected galaxy samples. The test yielded a K-S statistic of 0.21 and a corresponding p-value of 0.28. Since the $p$-value is above the significance threshold of 0.05, there is insufficient evidence to reject the null hypothesis that both samples originate from the same distribution. This result is consistent with previous findings \citep[see][]{2019A&A...624A..89N, 2020A&A...636A..47S}, which also reported no clear distinction between \heii\ emitting and non-emitting galaxies based on their UV magnitudes.

\subsection{Connection between \heii\ Emission and \lya\ Line Morphology}
\label{lya-heii}

\begin{figure}[ht!]
    \centering
\includegraphics[width=0.99\linewidth]
{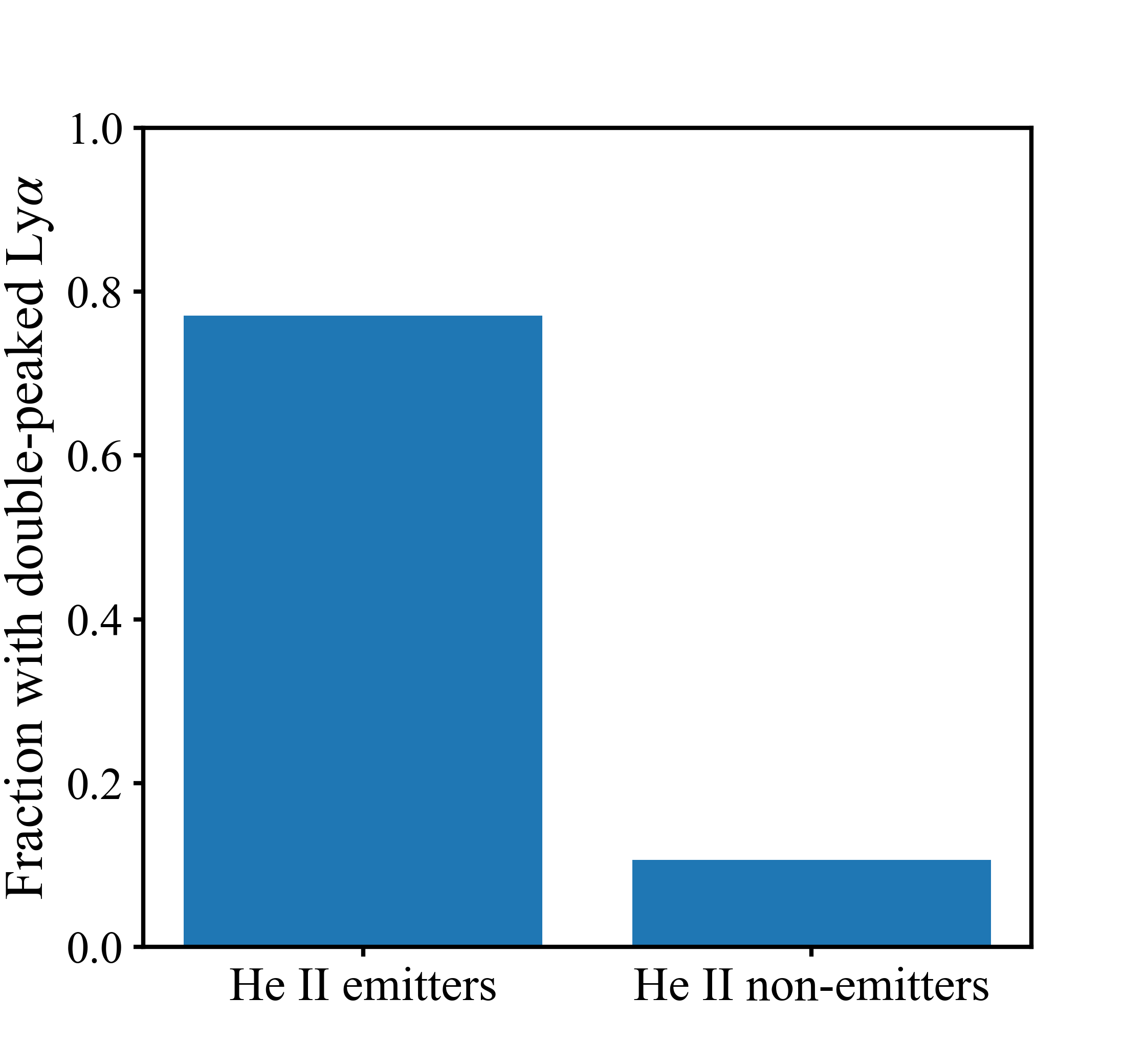}
    \caption{Fraction of galaxies exhibiting double-peaked \lya\ profiles among the \heii-emitting and non-\heii-emitting galaxy samples.}
    \label{fig:barchart}
\end{figure}

We find that a large fraction ($\sim 77\%$) of our \heii\ emitting galaxies exhibit double-peaked \lya\ profiles, which motivated us to explore the connection between \lya\ morphology and \heii\ emission. For comparison, we find that only $\sim11\%$ of the non-\heii-emitting galaxies in the MAGPI sample exhibit double-peaked \lya\ profiles. This substantially higher fraction among \heii\ emitters suggests a possible connection between \heii\ emission and the prevalence of double-peaked \lya\ profiles (see Fig.~\ref{fig:barchart}). For the following analyses, we adopt a conservative $S/N\geq3$ threshold for the \heii\ measurements, resulting in a sample of 10 sources. We therefore investigated a series of diagnostics of the characteristics of the \lya\ profile, such as peak separation, flux distribution, and kinematics relative to the strength and kinematics of the \heii\ line. The spectral parameters of \lya\ are derived in \cite{2026PASA...43...21M} and will also be presented in Mukherjee et al. (submitted).
 
 \subsubsection{Dependence of \lya\ Flux Components on \heii\ Emission}
\label{flux-flux}

\lya\ photons in typical galactic outflows escape more readily on the red side, leading to a dominant redshifted peak and a weak “blue bump” \citep{2006A&A...460..397V, 2026PASA...43...21M}. A detectable blueshifted peak can emerge when either the outflow has low opacity (e.g. lower N$_\mathrm{HI}$ or higher ionisation), allowing some photons to escape directly, or when photons scatter off of \hi\ in an inflowing circumgalactic medium \citep [CGM;][]{2006ApJ...649...14D}. Additionally, when accretion or inflow predominates over outflow, a strong blueshifted peak over a redshifted one arises \citep[see][]{2023A&A...680L...5M, Mukherjee2026BLUEPRINT}. Therefore, the relative flux and emergence of both the \lya\ red and blue peaks are governed by a complex interplay between the ionisation state, \hi\ abundance, and the overall gas kinematics of the ISM and CGM.

\begin{figure}[ht!]
    \centering
\includegraphics[width=1.03\linewidth]
{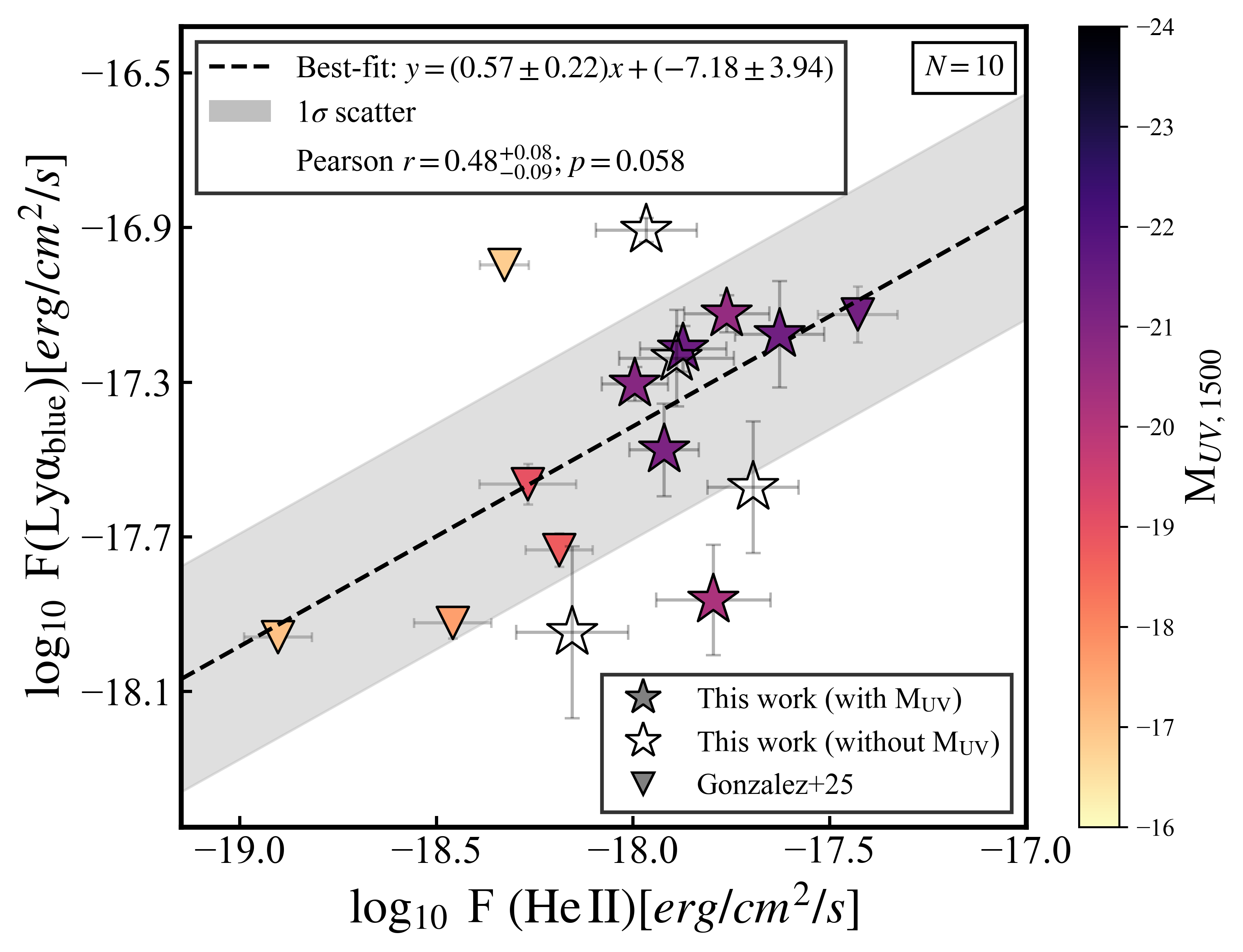}
    \caption{Relationship between the \lya$_\mathrm{blue}$ and the \heii\ fluxes for our sources, compared with the $z\sim 2-4$  sample from \citet[downward-pointing triangles]{2025A&A...702A...3G}. Our sources are shown as stars, with colours representing M$_\mathrm{UV, 1500}$ as indicated by the colour bar. The dashed black line represents the best fit to the combined data, with the grey shaded region denoting the $1\sigma$ intrinsic scatter of the fit. The best-fit linear equation, alongside the Pearson correlation coefficient are shown in the top-left.}
    \label{fig4:blue}
\end{figure}

Figure~\ref{fig4:blue} compares the measured fluxes of \heii\ emission with blueshifted \lya\ peak (\lya$_\mathrm{blue}$). Sources lacking M$_\mathrm{UV}$ measurements in our sample are shown as open stars. We also include a comparison sample from \citet{2025A&A...702A...3G}. A trend is seen where the \lya$_\mathrm{blue}$ flux increases with increasing \heii\ flux. To quantify this relationship, we performed a Pearson correlation test on the combined sample (our sources together with those from \citealt{2025A&A...702A...3G}). To account for measurement uncertainties in both axes, we employed a Monte Carlo resampling approach, obtaining  $r=0.484^{+0.082}_{-0.089}$ ($p=0.058$), indicating a significant, moderately positive correlation. The positive trend remains, suggesting that improved statistics of \heii\ sources would help confirm the relation.

We further performed a Least Trimmed Squares (LTS) regression using the \texttt{LtsFit\footnote{\url{https://pypi.org/project/ltsfit/}}} \citep{2013MNRAS.432.1709C} Python package on the combined data. This method accounts for measurement errors in all coordinates and incorporates intrinsic scatter into the resulting best-fit relation, which is given as:

\begin{equation}
\begin{aligned}
  \log_{10} F(\mathrm{Ly}\alpha_{\mathrm{blue}}) = (0.57\pm 0.22) \log_{10} F(\mathrm{He\,II}) + (-7.18 \pm 3.94)
\end{aligned}
\end{equation}

The grey shaded region in Fig.~\ref{fig4:blue} represents the 1$\sigma$ intrinsic scatter ($\sigma_\mathrm{int}=0.29\pm 0.09$ dex), which is consistent with the observed rms scatter ($0.3$ dex) around the relation. No outliers are identified by the clipping procedure (using a 4$\sigma$ threshold), suggesting that the combined sample follows a globally consistent trend.

%\textcolor{red}{As a robustness check, we also examined the \lya$_{\mathrm{blue}}$--\heii\ relation after restricting the sample to sources with $S/N \geq 3$ for \heii. The positive trend remains, with a Monte Carlo-derived Pearson coefficient of $r=0.484^{+0.082}_{-0.089}$ ($p=0.058$), comparable in strength to that obtained for the full sample. Although the correlation is marginal in the higher-$S/N$ subsample, the higher $p$-value may partly reflect the reduced sample size resulting from the more conservative $S/N$ criterion. The corresponding high-$S/N$ relation is shown in Appendix~...}

%These results indicate that galaxies exhibiting stronger \heii\ emission also tend to influence the ISM/CGM conditions in ways that facilitate the escape of \lya\ blueshifted photons thereby enhancing the \lya$_\mathrm{blue}$ flux. 

\subsubsection{\lya\ Peak Separation versus Equivalent Width of \heii}\

\begin{figure}[ht!]
    \centering
\includegraphics[width=0.98\linewidth]
{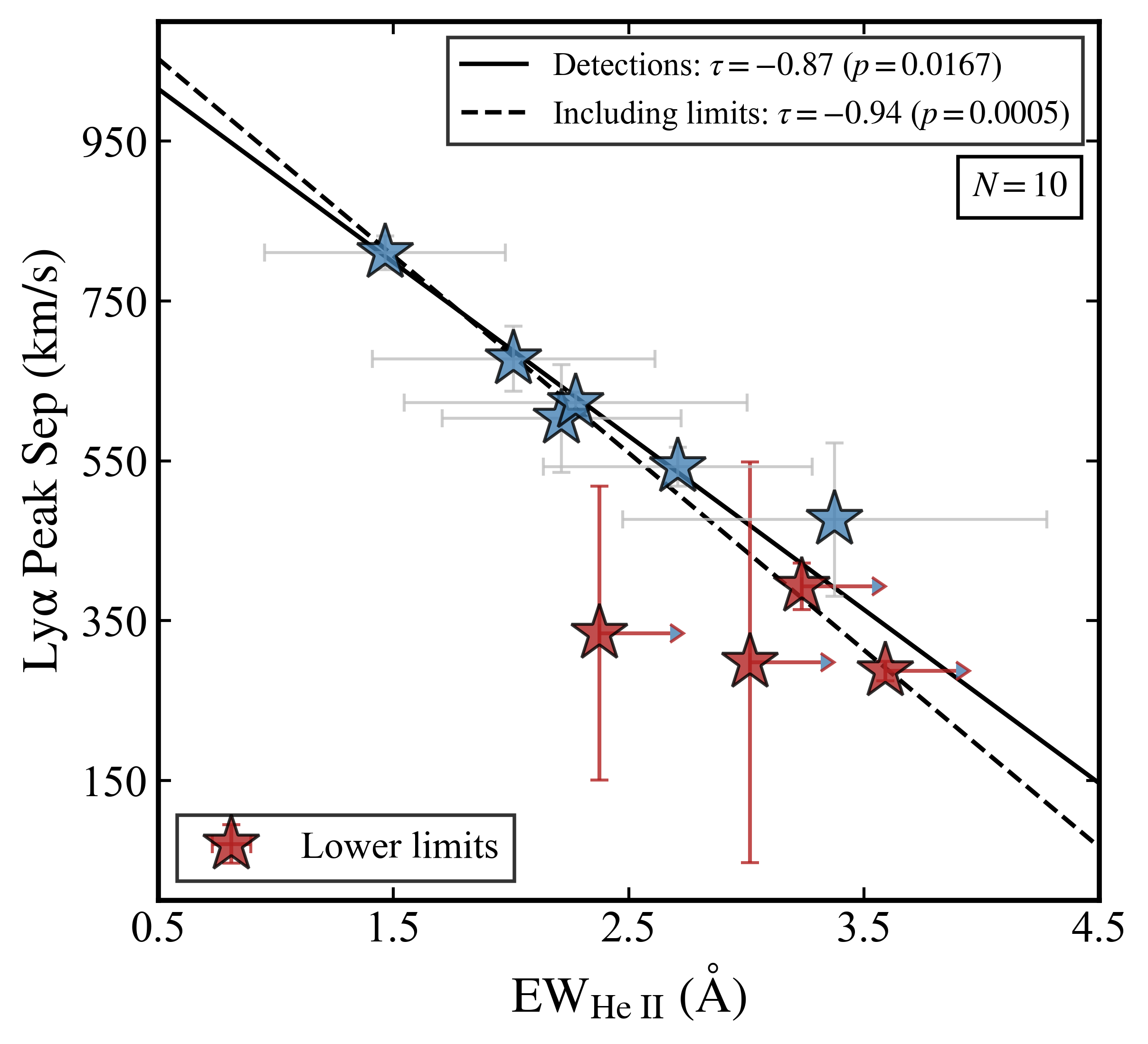}
    \caption{\lya\ peak separation as a function of the rest-frame EW$_\mathrm{He\,II}$. Red stars with right-pointing arrows denote lower limits for the EW$_\mathrm{He\,II}$. The solid black line represents the best-fit linear relation for the detections only, while the dashed black line corresponds to the fit obtained from the combined sample (including lower limits).}
    \label{fig5:EW_Lya}
\end{figure}

Figure~\ref{fig5:EW_Lya} illustrates the relationship between the rest-frame equivalent width of \heii\ (EW$_\mathrm{He\,II}$) and the velocity separation between the blue and red peaks of the double-peaked \lya\ profile for our sample. We find a significant anti-correlation between these two quantities. Since the sample includes censored data in the form of limits, we adopt the generalised Kendall's $\tau$ test, which accounts for censored data and is therefore more appropriate than the Pearson correlation test for the full sample. The generalised Kendall's $\tau$ test yields $\tau=-0.943$ ($p=4.99\times10^{-4}$), indicating a statistically significant trend where galaxies with larger peak separations exhibit lower \heii\ equivalent widths. Considering only the detections, the uncertainty-propagated Pearson correlation coefficient is $r=-0.693^{+0.303}{-0.178}$ ($p=0.0017$), providing further evidence for a strong negative correlation among sources with measured EW$_\mathrm{He,II}$ values.

We also modelled this trend using a robust LTS regression. The best-fit relation for the full sample (dashed line) is:

\begin{equation}
\begin{aligned}
  \Delta v_{\mathrm{peak}} = (-246.66 \pm 116.81)\ EW_\mathrm{He\,II}\ + \\(1176.61  \pm 266.99),  
\end{aligned}
\end{equation}

where $\Delta$$v_\mathrm{peak}$ is the separation between the blue and the red peaks of double-peaked \lya\ profiles.

The LTS procedure identified an observed rms scatter of 111 km\,s$^{-1}$, and no sources were identified as outliers by 4$\sigma$ clipping criterion. We note, however, that the current sample contains only a small number of galaxies with reliable EW measurements, so a larger statistical sample will be required to validate this anti-correlation. 

%The EW$_\mathrm{He\,II}$ is particularly sensitive to a hard ionising radiation field and the nature of the emitting source. On the other hand, the \lya\ peak separation is primarily governed by N$_\mathrm{HI}$ and the radiative transfer path length. A lower N$_\mathrm{HI}$ corresponds to a more optically transparent ISM/CGM through which both \lya\ and LyC photons can escape with fewer scatterings, leading to narrower \lya\ peak separation \citep[see][]{2015A&A...578A...7V,2018MNRAS.478L..60V, 2026PASA...43...21M}. 

%This relation could imply that systems with stronger EW$_\mathrm{He\,II}$ and thus a hard ionisation radiation field can ionise the surrounding ISM and reduce the neutral gas covering fraction, allowing \lya\ photons to escape with fewer resonant scatterings, resulting in narrower peak separation. 

\subsubsection{\heii\ FWHM vs \lya\ Peak Separation}

\begin{figure}[ht!]
    \centering
\includegraphics[width=1.0\linewidth]{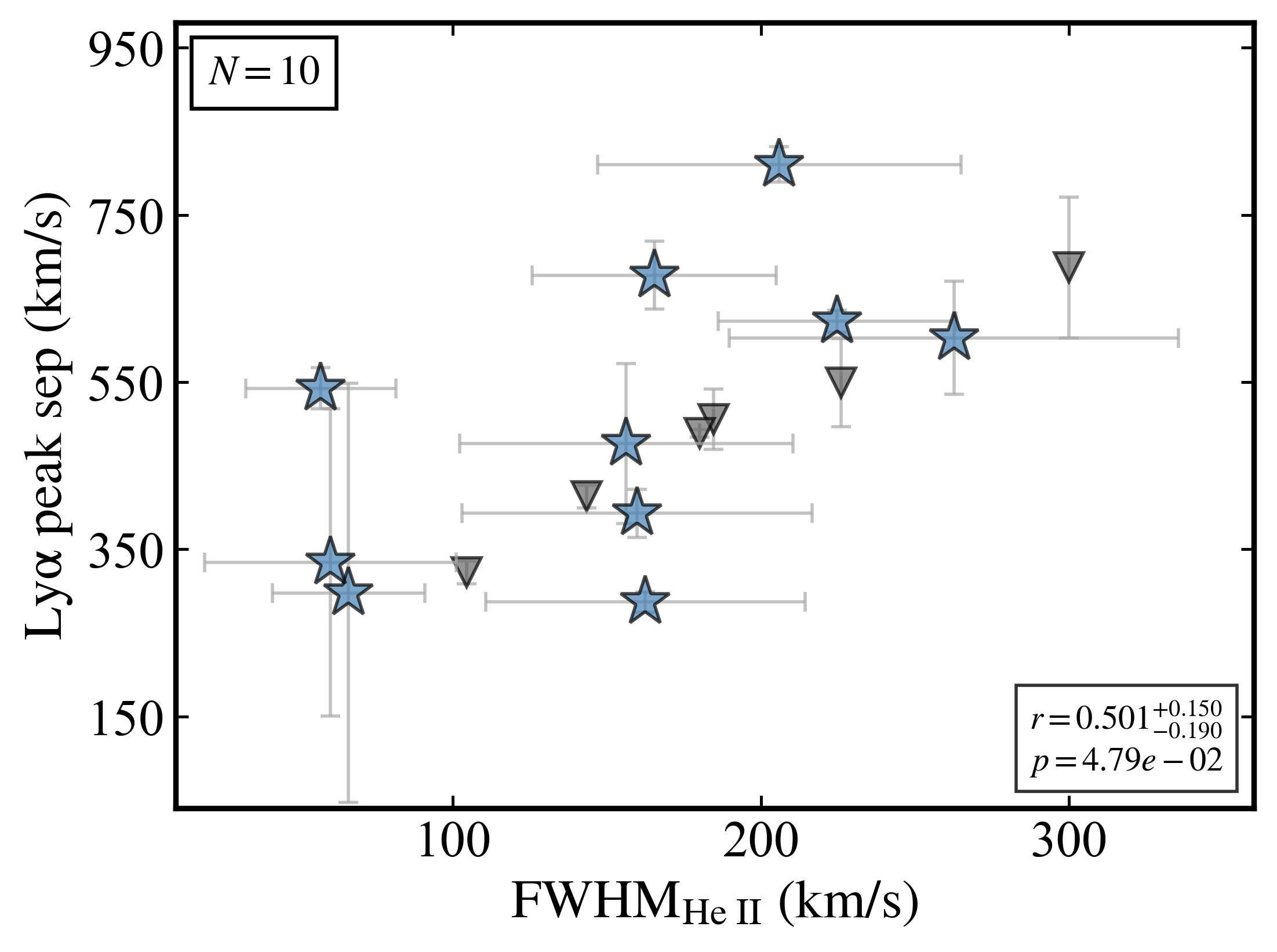}
    \caption{\lya\ peak separation versus the MUSE LSF-corrected FWHM$_\mathrm{He,II}$. The \heii\ emitters from \citet{2025A&A...702A...3G} are also shown for comparison.}
    \label{fig6:FWHM}
\end{figure}

\begin{table*}
\centering
\caption{Correlation
statistics for the \heii--\lya\ relations.}
\renewcommand{\arraystretch}{1.2}
\begin{tabular*}{\textwidth}{@{\extracolsep{\fill}}l @{\hspace{2em}} llc c@{}}
\hline
Relation & Sample & Analysis & Correlation & $p$-value \\
\hline

\multicolumn{5}{l}{\textit{\heii\ flux--\lya\ flux}} \\[2pt]

& Full sample ($N=13$) & Nominal & $r=0.521$ & 0.022 \\
& & 1$\sigma$ propagated & $r=0.473^{+0.084}_{-0.088}$ & 0.041 \\
\hline
& $S/N \geq 3$ ($N=10$) & Nominal & $r=0.519$ & 0.039 \\
& & 1$\sigma$ propagated & $r=0.484^{+0.082}_{-0.089}$ & 0.058 \\

\hline

\multicolumn{5}{l}{\textit{EW$_\mathrm{He\,II}$ --\lya\ peak separation}} \\[2pt]

& Full sample ($N=13$) & Generalised Kendall & $\tau=-0.957$ & $9.99\times10^{-5}$ \\
& Detections ($N=6$) & Nominal & $r=-0.966$ & 0.00175 \\
& & 1$\sigma$ propagated & $r=-0.69^{+0.300}_{-0.180}$ & 0.00175 \\
\hline
& $S/N \geq 3$, detections + censored ($N=10$) & Generalised Kendall & $\tau=-0.943$ & $4.99\times10^{-4}$ \\

\hline

\multicolumn{5}{l}{\textit{FWHM$_\mathrm{He\,II}$ --\lya\ peak separation}} \\[2pt]

& Full sample ($N=13$) & Nominal & $r=0.638$ & 0.0033 \\
& & 1$\sigma$ propagated & $r=0.464^{+0.148}_{-0.177}$ & 0.0455 \\
\hline
& $S/N \geq 3$ ($N=10$) & Nominal & $r=0.646$ & 0.0069 \\
& & 1$\sigma$ propagated & $r=0.501^{+0.150}_{-0.190}$ & 0.0479 \\

\hline
\end{tabular*}

\smallskip
\parbox{\linewidth}{\small \textbf{Note:} All correlation analyses include data from \citet{2025A&A...702A...3G}. For the Pearson correlations, ``Nominal'' refers to the standard correlation calculated from the unperturbed measured values, while ``1$\sigma$ propagated'' incorporates measurement uncertainties via Monte Carlo realisations. Note that FWHM uncertainties are not reported by \citet{2025A&A...702A...3G}; consequently, their FWHM values remain fixed during error propagation. For the EW$_\mathrm{He\,II}$--\lya\ peak-separation relation, the generalised Kendall test incorporates censored measurements (lower limits).}
\label{tab:correlations}
\end{table*}

In Figure~\ref{fig6:FWHM}, we explore the relationship between the kinematics of the ionised gas, traced by the FWHM$_\mathrm{He\,II}$ in km\,s$^{-1}$ and the \lya\ peak separation. The measured FWHM$_\mathrm{He\,II}$ values have been corrected for the instrumental line-spread function (LSF) of MUSE to account for instrumental broadening. The MUSE spectral resolving power varies from R$\sim1770$ at $4700$~\AA\ to R$\sim3590$ at $9350$~\AA, corresponding to an instrumental velocity resolution of approximately $169$--$84$ km\,s$^{-1}$ over the wavelength range of the observed \heii\ emission in our sample. The resulting intrinsic line widths are used in the analysis. For the combined sample (this work and \citealt{2025A&A...702A...3G}), we find a positive correlation. Accounting for the measurement uncertainties, the uncertainty-propagated Pearson correlation coefficient is $r=0.501^{+0.150}_{-0.190}$ ($p=0.0479$). We note that the FWHM uncertainties for the comparison sample from \citealt{2025A&A...702A...3G} are not available and therefore could not be propagated in this analysis. These results suggest that galaxies with broader \heii\ emission lines tend to exhibit larger \lya\ peak separations. Moreover, the correlation statistics for the \heii--\lya\ relations discussed above are summarised in Table~\ref{tab:correlations}.

\section{DISCUSSION}
\label{sec:discussion}
In this section, we discuss our results with particular emphasis on degeneracies between photoionisation model parameters and the physical connection between \heii\ and \lya\ radiative transfer.

\subsection{UV Line Ratio Diagnostics and Photoionisation Model}
\label{diagnostic}

Star-forming galaxies and AGN are known to occupy separate regions in the diagnostic diagram of \ciii/\heii\ versus \civ/\heii\, as their ionising sources produce noticeably different UV line ratios. Therefore, these diagnostics have been widely used to infer the dominant ionising source in high-$z$ galaxies \citep{2016MNRAS.456.3354F}. In addition, ratios involving these lines are sensitive to metallicity and ionisation conditions, with the \civ\ $\lambda\lambda1548,51$/\ciii\ $\lambda1907$ ratio in particular, being dependent on the ionisation parameter \citep[see][]{2006A&A...447..863N}. However, recent photoionisation models that incorporate binary evolution and a broader range of stellar ages and physical conditions, have shown that the separation between the stellar and AGN regions in UV line ratio space is not always clean \citep{2018MNRAS.477..904X}. In particular, such models demonstrate that metal-poor stellar populations with high ionisation parameters can extend into regions previously associated with AGN-like ionising spectra, leading to significant overlap between the two classes of models \citep{2018MNRAS.477..904X}. In our sample, only a few galaxies exhibit \civ\ in emission, therefore, we exclude it from the primary diagnostic analysis. Instead, we analyse the \ciii/\heii\ versus \oiii/\heii\ diagram (see Fig.~\ref{fig1:a}). Within this framework, AGN photoionisation usually produces \ciii/\heii\ < 1 \citep{2016MNRAS.456.3354F, 2016MNRAS.462.1757G}, a regime not occupied by our sources. While this disfavoures classical AGN-dominated photoionisation, we note that alternative mechanisms such as narrow-line AGN, radiative shocks, or high-mass X-ray binaries cannot be fully excluded on the basis of UV diagnostics alone.

The distribution of our sources in Fig.~\ref{fig1:a} suggests that these galaxies exhibit ionising conditions and metal contents similar to other high-redshift populations. The observed offset from local systems toward lower \ciii/\heii\ ratios indicates that our sample has relatively lower carbon abundances and less evolved chemical enrichment compared to $z\approx 0$ systems, consistent with expectations from galactic chemical evolution \citep[e.g.][]{2016ApJ...826..159S}. However, we note that there are  degeneracies between $Z$, log U$_\mathrm{s}$, and C/O ratio within the \citet{2016MNRAS.462.1757G} model grids, complicating the unique interpretation of the underlying physical conditions (see~\S \ref{U_Z} and \S \ref{Z_C/O}). Despite these degeneracies, the gas-phase oxygen abundance corresponding to the $Z=0.001$ track in \citet{2016MNRAS.462.1757G}, which encompasses our sources, assuming $\xi_\mathrm{d}=0.5$, is 12 + log\ (O/H) = 7.41 (see their Table 2). This is in good agreement with the empirical estimates of gas-phase abundance derived using the \citet{2020ApJ...893....1B} calibration (7.20 $\lesssim$ 12 + log\ (O/H) $\lesssim 7.71$; uncertainty of $\sim\pm0.14$), reinforcing a consistent low-metallicity interpretation for our sample.

Despite the overlap in metallicities, at a fixed metallicity, Fig.~\ref{fig1:b} is a good probe of the C/O abundance ratio of the sources, however metallicity itself influences these line ratios in complex ways. For instance, \citet{2016ApJ...833..136J} demonstrated that at very low-$Z$, where carbon is intrinsically less abundant, the \ciii\ flux may actually increase. This is because low-$Z$ environments exhibit harder ionising spectra and higher gas temperatures, both of which boost the collisional excitation efficiency of the \ciii\ transition. In addition, it should be noted that UV-based C/O values can vary up to $\sim$ 0.6 dex at low metallicity due to the dependence of carbon production on star formation history and supernova feedback \citep{2016ApJ...827..126B,2020ApJ...893....1B}.

In the \ciii /\oiii\ versus \siiii /\ciii\ diagnostic (Fig.~\ref{fig1:c}), the $Z=0.0001$ and $Z=0.001$ model tracks overlap substantially, limiting the ability to place tight constraints on metallicity using this diagram alone. However, consistent with findings from Figs.~\ref{fig1:a} and \ref{fig1:b}, the observed line ratios are well-reproduced by the $Z=0.001$ track, with ionisation parameters in the range log U$_\mathrm{s} \sim -3.5$ to $-2.5$ and C/O ratios of $\sim 0.18-0.25$. This demonstrates that the $Z=0.001$ model track provides a consistent description across all three diagnostics. Taken together, the line ratios of our sources are consistent with sub-solar metallicities of $Z\approx0.001$, moderately high ionisation parameters of log U$_\mathrm{s} \simeq -3.5$ to $-2.5$, and average C/O ratios in the range $\sim 0.15-0.30$. Therefore, while
the \citet{2016MNRAS.462.1757G} model grids indicate that the nebular emission is powered
by star formation, inherent degeneracies between the parameters prevent a unique determination of the physical conditions for individual sources.

%In the following section, we examine how variations in model parameters and inherent degeneracies in the line ratio grids influence these interpretations.

\subsection{Effect of Different Model Parameters on UV Line Ratios}

Here we briefly outline the impact of varying the adjustable parameters and degeneracies in the line ratio grids of \citet{2016MNRAS.462.1757G} photoionisation models on the emission line ratios.

\subsubsection{Upper mass cutoﬀ of the IMF (m$_\mathrm{up}$):} Keeping the other model parameters fixed and varying the m$_\mathrm{up}$ value from 100 M$\odot$ to 300 M$\odot$ produces a significant change in the model grids of the line ratio diagnostics concerning \heii\ (Figs.~\ref{fig1:a} and \ref{fig1:b}). A higher m$_\mathrm{up}$ hardens the ionising spectrum by boosting the production of photons with energies $\geq$ 54.4 eV, thereby enhancing the \heii\ emission relative to collisionally excited metal lines such as \ciii\ and \oiii.\ As a consequence, model tracks in line ratio normalised to \heii\ shift substantially toward lower values and the overall grid morphology changes. Moreover, low metallicity tracks in \heii-metal line ratio diagnostics become more compressed as m$_\mathrm{up}$ increases, leading to a partial overlap between model tracks.

\subsubsection{Hydrogen gas density (n$_\mathrm{H}$):} With all other model parameters fixed, increasing the hydrogen gas density causes the model tracks of the \ciii/\oiii\ vs \siiii/\ciii\  diagnostic (see Fig.~\ref{fig1:c}) to move toward the lower right, with the largest displacement occurring at higher metallicities and higher ionisation parameters (which increases from top to bottom). In contrast, the low metallicity tracks remain largely insensitive with increasing n$_\mathrm{H}$. This behaviour is driven by density dependent collisional de-excitation of the metal lines where, as the gas density increases, collisional processes become comparable to or dominate over radiative decay, leading to a partial suppression of semi-forbidden UV line emissivities \citep{2006agna.book.....O}. Moreover, this effect becomes particularly pronounced at higher metallicities, where cooling is mainly regulated by metal lines. In high-density (n$_\mathrm{H}$) environments, the increased frequency of collisions de-excites ions before they can emit UV photons, altering the relative strengths of the metal lines. Whereas \heii\, being a recombination line, remains only weakly dependent on gas density. Consequently, diagnostic ratios involving UV metal lines show larger shifts at high metallicities. A similar trend is observed in the \ciii /\heii\ vs \oiii /\heii\  (Fig.~\ref{fig1:a}) and \oiii/\heii\ vs \ciii/\siiii\ (Fig.~\ref{fig1:b}), with low-$Z$ tracks showing minimal change, while high-$Z$ tracks exhibit significant shifts. Despite this variation, our sources remain consistent with the sub-solar ($Z=0.001$) metallicity track.

\subsubsection{Dust to metal mass ratio ($\xi_\mathrm{d}$):}

Increasing $\xi_\mathrm{d}$ shifts the model tracks across all three diagnostics. In the \ciii/\oiii\ vs \siiii/\ciii\ (see Fig.~\ref{fig1:c})  plane, both high- and low-$Z$ tracks shift leftward and slightly downward. Silicon is a refractory element, as a result, \siiii\ intensity weakens relative to \ciii,\ driving the leftward shift. The downward displacement in \ciii/\oiii\ likely reflects reduced metal-line cooling at higher $\xi_\mathrm{d}$, which raises $T_\mathrm{e}$ and enhances \oiii\ relative to \ciii.\ In the \ciii /\heii\ vs \oiii /\heii\ (see Fig.~\ref{fig1:a})   plot, higher metallicity tracks move slightly in the upper-right direction as dust depletion suppresses cooling, increasing $T_\mathrm{e}$, and strengthening collisionally excited lines such as \ciii\ and \oiii\ \citep{2014MNRAS.445.3200S}. Whereas very low-$Z$ tracks move to the lower-left, as further depletion hardens the radiation field and pushes C$^{2+}$ and O$^{2+}$ into higher ionisation stages (e.g. \civ,\ \oiv),\ reducing both ratios. Finally, in the \oiii/\heii\ vs \ciii/\siiii\ diagram (see Fig.~\ref{fig1:b}), strong Si depletion drives a rightward shift at all metallicities. However, the \oiii/\heii\ ratio changes slightly differently with metallicity. At high-$Z$, high $T_\mathrm{e}$ promotes \oiii\ relative to \heii\, moving the tracks upward, whereas at very low-$Z$, increased metal depletion produces a harder radiation field, either boosting the production of \heii\ or pushing \oiii\ to \oiv. The \ciii /\heii\ vs \oiii /\heii\  line ratios of all our sources (see Fig.~\ref{fig1:a}) can be reproduced with $\xi_\mathrm{d}$ = 0.1 along the $Z=0.0002$ track, assuming ionising parameters (log U$_\mathrm{s} \sim$ -3.5 to -3.0). In contrast, the two diagnostics involving \siiii\ (see Fig.~\ref{fig1:b} and Fig.~\ref{fig1:c}) require $\xi_\mathrm{d}$ = 0.5 for the model grids to encompass the observed ratios, highlighting strong sensitivity of Si-based line ratios to $\xi_\mathrm{d}$.

\subsubsection{Carbon to oxygen abundance ratio (C/O): }
Increasing the C/O ratio enhances the gas-phase carbon abundance as compared to heavier $\alpha$-elements such as O and Si, leading to stronger \ciii\ emission. With other parameters fixed, this causes the model tracks in Fig.~\ref{fig1:c} to move top-left, driven by a simultaneous rise in \ciii/\oiii\ and a decline in \siiii/\ciii.\ Nevertheless, at higher metallicities, the tracks exhibit a slight downward shift. This is due to the fact that at high-$Z$, carbon becomes an important coolant, lowering the $T_\mathrm{e}$ and introducing a non-linear effect on \ciii\ emissivity \citep[e.g.][]{2016ApJ...833..136J}. A similar trend is observed in Fig.~\ref{fig1:a}, where increasing C/O primarily boosts \ciii\ relative to \heii\, shifting the tracks upward, with a modest decline at high-$Z$ caused by the carbon cooling. Whereas, the diagnostic presented in Fig.~\ref{fig1:b}, shows a rightward shift in the tracks as a function of C/O ratio.

\subsubsection{Ionisation parameter at the $Str\ddot{o}mgren$ radius (U$_\mathrm{s}$):} 

As the ionisation parameter increases, the gas is exposed to a higher flux of ionising photons and the \hii\ region becomes more compact, favouring the production of higher-ionisation lines. With other model parameters fixed, log U$_\mathrm{s}$, shown as a sequence of points along each track, increases from left to right until it reaches a maximum and subsequently declines at higher U$_\mathrm{s}$ in the  \ciii/\heii\ vs  \oiii/\heii\ diagram (Fig.~\ref{fig1:a}). This decline in both ratios specifically at higher log U$_\mathrm{s}$ values indicates that both \ciii\ (ionisation potential $\sim$ 24.4 eV) and \oiii\ (ionisation potential $\sim$ 35.1 eV) are pushed to higher ionisation stages, with a higher drop in \ciii\ emission fraction as compared to \oiii,\ owing to its lower ionisation potential. In the \oiii/\heii\ vs \ciii/\siiii\ plane (Fig.~\ref{fig1:b}), increasing log U$_\mathrm{s}$ enhances O$^{2+}$ production over \heii,\ while Si$^{2+}$ which has a lower ionisation potential ($\sim$ 16.4 eV) is progressively ionised to Si$^{3+}$, causing the points along the tracks to move up and rightward. Similarly, in the \ciii/\oiii\ vs \siiii/\ciii\  (Fig.~\ref{fig1:c}), higher log U$_\mathrm{s}$ values favour \oiii\ over \ciii\ and suppress \siiii\ by ionising it to higher states, resulting in a systemic movement of points along the tracks from the upper-right toward the lower-left. 

%Although our sources and the comparison samples occupy well-defined regions within the \citet{2016MNRAS.462.1757G} model grids, these positions are not uniquely tied to a single set of physical parameters. 
%As discussed earlier, adjusting different model parameters can often reproduce similar line ratios, and even for a fixed choice of parameters, the grids retain intrinsic degeneracies. 

\subsubsection{Ionisation parameter-metallicity degeneracy:}
\label{U_Z}

In Fig.~\ref{fig1:a}, the grids exhibit a degeneracy between U$_\mathrm{s}$ and metallicity, whereby a high-$Z$ track with lower log U$_\mathrm{s}$ can reproduce similar line ratios covered with a low-$Z$ track having higher ionisation parameter. This relation is in agreement with \citet{2010ApJ...719.1168E}, who highlighted that U$_\mathrm{s}$ increases significantly with decreasing metallicity due to the harder ionising spectra and weaker stellar winds of massive stars of low-$Z$. A similar interplay also appears in the \oiii/\heii\ versus \ciii/\siiii\ plot (Fig.~\ref{fig1:b}), further making it difficult to uniquely constrain the ISM conditions of our sources.

\subsubsection{Metallicity-C/O ratio degeneracy:}
\label{Z_C/O}

In addition, the model grids reveal a clear metallicity-C/O degeneracy, where variations in carbon abundance can partially mimic the effects of changing overall metallicity. Across all three diagnostic diagrams, a higher C/O ratio at lower metallicity can reproduce UV line ratios similar to those produced by a lower C/O at higher metallicity. As a result, different combinations of $Z$ and C/O occupy overlapping regions, preventing a unique determination of parameters.

While the model grids show some degeneracies between parameters, the positions of our sources within the diagnostics still allow parts of the parameter space to be ruled out, such as solar-metallicity models, which are disfavoured for our sample. In addition, most galaxies fall within the moderate to slightly high U$_\mathrm{s}$ values, partially constraining the ionisation state of the gas. To place more stringent constraints on the underlying stellar populations and the physical state of the ISM, rest-frame UV and optical absorption features originating from both the stellar continuum and the neutral and ionised gas should be accompanied with the nebular emission lines \citep[e.g.][]{2017MNRAS.470.3532V}. This combined approach facilitates a consistent comparison with the stellar population synthesis models. In recent years, several advanced spectral fitting frameworks have been developed to conduct spectro-photometric analyses of galaxies to put simultaneous constraints on stellar, nebular, and dust properties \cite [e.g.][]{2016MNRAS.462.1415C,2017ApJ...837..170L}. Nevertheless, studies concerning the Local  Universe show that models neglecting stellar rotation or binaries fail to reproduce key diagnostics like \heii\ at low $Z$ \citep{2017MNRAS.472.2608S}, highlighting the need for comprehensive stellar population models when analysing hard ionising radiation.

\subsection{Effects of Binary Stars}

While the \citet{2016MNRAS.462.1757G} models provide a broad framework for predicting nebular emission in high redshift star-forming galaxies, they are based on single-star SPS and neglect the effects of stellar rotation and binary interactions. However, observations in the Local Universe have established that a large fraction ($\sim$ 45 –75\%) of O-type stars are in close binaries, with the majority expected to undergo significant interaction over their lifetimes, influencing their structure and evolution \citep{Mason_2009, Sana_2012}. Furthermore, recent work by \citet{Sana_2025} suggests that stellar multiplicity is significantly independent of metallicity, implying that binary interactions remain ubiquitous even in the metal-poor conditions of the early Universe. Consequently, binary-driven evolution such as mass-transfer, envelope stripping, and mergers likely dominates the stellar populations at high-redshift. These processes can produce hot stripped stars and high-mass X-ray binaries, both capable of generating \heii-ionising photons. In this context, the Binary Population and Spectral Synthesis \citep[\texttt{BPASS};][]{10.1093/mnras/stv2661, 2017PASA...34...58E,Stanway_2018} models incorporate the effects of binary evolution and predict a harder, more sustained ionising spectrum, enhancing the production of WR stars and/or hot helium-rich stars even at low metallicities \citep{2019A&A...624A..89N}. 

\citet{2018MNRAS.477..904X} constructed photoionisation models using \texttt{BPASS} v2.1 \citep{2017PASA...34...58E} as the input radiation field to \texttt{CLOUDY} 13.03, enabling a self-consistent treatment of nebular emission. These models are computed for a single instantaneous starburst with ages spanning 1 Myr to 10 Gyr. The nebula is assumed to be dust-free, which is an important caveat, and is modelled as spherically symmetric, ionsation-bounded, with a constant n$_\mathrm{H}$. A key outcome of these models is that, with binary interactions included, the parameter space capable of producing
high-ionisation lines broadens, ruling out models with low ionisation parameters (log U$_\mathrm{s} < -2.5$) regardless of metallicity. We do not directly compare our data with the \citet{2018MNRAS.477..904X}
model grids, instead we utilise the \ciii/\heii\ versus \oiii/\heii\ diagnostic diagram presented in \citet{2019A&A...624A..89N}, which includes these models (see their Fig. 8). Given that their sample spans a similar redshift range ($z \sim 2-4$) and occupies a comparable line ratio space to our sources, this provides a consistent basis for indirect comparison from which we infer that our sources lie in a regime corresponding to log U$_\mathrm{s} \sim -2.0$ to -1.5 which is higher than the presumed value from \citet{2016MNRAS.462.1757G}, with a wide metallicity range ($Z=0.0001-0.02$, where assumed $Z_\odot =0.02$; \citealt{1993oee..conf...15G}) with strong degeneracies between $Z$, log U$_\mathrm{s}$, and age. Therefore, though \citet{2018MNRAS.477..904X} models support a hard ionising spectrum consistent with binary evolution, they do not place tight constraints on the underlying ISM properties of the sources.

\subsection{\lya\ Radiative Transfer and \heii\ Emission}
\label{Lya RT}

%We now discuss how the presence and strength of the \heii\ $\lambda1640$ \AA\ UV emission line correlate with the line morphology and flux distribution of double-peaked \lya\ profiles, and what these relations imply about the connection between ionising radiation field and the geometry of the neutral ISM.

The \lya$_\mathrm{blue}$ flux correlates positively with \heii\ flux (Fig.~\ref{fig4:blue}), although its escape is more sensitive to the detailed morphology and kinematics of the neutral gas. This trend suggests that harder ionising spectra develop a more porous environment on both  interstellar and circumgalactic scales. As demonstrated by \citet{2011ApJ...730....5H} in the context of Lyman-continuum escape, stellar feedback creates a "picket-fence" ISM geometry in which mechanical energy carves low-density channels that allow radiation to pass through more easily \citep[see also,][]{2025ApJ...987...12M}. Our systems are consistent with young, actively star forming galaxies with hard ionising spectra traced by \heii\ emission. In such environments, vigorous feedback (e.g., stellar winds, metal-poor starbursts, supernovae, radiation pressure, etc) can perturb not only the ISM but also the CGM \citep[see][and references within]{2015MNRAS.454.2691M}, lowering the effective neutral hydrogen covering fraction and creating extended low-density channels \citep{2013ApJ...777...39Z, 2020A&A...639A..85G}. These feedback-regulated geometries increase the
number of optically thin sightlines through which blueshifted
photons can escape despite the usual suppression by outflows. Another related pathway is the formation of density-bounded nebulae \citep{2025ApJ...992...91R}, where intense star formation photoionises the surrounding gas so thoroughly that the local ISM becomes almost optically transparent to LyC photons. Although LyC and \lya\ photons require different neutral \hi\ column densities to escape, both benefit from the same feedback-driven reduction in the effective \hi\ covering fraction. In particular, LyC photons escape only through extremely low column densities (N$_\mathrm{HI} \lesssim 10^{17}$cm$^{-2}$), while \lya\ photons can escape more readily even with higher N$_\mathrm{HI}$ \citep{2022ApJS..259...21K}. This facilitates the emergence of blueshifted \lya\ photons, which are otherwise easily suppressed by resonant scattering. 

Fig.~\ref{fig5:EW_Lya} investigates the relationship between the EW$_\mathrm{He\,II}$ and the \lya\ peak separation. The EW$_\mathrm{He\,II}$ is particularly sensitive to the hardness of ionising radiation field and the nature of the emitting source, while the \lya\ peak separation is primarily governed by N$_\mathrm{HI}$ and radiative transfer effects \citep{1965ApJ...142.1633G}. A lower N$_\mathrm{HI}$ corresponds to a more optically transparent ISM/CGM through which both \lya\ and LyC photons can escape with fewer resonant scattering events, leading to narrower \lya\ peak separations \citep[see][]{2015A&A...578A...7V,2018MNRAS.478L..60V, 2026PASA...43...21M}. Our analysis reveals a clear anti-correlation between these quantities (Fig.~\ref{fig5:EW_Lya}), where higher EW$_\mathrm{He\,II}$ and hence stronger \heii\ emission is associated with narrower peak separations. This trend suggests that systems with a harder ionising spectrum are more efficient at ionising their surrounding neutral gas reservoirs. This decreases the effective N$_\mathrm{HI}$, which lowers the \lya\ optical depth thereby reducing the frequency diffusion experienced by \lya\ photons, resulting in smaller \lya\ peak separations \citep{2015A&A...578A...7V}. Both \citet{2015ApJ...805...14R} in the Local Universe and \citet{Reddy_2022} at $z$ = 1.9 - 3.5 discovered that \lya\ preferentially escapes through low column density channels  in the ISM. Furthermore, \citet{2015ApJ...805...14R} reported an anti-correlation between the covering fraction and H$\alpha$ equivalent width, suggesting that stellar feedback  plays a key role in creating ionised pathways. 

%Together, these results support the notion that galaxies with harder ionising radiation fields can influence the neutral ISM/CGM in a way that facilitates both \lya\ escape and narrower peak separations observed in our sample.

The observed correlation between FWHM$_\mathrm{He\,II}$ and the \lya\ peak separation (Fig.~\ref{fig6:FWHM}) suggests a link between the kinematics of the ionised and neutral gas phases. In systems with stronger turbulence or more energetic feedback, the ionised gas exhibits larger velocity dispersions, resulting in broader \heii\ line widths.  Concurrently, this turbulence and kinematic complexity can dynamically stir the surrounding neutral gas up to the CGM scale \citep{2015MNRAS.454.2691M}, potentially enhancing \lya\ scattering and frequency diffusion through interactions with the surrounding gas. This can cause \lya\ photons to escape further from the line centre, producing larger peak separations. Sources from \citet{2025A&A...702A...3G} follow a similar trend in the \heii-\lya\ diagnostic space, in agreement with our data, suggesting that the observed relationship between \lya\ and \heii\ is not unique, but may instead reflect a more general connection between ionised gas kinematics and \lya\ radiative transfer. 

Altogether, the set of \lya-\heii\ diagnostics presents a coherent physical picture in which the hard ionising sources responsible for \heii\ emission are closely linked to both the production and escape of \lya\ photons. We suggest that galaxies with harder ionising radiation fields not only boost intrinsic \lya\ production but also change the morphology and kinematics of the ISM and CGM through strong feedback and photoionisation, affecting the \lya\ observables. However, \lya\ radiative transfer is quite complex and cannot be explained by a single physical mechanism acting in isolation. Several works \citep[e.g.][]{2019ApJ...887...85T, 2020ApJ...892...48R,2023MNRAS.520.5903H} have adopted multivariate frameworks to investigate how various galaxy properties collectively drive the radiative transfer and emergence of \lya\ photons. In  addition, distinguishing between the ionisation-bounded and density-bounded nebulae  generally requires optical diagnostics, such as O$_{32}  \equiv$ \Oiii\ $\lambda\lambda4959,5007$ / \oii\ $\lambda\lambda3726,3729$ line ratio and low-ionisation tracers like \sii\ $\lambda\lambda6717,6731$, which are sensitive to the presence of globally low N$_\mathrm{HI}$ and the escape of ionising radiation. However, these diagnostics are beyond the scope of this work due to the lack of optical emission spectral coverage. Therefore, while our diagnostics point toward a consistent physical interpretation, a larger sample with complementary optical data is essential to confirm the robustness of these correlations.

\subsection{Sources with \heii+\lya\ only: Extremely Metal-Poor Starbursts?}
\label{metal-poor}

A subset of our sample shows \lya\ and \heii\ emission without any detectable accompanying rest-frame UV metal lines (Figs.~\ref{fig:21}-\ref{fig:24}). Such systems are particularly intriguing, because the absence of metal lines possibly indicates extremely metal-poor ionising sources or galaxies in the early stages of chemical enrichment. In very low-metallicity environments, massive stars produce harder ionising radiation capable of doubly ionising He, while the weakness or absence of carbon and nitrogen lines reflects the limited metal content of the ISM. Similar systems exhibiting strong \heii\ emission without accompanying UV metal lines have been interpreted as potential candidates for  galaxies hosting extremely metal-poor or Pop\,III-like stellar populations \citep[e.g.][]{2024ApJ...967L..42W, 2025ApJ...988..171M, maiolino2026searchpopulationiiiconfirmation}. Nevertheless, we emphasise that the non-detection of metal lines in our sample is limited by the depth of the current MAGPI observations, and a detailed assessment
of the underlying ionising sources is beyond the scope of this work. As demonstrated by landmark studies such as the analysis of CR7 \citep{Sobral_2015}, deep spectroscopic follow-up can reveal subtle metal lines in systems that appear primitive in current data. Therefore, while our four sources exhibit only \lya\ and \heii\ consistent with pristine or very metal-poor environments, deeper observations are required to definitively constrain their metallicities.

\subsection{Possible ionising Sources Powering the \heii\ Emission}

The origin of nebular \heii\ emission is a long-standing problem, as it requires sources capable of  producing photons with energies $\geq$ 54.4 eV, which are not easily generated in typical star-forming environments. Consequently, several studies have proposed various stellar and non-stellar mechanisms to account for these high energy photons. In this section, we discuss the mechanisms to assess the plausible ionising source of \heii\ emitters. However, due to the lack of rest-frame optical coverage for our sample, we refrain from drawing strong conclusions about the dominant source.

%\subsubsection{AGN}
%\label{AGN}

Within our parent sample, three objects exhibit quasar-like spectral features, and are therefore classified as AGN. The remaining \heii\ emitters show no obvious AGN signatures. In the \ciii/\heii\ vs \oiii/\heii\ diagram (Fig.~\ref{fig1:a}), all sources occupy regions consistent with star-forming galaxies, and none display AGN-like signatures such as strong \nv\ emission. Although a sub-dominant contribution from AGN cannot be completely ruled out, our current analysis suggests that the our \heii\ emitters are unlikely to be AGN dominated. 

Moreover, in our sample, three galaxies exhibit clear \civ\ P-Cygni profiles, indicative of strong stellar winds and a possible contribution from WR-like massive stars. However, our sources exhibit narrow rest-frame \heii\ emission, with FWHM $\sim$110--780 km s$^{-1}$, whereas stellar-wind emission from classical WR populations produces broad \heii\ emission with FWHM $\gtrsim$1000 km s$^{-1}$ \citep[e.g.][]{2020MNRAS.496.3796S}. 
The absence of such broad \heii\ emission, together with our line-ratio diagnostics indicating sub-solar metallicities ($Z\sim0.001$), suggests that classical WR stars are unlikely to dominate the observed \heii\ emission across the full sample. Nevertheless, very massive WNh stars at low metallicity are predicted to produce substantial \heii-ionising flux while driving relatively weaker winds ($\gtrsim$300 km s$^{-1}$), potentially generating \heii\ emission without the prominent broad WR signatures \citep{2015A&A...578L...2G, 2018MNRAS.480.1081K}. In addition, their evolutionary pathways may be influenced by enhanced stellar rotation and binary mass transfer, which could further contribute to He II emission without producing prominent broad WR features. Thus, while classical WR stars are disfavoured as the dominant source of \heii\ emission, a contribution from low-$Z$ WN/WNh stars cannot be completely excluded for a subset of our sample. Additional optical observations would be required to search for characteristic WR bumps and provide further evidence for a possible WR contribution.

The inclusion of binary interactions and stellar rotation plays an essential role in producing high-energy photons in high-redshift galaxies and their local analogues \citep{2025arXiv251116752H}. Moreover, \citet{2025MNRAS.543.3172G} showed that incorporating WR stars into stellar population models can significantly enhance the hardness of the ionising radiation field and can even account for some extreme emission lines such as \niii\ $\lambda1750$ \AA\ to some extent. At low metallicities, binary evolutionary processes can produce hot, stripped stars, and WR-like objects at lower initial masses than predicted by single-star evolution, thereby enhancing the generation of hard ionising radiation. The sample from \citet{2019A&A...624A..89N} spanning a redshift range similiar to ours ($z \sim$ 2-4), reports that QHE alone allows BPASS models to produce > 0.1 dex more ionising photons at $Z \leq $ 0.004 (see their Fig. 12). Although we do not perform spectral synthesis using binary population models (e.g., BPASS) in this work, binary interactions may still contribute to the observed nebular \heii\ emission in our sample. 

A non-stellar candidate for powering \heii\ emission in star-forming galaxies is X-ray binaries \citep[XRBs;][]{1991ApJ...373..458G, 2019A&A...622L..10S}. Their contribution is expected to increase in low-$Z$ environments, both at high redshift \citep{2019ApJ...885...65F} and in the Local Universe 
\citep{2015A&A...579A..44D, 2016MNRAS.457.4081B}. However, due to the inherent difficulty in detecting X-ray signatures around $z \sim$ 2 - 4, our study is subject to observational constraints, preventing us from quantifying the extent to which XRBs contribute to the observed \heii\ emission in our current dataset. Radiative shocks are another potential non-stellar contributor to the observed \heii\ budget \citep{Allen_2008,2016ApJ...833..136J}. However, given that our \heii\ detections have $S/N < 10$  and lack rest-frame optical diagnostics, we cannot separate the contributions of shocks from other ionising mechanisms. Future high $S/N$ spectroscopy will be essential  for isolating the role of shocks in powering \heii\ emission.

\section{SUMMARY AND CONCLUSIONS}
\label{conclusion}
This paper presents a sample of prominent \heii\ $\lambda1640$ \AA\ emitting galaxies at $2 \lesssim z \lesssim 4$ identified using data from the MAGPI survey. We selected \heii\ candidates via visual inspection of 1D  and 3D spectral data of sources, shortlisting 28 \heii\ emitters with additional UV emission lines present. We measured key spectral properties including flux, FWHM, and EW with their associated uncertainties, and used UV emission line ratio diagnostics to constrain the underlying ionising mechanisms of our sources. We summarise our key results as follows.

\begin{itemize}

\item Using the \citet{2020ApJ...893....1B} calibration, we derive gas-phase oxygen abundances of 7.20 $\lesssim$ 12 + log\ (O/H) $\lesssim $ 7.71 ($\pm 0.12$ dex), with one source at $\sim 7.2$, consistent with very low metallicity environments.

\item UV emission line diagnostics show that our sources are consistent with star-forming galaxies, with no strong evidence of AGN-dominated activity.
\item The observed UV line ratios are well reproduced by sub-solar metallicities ($Z=0.001$) with a moderately high ionisation parameter (log U$_\mathrm{s} \sim$ -3.5 to -2.5).

\item The oxygen abundance of 12 + log\ (O/H) = 7.41, estimated from the \citet{2016MNRAS.462.1757G} models, is consistent with the mean value ($\sim7.46 \pm0.12$) derived from the calibration of \citet{2020ApJ...893....1B}.

\item 
We interpret our line ratios using the diagnostic framework of \citet{2019A&A...624A..89N} which incorporates the \citet{2018MNRAS.477..904X} \texttt{BPASS} models. The inferred ionisation parameters (log U$_\mathrm{s} \sim -2.0$ to $-1.5$) suggest a harder ionising spectrum than predicted by SPS models \citep{2016MNRAS.462.1757G}, highlighting the importance of binary interactions, though degeneracies between metallicity, age, and ionisation conditions preclude tighter constraints.

\item Our \heii\ emitter sample spans rest-frame UV absolute magnitudes of $-22.95 \lesssim M_\mathrm{UV,1500} \lesssim -19.64$, and statistical tests indicate no significant difference in the M$_\mathrm{UV}$ distributions between \heii-detected and non-detected galaxies in the parent sample.

\item Interestingly, we find that approximately 77\% of \heii\ emitters are associated with double-peaked \lya\ profiles, indicating a connection between hard ionising radiation fields and the neutral gas geometry and kinematics governing \lya\ radiative transfer.

\item We find a positive correlation between the \heii\ flux and \lya$_\mathrm{blue}$ flux,  suggesting that hard ionising sources associated with strong stellar feedback can significantly reshape the ISM/CGM. Such feedback can carve low-opacity channels  extended into the CGM, creating a perforated gas geometry, while intense star formation can also photoionise surrounding gas, producing density-bounded nebulae. Both effects act to reduce the N$_\mathrm{HI}$ along sightlines, allowing the blueshifted \lya\ photons to escape more readily into the IGM.

\item We also observe a statistically significant anti-correlation between EW$_\mathrm{He\,II}$ and \lya\ peak separation, implying that harder ionising spectra effectively reduce the N$_\mathrm{HI}$, facilitating the escape of \lya\ photons, and leading to narrower peak separations.

\item A positive trend between FWHM$_\mathrm{He\,II}$ and \lya\ peak separation points to a shared kinematic influence between the ionised and neutral gas. Stronger feedback, traced by broader \heii\ emission, likely increases turbulence in the ISM/CGM, enhancing \lya\ scattering and shifting photon escape further from the line centre, thereby widening peak separation.

\item Finally, we identify four sources exhibiting \lya\ and \heii\ emission without any UV metal lines. While this may suggest chemically primitive systems with hard ionising radiation fields, deeper observations are required to confirm the absence of faint metal-line emission.

\end{itemize}

Our analysis suggests that the \heii\ emission in our sample is primarily driven by star formation in a low-$Z$ regime. While we can rule out the dominance of classical WR stars due to the absence of broad-line signatures, lack of optical diagnostics, and their scarcity in low-$Z$ environments, we cannot fully exclude contributions from low-metallicity WN/WNh-type without broad features. Additionally, contributions from XRBs or radiative shocks remain a possibility. 

In conclusion, our study highlights a clear connection between \heii\ emission and \lya\ radiative transfer conditions. Overall, the \heii-\lya\ diagnostics suggest that physical environments that favour the production of \heii\ also influence the kinematics and structure of the surrounding ISM/CGM, thus shaping the escape pathways of \lya. However, to confirm these relations and further explore the underlying physics, a larger dataset with rest-frame optical spectroscopic coverage is required.

\begin{acknowledgement}

We thank the ESO staff, and in particular the staff at Paranal Observatory, for carrying out the MAGPI observations. MAGPI targets were selected from GAMA. GAMA is a joint European-Australasian project based around a spectroscopic campaign using the Anglo-Australian Telescope. GAMA was funded by the STFC (UK), the ARC (Australia), the AAO, and the participating institutions. GAMA photometry is based on observations made with ESO Telescopes at the La Silla Paranal Observatory under programme ID 179.A-2004, ID 177.A-3016. The MAGPI team acknowledge support from the Australian Research Council Centre of Excellence for All Sky Astrophysics in 3 Dimensions (ASTRO 3D), through project number CE170100013. TN thanks support through ARC Discovery Project Grant DP230103161. LMV acknowledges support by the German Academic Scholarship Foundation (Studienstiftung des deutschen Volkes) and the Marianne-Plehn-Program of the Elite Network of Bavaria. I.B. has received funding from the European Union's Horizon 2020 research and innovation programme under the Marie Sklodowska Curie Grant agreement ID n.º 101059532. This project was extended for 6 months by the Franziska Seidl Funding Program of the University of Vienna.

\end{acknowledgement}

\textbf{Data Availability Statement:} All MUSE data used in this work are publicly accessible through the ESO archive (\href{http://archive.eso.org/cms.html}{http://archive.eso.org/cms.html}). Data products such as fully reduced data cubes and emission line fits will be made available as part of an upcoming MAGPI team data release (Mendel et al. in prep).

% PASA uses footnotes, not endnotes. \endnote in this template will behave like \footnote; and \printendnotes will not output anything.
% \printendnotes

\bibliography{bibtemplate}

@INPROCEEDINGS{2010SPIE.7735E..08B,
       author = {{Bacon}, R. and {Accardo}, M. and {Adjali}, L. and {Anwand}, H. and {Bauer}, S. and {Biswas}, I. and {Blaizot}, J. and {Boudon}, D. and {Brau-Nogue}, S. and {Brinchmann}, J. and {Caillier}, P. and {Capoani}, L. and {Carollo}, C.~M. and {Contini}, T. and {Couderc}, P. and {Daguis{\'e}}, E. and {Deiries}, S. and {Delabre}, B. and {Dreizler}, S. and {Dubois}, J. and {Dupieux}, M. and {Dupuy}, C. and {Emsellem}, E. and {Fechner}, T. and {Fleischmann}, A. and {Fran{\c{c}}ois}, M. and {Gallou}, G. and {Gharsa}, T. and {Glindemann}, A. and {Gojak}, D. and {Guiderdoni}, B. and {Hansali}, G. and {Hahn}, T. and {Jarno}, A. and {Kelz}, A. and {Koehler}, C. and {Kosmalski}, J. and {Laurent}, F. and {Le Floch}, M. and {Lilly}, S.~J. and {Lizon}, J.-L. and {Loupias}, M. and {Manescau}, A. and {Monstein}, C. and {Nicklas}, H. and {Olaya}, J.-C. and {Pares}, L. and {Pasquini}, L. and {P{\'e}contal-Rousset}, A. and {Pell{\'o}}, R. and {Petit}, C. and {Popow}, E. and {Reiss}, R. and {Remillieux}, A. and {Renault}, E. and {Roth}, M. and {Rupprecht}, G. and {Serre}, D. and {Schaye}, J. and {Soucail}, G. and {Steinmetz}, M. and {Streicher}, O. and {Stuik}, R. and {Valentin}, H. and {Vernet}, J. and {Weilbacher}, P. and {Wisotzki}, L. and {Yerle}, N.},
        title = "{The MUSE second-generation VLT instrument}",
    booktitle = {Ground-based and Airborne Instrumentation for Astronomy III},
         year = 2010,
       editor = {{McLean}, Ian S. and {Ramsay}, Suzanne K. and {Takami}, Hideki},
       series = {Society of Photo-Optical Instrumentation Engineers (SPIE) Conference Series},
       volume = {7735},
        month = jul,
          eid = {773508},
        pages = {773508},
          doi = {10.1117/12.856027},
archivePrefix = {arXiv},
       eprint = {2211.16795},
 primaryClass = {astro-ph.IM},
       adsurl = {https://ui.adsabs.harvard.edu/abs/2010SPIE.7735E..08B}
}

@article{Schneider_2018,
   title={An excess of massive stars in the local 30 Doradus starburst},
   volume={359},
   ISSN={1095-9203},
   url={http://dx.doi.org/10.1126/science.aan0106},
   DOI={10.1126/science.aan0106},
   number={6371},
   journal={Science},
   publisher={American Association for the Advancement of Science (AAAS)},
   author={Schneider, F. R. N. and Sana, H. and Evans, C. J. and Bestenlehner, J. M. and Castro, N. and Fossati, L. and Gräfener, G. and Langer, N. and Ramírez-Agudelo, O. H. and Sabín-Sanjulián, C. and Simón-Díaz, S. and Tramper, F. and Crowther, P. A. and de Koter, A. and de Mink, S. E. and Dufton, P. L. and Garcia, M. and Gieles, M. and Hénault-Brunet, V. and Herrero, A. and Izzard, R. G. and Kalari, V. and Lennon, D. J. and Maíz Apellániz, J. and Markova, N. and Najarro, F. and Podsiadlowski, Ph. and Puls, J. and Taylor, W. D. and van Loon, J. Th. and Vink, J. S. and Norman, C.},
   year={2018},
   month=Jan, pages={69–71} }

@article{Crowther_2010,
   title={The R136 star cluster hosts several stars whose individual masses greatly exceed the accepted 150 M⊙ stellar mass limit: Very massive stars in R136 and NGC 3603},
   volume={408},
   ISSN={0035-8711},
   url={http://dx.doi.org/10.1111/j.1365-2966.2010.17167.x},
   DOI={10.1111/j.1365-2966.2010.17167.x},
   number={2},
   journal={Monthly Notices of the Royal Astronomical Society},
   publisher={Oxford University Press (OUP)},
   author={Crowther, Paul A. and Schnurr, Olivier and Hirschi, Raphael and Yusof, Norhasliza and Parker, Richard J. and Goodwin, Simon P. and Kassim, Hasan Abu},
   year={2010},
   month=July, pages={731–751} }

@software{2016ascl.soft11003B,
       author = {{Bacon}, Roland and {Piqueras}, Laure and {Conseil}, Simon and {Richard}, Johan and {Shepherd}, Martin},
        title = "{MPDAF: MUSE Python Data Analysis Framework}",
 howpublished = {Astrophysics Source Code Library, record ascl:1611.003},
         year = 2016,
        month = nov,
          eid = {ascl:1611.003},
archivePrefix = {ascl},
       eprint = {1611.003},
       adsurl = {https://ui.adsabs.harvard.edu/abs/2016ascl.soft11003B}
}

@article{Mukherjee2026BLUEPRINT,
  author  = {Mukherjee, Tamal and
             Li, Zhihui and
             Zafar, Tayyaba and
             Nanayakkara, Themiya and
             Tornotti, Davide and
             Costantin, Luca and
             Aalia, Imam Uzma},
  title   = {{BLUEPRINT}: Blue-dominant {Lyman-{\alpha}} emission as evidence of gas inflow in ultra-low-mass galaxies at {z} = 3},
  journal = {Astronomy \& Astrophysics},
  year    = {2026},
  volume  = {710},
  pages   = {A331},
  doi     = {10.1051/0004-6361/202659780},
}

@article{Battisti2026MAGPI,
  author  = {Battisti, A. J. and
             Muller, E. G. M. and
             Wisnioski, E. and
             Mendel, J. T. and
             Foster, C. and
             Lagos, C. D. P. and
             Harborne, K. E. and
             Aalia, I. U. and
             Barsanti, S. and
             Breda, I. and
             Calzetti, D. and
             Croom, S. M. and
             Das, P. K. and
             Ferr{\'e}-Mateu, A. and
             Gao, T. and
             Gjergo, E. and
             Grasha, K. and
             Mai, Y. and
             Mailvaganam, A. and
             Mukherjee, T. and
             Mun, M. and
             Remus, R.-S. and
             Sharma, G. and
             Sweet, S. M. and
             Thater, S. and
             Valenzuela, L. M. and
             van de Sande, J. and
             van de Ven, G. and
             Zafar, T. and
             Ziegler, B.},
  title   = {The {MAGPI} Survey: Emission Line Products Data Release and the Role of Spectroscopic Aperture Covering Fraction on the Balmer Decrement-Stellar Mass Relation},
  journal = {Monthly Notices of the Royal Astronomical Society},
  year    = {2026},
  volume  = {552},
  number  = {1},
  pages   = {stag1585},
  doi     = {10.1093/mnras/stag1585},
}

@ARTICLE{2017arXiv171003554P,
       author = {{Piqueras}, Laure and {Conseil}, Simon and {Shepherd}, Martin and {Bacon}, Roland and {Leclercq}, Floriane and {Richard}, Johan},
        title = "{MPDAF - A Python package for the analysis of VLT/MUSE data}",
      journal = {arXiv e-prints},
         year = 2017,
        month = oct,
          eid = {arXiv:1710.03554},
        pages = {arXiv:1710.03554},
          doi = {10.48550/arXiv.1710.03554},
archivePrefix = {arXiv},
       eprint = {1710.03554},
 primaryClass = {astro-ph.IM},
       adsurl = {https://ui.adsabs.harvard.edu/abs/2017arXiv171003554P}
}

@article{article,
author = {Ott, Thomas},
year = {2012},
month = {10},
pages = {10019-},
title = {QFitsView: FITS file viewer},
journal = {Astrophysics Source Code Library}
}

@ARTICLE{2021PASA...38...31F,
       author = {{Foster}, C. and {Mendel}, J.~T. and {Lagos}, C.~D.~P. and {Wisnioski}, E. and {Yuan}, T. and {D'Eugenio}, F. and {Barone}, T.~M. and {Harborne}, K.~E. and {Vaughan}, S.~P. and {Schulze}, F. and {Remus}, R.-S. and {Gupta}, A. and {Collacchioni}, F. and {Khim}, D.~J. and {Taylor}, P. and {Bassett}, R. and {Croom}, S.~M. and {McDermid}, R.~M. and {Poci}, A. and {Battisti}, A.~J. and {Bland-Hawthorn}, J. and {Bellstedt}, S. and {Colless}, M. and {Davies}, L.~J.~M. and {Derkenne}, C. and {Driver}, S. and {Ferr{\'e}-Mateu}, A. and {Fisher}, D.~B. and {Gjergo}, E. and {Johnston}, E.~J. and {Khalid}, A. and {Kobayashi}, C. and {Oh}, S. and {Peng}, Y. and {Robotham}, A.~S.~G. and {Sharda}, P. and {Sweet}, S.~M. and {Taylor}, E.~N. and {Tran}, K.-V.~H. and {Trayford}, J.~W. and {van de Sande}, J. and {Yi}, S.~K. and {Zanisi}, L.},
        title = "{The MAGPI survey: Science goals, design, observing strategy, early results and theoretical framework}",
      journal = {\pasa},
         year = 2021,
        month = jul,
       volume = {38},
          eid = {e031},
        pages = {e031},
          doi = {10.1017/pasa.2021.25},
archivePrefix = {arXiv},
       eprint = {2011.13567},
 primaryClass = {astro-ph.GA},
       adsurl = {https://ui.adsabs.harvard.edu/abs/2021PASA...38...31F}
}

@ARTICLE{2016MNRAS.462.1757G,
       author = {{Gutkin}, Julia and {Charlot}, St{\'e}phane and {Bruzual}, Gustavo},
        title = "{Modelling the nebular emission from primeval to present-day star-forming galaxies}",
      journal = {\mnras},
         year = 2016,
        month = oct,
       volume = {462},
       number = {2},
        pages = {1757-1774},
          doi = {10.1093/mnras/stw1716},
archivePrefix = {arXiv},
       eprint = {1607.06086},
 primaryClass = {astro-ph.GA},
       adsurl = {https://ui.adsabs.harvard.edu/abs/2016MNRAS.462.1757G}
}

@ARTICLE{2003MNRAS.344.1000B,
       author = {{Bruzual}, G. and {Charlot}, S.},
        title = "{Stellar population synthesis at the resolution of 2003}",
      journal = {\mnras},
         year = 2003,
        month = oct,
       volume = {344},
       number = {4},
        pages = {1000-1028},
          doi = {10.1046/j.1365-8711.2003.06897.x},
archivePrefix = {arXiv},
       eprint = {astro-ph/0309134},
 primaryClass = {astro-ph},
       adsurl = {https://ui.adsabs.harvard.edu/abs/2003MNRAS.344.1000B}
}

@ARTICLE{2013RMxAA..49..137F,
       author = {{Ferland}, G.~J. and {Porter}, R.~L. and {van Hoof}, P.~A.~M. and {Williams}, R.~J.~R. and {Abel}, N.~P. and {Lykins}, M.~L. and {Shaw}, G. and {Henney}, W.~J. and {Stancil}, P.~C.},
        title = "{The 2013 Release of Cloudy}",
      journal = {\rmxaa},
         year = 2013,
        month = apr,
       volume = {49},
        pages = {137-163},
          doi = {10.48550/arXiv.1302.4485},
archivePrefix = {arXiv},
       eprint = {1302.4485},
 primaryClass = {astro-ph.GA},
       adsurl = {https://ui.adsabs.harvard.edu/abs/2013RMxAA..49..137F}
}

@ARTICLE{2001MNRAS.323..887C,
       author = {{Charlot}, St{\'e}phane and {Longhetti}, Marcella},
        title = "{Nebular emission from star-forming galaxies}",
      journal = {\mnras},
         year = 2001,
        month = may,
       volume = {323},
       number = {4},
        pages = {887-903},
          doi = {10.1046/j.1365-8711.2001.04260.x},
archivePrefix = {arXiv},
       eprint = {astro-ph/0101097},
 primaryClass = {astro-ph},
       adsurl = {https://ui.adsabs.harvard.edu/abs/2001MNRAS.323..887C}
}

@ARTICLE{2018ApJ...858...99S,
       author = {{Sanders}, Ryan L. and {Shapley}, Alice E. and {Kriek}, Mariska and {Freeman}, William R. and {Reddy}, Naveen A. and {Siana}, Brian and {Coil}, Alison L. and {Mobasher}, Bahram and {Dav{\'e}}, Romeel and {Shivaei}, Irene and {Azadi}, Mojegan and {Price}, Sedona H. and {Leung}, Gene and {Fetherolf}, Tara and {de Groot}, Laura and {Zick}, Tom and {Fornasini}, Francesca M. and {Barro}, Guillermo},
        title = "{The MOSDEF Survey: A Stellar Mass-SFR-Metallicity Relation Exists at z {\ensuremath{\sim}} 2.3}",
      journal = {\apj},
         year = 2018,
        month = may,
       volume = {858},
       number = {2},
          eid = {99},
        pages = {99},
          doi = {10.3847/1538-4357/aabcbd},
archivePrefix = {arXiv},
       eprint = {1711.00224},
 primaryClass = {astro-ph.GA},
       adsurl = {https://ui.adsabs.harvard.edu/abs/2018ApJ...858...99S}
}

@ARTICLE{2025A&A...702A...3G,
       author = {{Gonz{\'a}lez-D{\'\i}az}, R. and {V{\'\i}lchez}, J.~M. and {Kehrig}, C. and {del Moral-Castro}, I. and {Iglesias-P{\'a}ramo}, J.},
        title = "{He II emitters at cosmic noon and beyond: Characterising the He II {\ensuremath{\lambda}}1640 emission with MUSE and JWST/NIRSpec}",
      journal = {\aap},
         year = 2025,
        month = oct,
       volume = {702},
          eid = {A3},
        pages = {A3},
          doi = {10.1051/0004-6361/202555935},
archivePrefix = {arXiv},
       eprint = {2506.11685},
 primaryClass = {astro-ph.GA},
       adsurl = {https://ui.adsabs.harvard.edu/abs/2025A&A...702A...3G}
}

@ARTICLE{2018ApJ...859..164B,
       author = {{Berg}, Danielle A. and {Erb}, Dawn K. and {Auger}, Matthew W. and {Pettini}, Max and {Brammer}, Gabriel B.},
        title = "{A Window on the Earliest Star Formation: Extreme Photoionization Conditions of a High-ionization, Low-metallicity Lensed Galaxy at z {\ensuremath{\sim}} 2*}",
      journal = {\apj},
         year = 2018,
        month = jun,
       volume = {859},
       number = {2},
          eid = {164},
        pages = {164},
          doi = {10.3847/1538-4357/aab7fa},
archivePrefix = {arXiv},
       eprint = {1803.02340},
 primaryClass = {astro-ph.GA},
       adsurl = {https://ui.adsabs.harvard.edu/abs/2018ApJ...859..164B}
}

@ARTICLE{2011MNRAS.413..971D,
       author = {{Driver}, S.~P. and {Hill}, D.~T. and {Kelvin}, L.~S. and {Robotham}, A.~S.~G. and {Liske}, J. and {Norberg}, P. and {Baldry}, I.~K. and {Bamford}, S.~P. and {Hopkins}, A.~M. and {Loveday}, J. and {Peacock}, J.~A. and {Andrae}, E. and {Bland-Hawthorn}, J. and {Brough}, S. and {Brown}, M.~J.~I. and {Cameron}, E. and {Ching}, J.~H.~Y. and {Colless}, M. and {Conselice}, C.~J. and {Croom}, S.~M. and {Cross}, N.~J.~G. and {de Propris}, R. and {Dye}, S. and {Drinkwater}, M.~J. and {Ellis}, S. and {Graham}, Alister W. and {Grootes}, M.~W. and {Gunawardhana}, M. and {Jones}, D.~H. and {van Kampen}, E. and {Maraston}, C. and {Nichol}, R.~C. and {Parkinson}, H.~R. and {Phillipps}, S. and {Pimbblet}, K. and {Popescu}, C.~C. and {Prescott}, M. and {Roseboom}, I.~G. and {Sadler}, E.~M. and {Sansom}, A.~E. and {Sharp}, R.~G. and {Smith}, D.~J.~B. and {Taylor}, E. and {Thomas}, D. and {Tuffs}, R.~J. and {Wijesinghe}, D. and {Dunne}, L. and {Frenk}, C.~S. and {Jarvis}, M.~J. and {Madore}, B.~F. and {Meyer}, M.~J. and {Seibert}, M. and {Staveley-Smith}, L. and {Sutherland}, W.~J. and {Warren}, S.~J.},
        title = "{Galaxy and Mass Assembly (GAMA): survey diagnostics and core data release}",
      journal = {\mnras},
         year = 2011,
        month = may,
       volume = {413},
       number = {2},
        pages = {971-995},
          doi = {10.1111/j.1365-2966.2010.18188.x},
archivePrefix = {arXiv},
       eprint = {1009.0614},
 primaryClass = {astro-ph.CO},
       adsurl = {https://ui.adsabs.harvard.edu/abs/2011MNRAS.413..971D}
}

@ARTICLE{2016MNRAS.456.3354F,
       author = {{Feltre}, A. and {Charlot}, S. and {Gutkin}, J.},
        title = "{Nuclear activity versus star formation: emission-line diagnostics at ultraviolet and optical wavelengths}",
      journal = {\mnras},
         year = 2016,
        month = mar,
       volume = {456},
       number = {3},
        pages = {3354-3374},
          doi = {10.1093/mnras/stv2794},
archivePrefix = {arXiv},
       eprint = {1511.08217},
 primaryClass = {astro-ph.GA},
       adsurl = {https://ui.adsabs.harvard.edu/abs/2016MNRAS.456.3354F}
}

@ARTICLE{2016ApJ...833..136J,
       author = {{Jaskot}, A.~E. and {Ravindranath}, S.},
        title = "{Photoionization Models for the Semi-forbidden C III] 1909 Emission in Star-forming Galaxies}",
      journal = {\apj},
         year = 2016,
        month = dec,
       volume = {833},
       number = {2},
          eid = {136},
        pages = {136},
          doi = {10.3847/1538-4357/833/2/136},
archivePrefix = {arXiv},
       eprint = {1610.03778},
 primaryClass = {astro-ph.GA},
       adsurl = {https://ui.adsabs.harvard.edu/abs/2016ApJ...833..136J}
}

@ARTICLE{2018A&A...612A..94N,
       author = {{Nakajima}, K. and {Schaerer}, D. and {Le F{\`e}vre}, O. and {Amor{\'\i}n}, R. and {Talia}, M. and {Lemaux}, B.~C. and {Tasca}, L.~A.~M. and {Vanzella}, E. and {Zamorani}, G. and {Bardelli}, S. and {Grazian}, A. and {Guaita}, L. and {Hathi}, N.~P. and {Pentericci}, L. and {Zucca}, E.},
        title = "{The VIMOS Ultra Deep Survey: Nature, ISM properties, and ionizing spectra of CIII]{\ensuremath{\lambda}}1909 emitters at z = 2-4}",
      journal = {\aap},
         year = 2018,
        month = may,
       volume = {612},
          eid = {A94},
        pages = {A94},
          doi = {10.1051/0004-6361/201731935},
archivePrefix = {arXiv},
       eprint = {1709.03990},
 primaryClass = {astro-ph.GA},
       adsurl = {https://ui.adsabs.harvard.edu/abs/2018A&A...612A..94N}
}

@ARTICLE{2024ApJ...962...95M,
       author = {{Mingozzi}, Matilde and {James}, Bethan L. and {Berg}, Danielle A. and {Arellano-C{\'o}rdova}, Karla Z. and {Plat}, Adele and {Scarlata}, Claudia and {Aloisi}, Alessandra and {Amor{\'\i}n}, Ricardo O. and {Brinchmann}, Jarle and {Charlot}, St{\'e}phane and {Chisholm}, John and {Feltre}, Anna and {Gazagnes}, Simon and {Hayes}, Matthew and {Heckman}, Timothy and {Hernandez}, Svea and {Kewley}, Lisa J. and {Kumari}, Nimisha and {Leitherer}, Claus and {Martin}, Crystal L. and {Maseda}, Michael and {Nanayakkara}, Themiya and {Ravindranath}, Swara and {Rigby}, Jane R. and {Senchyna}, Peter and {Skillman}, Evan D. and {Sugahara}, Yuma and {Wilkins}, Stephen M. and {Wofford}, Aida and {Xu}, Xinfeng},
        title = "{CLASSY. VIII. Exploring the Source of Ionization with UV Interstellar Medium Diagnostics in Local High-z Analogs}",
      journal = {\apj},
         year = 2024,
        month = feb,
       volume = {962},
       number = {1},
          eid = {95},
        pages = {95},
          doi = {10.3847/1538-4357/ad1033},
archivePrefix = {arXiv},
       eprint = {2306.15062},
 primaryClass = {astro-ph.GA},
       adsurl = {https://ui.adsabs.harvard.edu/abs/2024ApJ...962...95M}
}

@ARTICLE{2025ApJ...982...14H,
       author = {{Hayes}, Matthew J. and {Saldana-Lopez}, Alberto and {Citro}, Annalisa and {James}, Bethan L. and {Mingozzi}, Matilde and {Scarlata}, Claudia and {Martinez}, Zorayda and {Berg}, Danielle A.},
        title = "{On the Average Ultraviolet Emission-line Spectra of High-redshift Galaxies: Hot and Cold, Carbon-poor, Nitrogen Modest, and Oozing Ionizing Photons}",
      journal = {\apj},
         year = 2025,
        month = mar,
       volume = {982},
       number = {1},
          eid = {14},
        pages = {14},
          doi = {10.3847/1538-4357/adaea1},
archivePrefix = {arXiv},
       eprint = {2411.09262},
 primaryClass = {astro-ph.GA},
       adsurl = {https://ui.adsabs.harvard.edu/abs/2025ApJ...982...14H}
}

@ARTICLE{2010ApJ...719.1168E,
       author = {{Erb}, Dawn K. and {Pettini}, Max and {Shapley}, Alice E. and {Steidel}, Charles C. and {Law}, David R. and {Reddy}, Naveen A.},
        title = "{Physical Conditions in a Young, Unreddened, Low-metallicity Galaxy at High Redshift}",
      journal = {\apj},
         year = 2010,
        month = aug,
       volume = {719},
       number = {2},
        pages = {1168-1190},
          doi = {10.1088/0004-637X/719/2/1168},
archivePrefix = {arXiv},
       eprint = {1006.5456},
 primaryClass = {astro-ph.CO},
       adsurl = {https://ui.adsabs.harvard.edu/abs/2010ApJ...719.1168E}
}

@article{10.1093/mnras/stv2859,
    author = {Patrício, Vera and Richard, Johan and Verhamme, Anne and Wisotzki, Lutz and Brinchmann, Jarle and Turner, Monica L. and Christensen, Lise and Weilbacher, Peter M. and Blaizot, Jérémy and Bacon, Roland and Contini, Thierry and Lagattuta, David and Cantalupo, Sebastiano and Clément, Benjamin and Soucail, Geneviève},
    title = {A young star-forming galaxy at z = 3.5 with an extended Lyman α halo seen with MUSE},
    journal = {Monthly Notices of the Royal Astronomical Society},
    volume = {456},
    number = {4},
    pages = {4191-4208},
    year = {2016},
    month = {01},
    issn = {0035-8711},
    doi = {10.1093/mnras/stv2859},
    url = {https://doi.org/10.1093/mnras/stv2859},
    eprint = {https://academic.oup.com/mnras/article-pdf/456/4/4191/9380736/stv2859.pdf},
}

@ARTICLE{2017MNRAS.472.2608S,
       author = {{Senchyna}, Peter and {Stark}, Daniel P. and {Vidal-Garc{\'\i}a}, Alba and {Chevallard}, Jacopo and {Charlot}, St{\'e}phane and {Mainali}, Ramesh and {Jones}, Tucker and {Wofford}, Aida and {Feltre}, Anna and {Gutkin}, Julia},
        title = "{Ultraviolet spectra of extreme nearby star-forming regions - approaching a local reference sample for JWST}",
      journal = {\mnras},
         year = 2017,
        month = dec,
       volume = {472},
       number = {3},
        pages = {2608-2632},
          doi = {10.1093/mnras/stx2059},
archivePrefix = {arXiv},
       eprint = {1706.00881},
 primaryClass = {astro-ph.GA},
       adsurl = {https://ui.adsabs.harvard.edu/abs/2017MNRAS.472.2608S}
}

@ARTICLE{2016ApJ...827..126B,
       author = {{Berg}, Danielle A. and {Skillman}, Evan D. and {Henry}, Richard B.~C. and {Erb}, Dawn K. and {Carigi}, Leticia},
        title = "{Carbon and Oxygen Abundances in Low Metallicity Dwarf Galaxies}",
      journal = {\apj},
         year = 2016,
        month = aug,
       volume = {827},
       number = {2},
          eid = {126},
        pages = {126},
          doi = {10.3847/0004-637X/827/2/126},
archivePrefix = {arXiv},
       eprint = {1605.06152},
 primaryClass = {astro-ph.GA},
       adsurl = {https://ui.adsabs.harvard.edu/abs/2016ApJ...827..126B}
}

@ARTICLE{2019A&A...624A..89N,
       author = {{Nanayakkara}, Themiya and {Brinchmann}, Jarle and {Boogaard}, Leindert and {Bouwens}, Rychard and {Cantalupo}, Sebastiano and {Feltre}, Anna and {Kollatschny}, Wolfram and {Marino}, Raffaella Anna and {Maseda}, Michael and {Matthee}, Jorryt and {Paalvast}, Mieke and {Richard}, Johan and {Verhamme}, Anne},
        title = "{Exploring He II {\ensuremath{\lambda}}1640 emission line properties at z {\ensuremath{\sim}}2-4}",
      journal = {\aap},
         year = 2019,
        month = apr,
       volume = {624},
          eid = {A89},
        pages = {A89},
          doi = {10.1051/0004-6361/201834565},
archivePrefix = {arXiv},
       eprint = {1902.05960},
 primaryClass = {astro-ph.GA},
       adsurl = {https://ui.adsabs.harvard.edu/abs/2019A&A...624A..89N}
}

@ARTICLE{2014MNRAS.445.3200S,
       author = {{Stark}, Daniel P. and {Richard}, Johan and {Siana}, Brian and {Charlot}, St{\'e}phane and {Freeman}, William R. and {Gutkin}, Julia and {Wofford}, Aida and {Robertson}, Brant and {Amanullah}, Rahman and {Watson}, Darach and {Milvang-Jensen}, Bo},
        title = "{Ultraviolet emission lines in young low-mass galaxies at z ≃ 2: physical properties and implications for studies at z > 7}",
      journal = {\mnras},
         year = 2014,
        month = dec,
       volume = {445},
       number = {3},
        pages = {3200-3220},
          doi = {10.1093/mnras/stu1618},
archivePrefix = {arXiv},
       eprint = {1408.1420},
 primaryClass = {astro-ph.GA},
       adsurl = {https://ui.adsabs.harvard.edu/abs/2014MNRAS.445.3200S}
}

@ARTICLE{2020A&A...636A..47S,
       author = {{Saxena}, A. and {Pentericci}, L. and {Mirabelli}, M. and {Schaerer}, D. and {Schneider}, R. and {Cullen}, F. and {Amorin}, R. and {Bolzonella}, M. and {Bongiorno}, A. and {Carnall}, A.~C. and {Castellano}, M. and {Cucciati}, O. and {Fontana}, A. and {Fynbo}, J.~P.~U. and {Garilli}, B. and {Gargiulo}, A. and {Guaita}, L. and {Hathi}, N.~P. and {Hutchison}, T.~A. and {Koekemoer}, A.~M. and {Marchi}, F. and {McLeod}, D.~J. and {McLure}, R.~J. and {Papovich}, C. and {Pozzetti}, L. and {Talia}, M. and {Zamorani}, G.},
        title = "{The properties of He II{\ensuremath{\lambda}}1640 emitters at z {\ensuremath{\sim}} 2.5-5 from the VANDELS survey}",
      journal = {\aap},
         year = 2020,
        month = apr,
       volume = {636},
          eid = {A47},
        pages = {A47},
          doi = {10.1051/0004-6361/201937170},
archivePrefix = {arXiv},
       eprint = {1911.09999},
 primaryClass = {astro-ph.GA},
       adsurl = {https://ui.adsabs.harvard.edu/abs/2020A&A...636A..47S}
}

@ARTICLE{2015ApJ...803...34B,
       author = {{Bouwens}, R.~J. and {Illingworth}, G.~D. and {Oesch}, P.~A. and {Trenti}, M. and {Labb{\'e}}, I. and {Bradley}, L. and {Carollo}, M. and {van Dokkum}, P.~G. and {Gonzalez}, V. and {Holwerda}, B. and {Franx}, M. and {Spitler}, L. and {Smit}, R. and {Magee}, D.},
        title = "{UV Luminosity Functions at Redshifts z {\ensuremath{\sim}} 4 to z {\ensuremath{\sim}} 10: 10,000 Galaxies from HST Legacy Fields}",
      journal = {\apj},
         year = 2015,
        month = apr,
       volume = {803},
       number = {1},
          eid = {34},
        pages = {34},
          doi = {10.1088/0004-637X/803/1/34},
archivePrefix = {arXiv},
       eprint = {1403.4295},
 primaryClass = {astro-ph.CO},
       adsurl = {https://ui.adsabs.harvard.edu/abs/2015ApJ...803...34B}
}

@ARTICLE{2015ApJ...802L..19R,
       author = {{Robertson}, Brant E. and {Ellis}, Richard S. and {Furlanetto}, Steven R. and {Dunlop}, James S.},
        title = "{Cosmic Reionization and Early Star-forming Galaxies: A Joint Analysis of New Constraints from Planck and the Hubble Space Telescope}",
      journal = {\apjl},
         year = 2015,
        month = apr,
       volume = {802},
       number = {2},
          eid = {L19},
        pages = {L19},
          doi = {10.1088/2041-8205/802/2/L19},
archivePrefix = {arXiv},
       eprint = {1502.02024},
 primaryClass = {astro-ph.CO},
       adsurl = {https://ui.adsabs.harvard.edu/abs/2015ApJ...802L..19R}
}

@ARTICLE{2010Natur.468...49R,
       author = {{Robertson}, Brant E. and {Ellis}, Richard S. and {Dunlop}, James S. and {McLure}, Ross J. and {Stark}, Daniel P.},
        title = "{Early star-forming galaxies and the reionization of the Universe}",
      journal = {\nat},
         year = 2010,
        month = nov,
       volume = {468},
       number = {7320},
        pages = {49-55},
          doi = {10.1038/nature09527},
archivePrefix = {arXiv},
       eprint = {1011.0727},
 primaryClass = {astro-ph.CO},
       adsurl = {https://ui.adsabs.harvard.edu/abs/2010Natur.468...49R}
}

@ARTICLE{2004ApJ...605..579Y,
       author = {{Yoshida}, Naoki and {Bromm}, Volker and {Hernquist}, Lars},
        title = "{The Era of Massive Population III Stars: Cosmological Implications and Self-Termination}",
      journal = {\apj},
         year = 2004,
        month = apr,
       volume = {605},
       number = {2},
        pages = {579-590},
          doi = {10.1086/382499},
archivePrefix = {arXiv},
       eprint = {astro-ph/0310443},
 primaryClass = {astro-ph},
       adsurl = {https://ui.adsabs.harvard.edu/abs/2004ApJ...605..579Y}
}

@ARTICLE{2002A&A...382...28S,
       author = {{Schaerer}, D.},
        title = "{On the properties of massive Population III stars and metal-free stellar populations}",
      journal = {\aap},
         year = 2002,
        month = jan,
       volume = {382},
        pages = {28-42},
          doi = {10.1051/0004-6361:20011619},
archivePrefix = {arXiv},
       eprint = {astro-ph/0110697},
 primaryClass = {astro-ph},
       adsurl = {https://ui.adsabs.harvard.edu/abs/2002A&A...382...28S}
}

@ARTICLE{2000ApJ...528L..65T,
       author = {{Tumlinson}, Jason and {Shull}, J. Michael},
        title = "{Zero-Metallicity Stars and the Effects of the First Stars on Reionization}",
      journal = {\apjl},
         year = 2000,
        month = jan,
       volume = {528},
       number = {2},
        pages = {L65-L68},
          doi = {10.1086/312432},
archivePrefix = {arXiv},
       eprint = {astro-ph/9911339},
 primaryClass = {astro-ph},
       adsurl = {https://ui.adsabs.harvard.edu/abs/2000ApJ...528L..65T}
}

@ARTICLE{2001PhR...349..125B,
       author = {{Barkana}, R. and {Loeb}, A.},
        title = "{In the beginning: the first sources of light and the reionization of the universe}",
      journal = {\physrep},
         year = 2001,
        month = jul,
       volume = {349},
       number = {2},
        pages = {125-238},
          doi = {10.1016/S0370-1573(01)00019-9},
archivePrefix = {arXiv},
       eprint = {astro-ph/0010468},
 primaryClass = {astro-ph},
       adsurl = {https://ui.adsabs.harvard.edu/abs/2001PhR...349..125B}
}

@ARTICLE{2001ApJ...550L...1T,
       author = {{Tumlinson}, Jason and {Giroux}, Mark L. and {Shull}, J. Michael},
        title = "{Probing the First Stars with Hydrogen and Helium Recombination Emission}",
      journal = {\apjl},
         year = 2001,
        month = mar,
       volume = {550},
       number = {1},
        pages = {L1-L5},
          doi = {10.1086/319477},
archivePrefix = {arXiv},
       eprint = {astro-ph/0011303},
 primaryClass = {astro-ph},
       adsurl = {https://ui.adsabs.harvard.edu/abs/2001ApJ...550L...1T}
}

@ARTICLE{2012ApJ...745...50W,
       author = {{Wise}, John H. and {Turk}, Matthew J. and {Norman}, Michael L. and {Abel}, Tom},
        title = "{The Birth of a Galaxy: Primordial Metal Enrichment and Stellar Populations}",
      journal = {\apj},
         year = 2012,
        month = jan,
       volume = {745},
       number = {1},
          eid = {50},
        pages = {50},
          doi = {10.1088/0004-637X/745/1/50},
archivePrefix = {arXiv},
       eprint = {1011.2632},
 primaryClass = {astro-ph.CO},
       adsurl = {https://ui.adsabs.harvard.edu/abs/2012ApJ...745...50W}
}

@ARTICLE{1996MNRAS.283.1388M,
       author = {{Madau}, Piero and {Ferguson}, Henry C. and {Dickinson}, Mark E. and {Giavalisco}, Mauro and {Steidel}, Charles C. and {Fruchter}, Andrew},
        title = "{High-redshift galaxies in the Hubble Deep Field: colour selection and star formation history to z\raisebox{-0.5ex}\textasciitilde4}",
      journal = {\mnras},
         year = 1996,
        month = dec,
       volume = {283},
       number = {4},
        pages = {1388-1404},
          doi = {10.1093/mnras/283.4.1388},
archivePrefix = {arXiv},
       eprint = {astro-ph/9607172},
 primaryClass = {astro-ph},
       adsurl = {https://ui.adsabs.harvard.edu/abs/1996MNRAS.283.1388M}
}

@ARTICLE{2013ApJ...776L..27H,
       author = {{Henry}, Alaina and {Scarlata}, Claudia and {Dom{\'\i}nguez}, Alberto and {Malkan}, Matthew and {Martin}, Crystal L. and {Siana}, Brian and {Atek}, Hakim and {Bedregal}, Alejandro G. and {Colbert}, James W. and {Rafelski}, Marc and {Ross}, Nathaniel and {Teplitz}, Harry and {Bunker}, Andrew J. and {Dressler}, Alan and {Hathi}, Nimish and {Masters}, Daniel and {McCarthy}, Patrick and {Straughn}, Amber},
        title = "{Low Masses and High Redshifts: The Evolution of the Mass-Metallicity Relation}",
      journal = {\apjl},
         year = 2013,
        month = oct,
       volume = {776},
       number = {2},
          eid = {L27},
        pages = {L27},
          doi = {10.1088/2041-8205/776/2/L27},
archivePrefix = {arXiv},
       eprint = {1309.4458},
 primaryClass = {astro-ph.CO},
       adsurl = {https://ui.adsabs.harvard.edu/abs/2013ApJ...776L..27H}
}

@ARTICLE{2014ApJ...795..165S,
       author = {{Steidel}, Charles C. and {Rudie}, Gwen C. and {Strom}, Allison L. and {Pettini}, Max and {Reddy}, Naveen A. and {Shapley}, Alice E. and {Trainor}, Ryan F. and {Erb}, Dawn K. and {Turner}, Monica L. and {Konidaris}, Nicholas P. and {Kulas}, Kristin R. and {Mace}, Gregory and {Matthews}, Keith and {McLean}, Ian S.},
        title = "{Strong Nebular Line Ratios in the Spectra of z \raisebox{-0.5ex}\textasciitilde 2-3 Star Forming Galaxies: First Results from KBSS-MOSFIRE}",
      journal = {\apj},
         year = 2014,
        month = nov,
       volume = {795},
       number = {2},
          eid = {165},
        pages = {165},
          doi = {10.1088/0004-637X/795/2/165},
archivePrefix = {arXiv},
       eprint = {1405.5473},
 primaryClass = {astro-ph.GA},
       adsurl = {https://ui.adsabs.harvard.edu/abs/2014ApJ...795..165S}
}

@ARTICLE{2014MNRAS.440.2300C,
       author = {{Cullen}, F. and {Cirasuolo}, M. and {McLure}, R.~J. and {Dunlop}, J.~S. and {Bowler}, R.~A.~A.},
        title = "{The mass-metallicity-star formation rate relation at z {\ensuremath{\gtrsim}} 2 with 3D Hubble Space Telescope}",
      journal = {\mnras},
         year = 2014,
        month = may,
       volume = {440},
       number = {3},
        pages = {2300-2312},
          doi = {10.1093/mnras/stu443},
archivePrefix = {arXiv},
       eprint = {1310.0816},
 primaryClass = {astro-ph.CO},
       adsurl = {https://ui.adsabs.harvard.edu/abs/2014MNRAS.440.2300C}
}

@ARTICLE{2019MNRAS.487.2038C,
       author = {{Cullen}, F. and {McLure}, R.~J. and {Dunlop}, J.~S. and {Khochfar}, S. and {Dav{\'e}}, R. and {Amor{\'\i}n}, R. and {Bolzonella}, M. and {Carnall}, A.~C. and {Castellano}, M. and {Cimatti}, A. and {Cirasuolo}, M. and {Cresci}, G. and {Fynbo}, J.~P.~U. and {Fontanot}, F. and {Gargiulo}, A. and {Garilli}, B. and {Guaita}, L. and {Hathi}, N. and {Hibon}, P. and {Mannucci}, F. and {Marchi}, F. and {McLeod}, D.~J. and {Pentericci}, L. and {Pozzetti}, L. and {Shapley}, A.~E. and {Talia}, M. and {Zamorani}, G.},
        title = "{The VANDELS survey: the stellar metallicities of star-forming galaxies at 2.5 < z < 5.0}",
      journal = {\mnras},
         year = 2019,
        month = aug,
       volume = {487},
       number = {2},
        pages = {2038-2060},
          doi = {10.1093/mnras/stz1402},
archivePrefix = {arXiv},
       eprint = {1903.11081},
 primaryClass = {astro-ph.GA},
       adsurl = {https://ui.adsabs.harvard.edu/abs/2019MNRAS.487.2038C}
}

@ARTICLE{2017NatAs...1E..52A,
       author = {{Amor{\'\i}n}, Ricardo and {Fontana}, Adriano and {P{\'e}rez-Montero}, Enrique and {Castellano}, Marco and {Guaita}, Lucia and {Grazian}, Andrea and {Le F{\`e}vre}, Olivier and {Ribeiro}, Bruno and {Schaerer}, Daniel and {Tasca}, Lidia A.~M. and {Thomas}, Romain and {Bardelli}, Sandro and {Cassar{\`a}}, Letizia and {Cassata}, Paolo and {Cimatti}, Andrea and {Contini}, Thierry and {de Barros}, Stephane and {Garilli}, Bianca and {Giavalisco}, Mauro and {Hathi}, Nimish and {Koekemoer}, Anton and {Le Brun}, Vincent and {Lemaux}, Brian C. and {Maccagni}, Dario and {Pentericci}, Laura and {Pforr}, Janine and {Talia}, Margherita and {Tresse}, Laurence and {Vanzella}, Eros and {Vergani}, Daniela and {Zamorani}, Giovanni and {Zucca}, Elena and {Merlin}, Emiliano},
        title = "{Analogues of primeval galaxies two billion years after the Big Bang}",
      journal = {Nature Astronomy},
         year = 2017,
        month = mar,
       volume = {1},
          eid = {0052},
        pages = {0052},
          doi = {10.1038/s41550-017-0052},
archivePrefix = {arXiv},
       eprint = {1701.04416},
 primaryClass = {astro-ph.GA},
       adsurl = {https://ui.adsabs.harvard.edu/abs/2017NatAs...1E..52A}
}

@article{Sanders_2018,
doi = {10.3847/1538-4357/aabcbd},
url = {https://doi.org/10.3847/1538-4357/aabcbd},
year = {2018},
month = {may},
publisher = {The American Astronomical Society},
volume = {858},
number = {2},
pages = {99},
author = {Sanders, Ryan L. and Shapley, Alice E. and Kriek, Mariska and Freeman, William R. and Reddy, Naveen A. and Siana, Brian and Coil, Alison L. and Mobasher, Bahram and Dav√©, Romeel and Shivaei, Irene and Azadi, Mojegan and Price, Sedona H. and Leung, Gene and Fetherholf, Tara and Groot, Laura de and Zick, Tom and Fornasini, Francesca M. and Barro, Guillermo},
title = {The MOSDEF Survey: A Stellar Mass‚ÄìSFR‚ÄìMetallicity Relation Exists at z¬†‚àº¬†2.3‚àó},
journal = {The Astrophysical Journal},
}

@ARTICLE{2016ApJ...826..159S,
       author = {{Steidel}, Charles C. and {Strom}, Allison L. and {Pettini}, Max and {Rudie}, Gwen C. and {Reddy}, Naveen A. and {Trainor}, Ryan F.},
        title = "{Reconciling the Stellar and Nebular Spectra of High-redshift Galaxies}",
      journal = {\apj},
         year = 2016,
        month = aug,
       volume = {826},
       number = {2},
          eid = {159},
        pages = {159},
          doi = {10.3847/0004-637X/826/2/159},
archivePrefix = {arXiv},
       eprint = {1605.07186},
 primaryClass = {astro-ph.GA},
       adsurl = {https://ui.adsabs.harvard.edu/abs/2016ApJ...826..159S}
}

@ARTICLE{2011ARA&A..49..373B,
       author = {{Bromm}, Volker and {Yoshida}, Naoki},
        title = "{The First Galaxies}",
      journal = {\araa},
         year = 2011,
        month = sep,
       volume = {49},
       number = {1},
        pages = {373-407},
          doi = {10.1146/annurev-astro-081710-102608},
archivePrefix = {arXiv},
       eprint = {1102.4638},
 primaryClass = {astro-ph.CO},
       adsurl = {https://ui.adsabs.harvard.edu/abs/2011ARA&A..49..373B}
}

@ARTICLE{2013A&A...556A..68C,
       author = {{Cassata}, P. and {Le F{\`e}vre}, O. and {Charlot}, S. and {Contini}, T. and {Cucciati}, O. and {Garilli}, B. and {Zamorani}, G. and {Adami}, C. and {Bardelli}, S. and {Le Brun}, V. and {Lemaux}, B. and {Maccagni}, D. and {Pollo}, A. and {Pozzetti}, L. and {Tresse}, L. and {Vergani}, D. and {Zanichelli}, A. and {Zucca}, E.},
        title = "{He II emitters in the VIMOS VLT Deep Survey: Population III star formation or peculiar stellar populations in galaxies at 2 < z < 4.6?}",
      journal = {\aap},
         year = 2013,
        month = aug,
       volume = {556},
          eid = {A68},
        pages = {A68},
          doi = {10.1051/0004-6361/201220969},
archivePrefix = {arXiv},
       eprint = {1212.5270},
 primaryClass = {astro-ph.CO},
       adsurl = {https://ui.adsabs.harvard.edu/abs/2013A&A...556A..68C}
}

@ARTICLE{2018MNRAS.480.1081K,
       author = {{Kehrig}, C. and {V{\'\i}lchez}, J.~M. and {Guerrero}, M.~A. and {Iglesias-P{\'a}ramo}, J. and {Hunt}, L.~K. and {Duarte-Puertas}, S. and {Ramos-Larios}, G.},
        title = "{The extended He II {\ensuremath{\lambda}}4686 emission in the extremely metal-poor galaxy SBS 0335 - 052E seen with MUSE}",
      journal = {\mnras},
         year = 2018,
        month = oct,
       volume = {480},
       number = {1},
        pages = {1081-1095},
          doi = {10.1093/mnras/sty1920},
archivePrefix = {arXiv},
       eprint = {1807.09307},
 primaryClass = {astro-ph.GA},
       adsurl = {https://ui.adsabs.harvard.edu/abs/2018MNRAS.480.1081K}
}

@ARTICLE{2023ARA&A..61...65K,
       author = {{Klessen}, Ralf S. and {Glover}, Simon C.~O.},
        title = "{The First Stars: Formation, Properties, and Impact}",
      journal = {\araa},
         year = 2023,
        month = aug,
       volume = {61},
        pages = {65-130},
          doi = {10.1146/annurev-astro-071221-053453},
archivePrefix = {arXiv},
       eprint = {2303.12500},
 primaryClass = {astro-ph.CO},
       adsurl = {https://ui.adsabs.harvard.edu/abs/2023ARA&A..61...65K}
}

@ARTICLE{2020MNRAS.497.2839L,
       author = {{Liu}, Boyuan and {Bromm}, Volker},
        title = "{When did Population III star formation end?}",
      journal = {\mnras},
         year = 2020,
        month = sep,
       volume = {497},
       number = {3},
        pages = {2839-2854},
          doi = {10.1093/mnras/staa2143},
archivePrefix = {arXiv},
       eprint = {2006.15260},
 primaryClass = {astro-ph.GA},
       adsurl = {https://ui.adsabs.harvard.edu/abs/2020MNRAS.497.2839L}
}

@ARTICLE{2022ApJ...935..174S,
       author = {{Sarmento}, Richard and {Scannapieco}, Evan},
        title = "{The Effects of Radiative Feedback and Supernova-induced Turbulence on Early Galaxies}",
      journal = {\apj},
         year = 2022,
        month = aug,
       volume = {935},
       number = {2},
          eid = {174},
        pages = {174},
          doi = {10.3847/1538-4357/ac815c},
archivePrefix = {arXiv},
       eprint = {2207.07161},
 primaryClass = {astro-ph.GA},
       adsurl = {https://ui.adsabs.harvard.edu/abs/2022ApJ...935..174S}
}

@ARTICLE{2023MNRAS.522.3809V,
       author = {{Venditti}, Alessandra and {Graziani}, Luca and {Schneider}, Raffaella and {Pentericci}, Laura and {Di Cesare}, Claudia and {Maio}, Umberto and {Omukai}, Kazuyuki},
        title = "{A needle in a haystack? Catching Population III stars in the epoch of reionization: I. Population III star-forming environments}",
      journal = {\mnras},
         year = 2023,
        month = jul,
       volume = {522},
       number = {3},
        pages = {3809-3830},
          doi = {10.1093/mnras/stad1201},
archivePrefix = {arXiv},
       eprint = {2301.10259},
 primaryClass = {astro-ph.GA},
       adsurl = {https://ui.adsabs.harvard.edu/abs/2023MNRAS.522.3809V}
}

@article{Visbal_2020,
doi = {10.3847/1538-4357/ab994e},
url = {https://doi.org/10.3847/1538-4357/ab994e},
year = {2020},
month = {jul},
publisher = {The American Astronomical Society},
volume = {897},
number = {1},
pages = {95},
author = {Visbal, Eli and Bryan, Greg L. and Haiman, Zolt√°n},
title = {Self-consistent Semianalytic Modeling of Feedback during Primordial Star Formation and Reionization},
journal = {The Astrophysical Journal},
}

@ARTICLE{2003A&A...397..527S,
       author = {{Schaerer}, D.},
        title = "{The transition from Population III to normal galaxies: Lyalpha and He II emission and the ionising properties of high redshift starburst galaxies}",
      journal = {\aap},
         year = 2003,
        month = jan,
       volume = {397},
        pages = {527-538},
          doi = {10.1051/0004-6361:20021525},
archivePrefix = {arXiv},
       eprint = {astro-ph/0210462},
 primaryClass = {astro-ph},
       adsurl = {https://ui.adsabs.harvard.edu/abs/2003A&A...397..527S}
}

@ARTICLE{2010A&A...523A..64R,
       author = {{Raiter}, A. and {Schaerer}, D. and {Fosbury}, R.~A.~E.},
        title = "{Predicted UV properties of very metal-poor starburst galaxies}",
      journal = {\aap},
         year = 2010,
        month = nov,
       volume = {523},
          eid = {A64},
        pages = {A64},
          doi = {10.1051/0004-6361/201015236},
archivePrefix = {arXiv},
       eprint = {1008.2114},
 primaryClass = {astro-ph.CO},
       adsurl = {https://ui.adsabs.harvard.edu/abs/2010A&A...523A..64R}
}

@ARTICLE{1996MNRAS.281..163S,
       author = {{Smith}, Lindsey F. and {Shara}, Michael M. and {Moffat}, Anthony F.~J.},
        title = "{A three-dimensional classification for WN stars}",
      journal = {\mnras},
         year = 1996,
        month = jul,
       volume = {281},
       number = {1},
        pages = {163-191},
          doi = {10.1093/mnras/281.1.163},
       adsurl = {https://ui.adsabs.harvard.edu/abs/1996MNRAS.281..163S}
}

@ARTICLE{1998MNRAS.296..367C,
       author = {{Crowther}, P.~A. and {De Marco}, Orsola and {Barlow}, M.~J.},
        title = "{Quantitative classification of WC and WO stars}",
      journal = {\mnras},
         year = 1998,
        month = may,
       volume = {296},
       number = {2},
        pages = {367-378},
          doi = {10.1046/j.1365-8711.1998.01360.x},
       adsurl = {https://ui.adsabs.harvard.edu/abs/1998MNRAS.296..367C}
}

@ARTICLE{2022arXiv221105424S,
       author = {{Sander}, Andreas A.~C.},
        title = "{Massive Stars in the Far and Extreme Ultraviolet}",
      journal = {arXiv e-prints},
         year = 2022,
        month = nov,
          eid = {arXiv:2211.05424},
        pages = {arXiv:2211.05424},
          doi = {10.48550/arXiv.2211.05424},
archivePrefix = {arXiv},
       eprint = {2211.05424},
 primaryClass = {astro-ph.SR},
       adsurl = {https://ui.adsabs.harvard.edu/abs/2022arXiv221105424S}
}

@ARTICLE{2005A&A...442..587V,
       author = {{Vink}, Jorick S. and {de Koter}, A.},
        title = "{On the metallicity dependence of Wolf-Rayet winds}",
      journal = {\aap},
         year = 2005,
        month = nov,
       volume = {442},
       number = {2},
        pages = {587-596},
          doi = {10.1051/0004-6361:20052862},
archivePrefix = {arXiv},
       eprint = {astro-ph/0507352},
 primaryClass = {astro-ph},
       adsurl = {https://ui.adsabs.harvard.edu/abs/2005A&A...442..587V}
}

@ARTICLE{2023A&A...670A..83S,
       author = {{Sander}, A.~A.~C. and {Lefever}, R.~R. and {Poniatowski}, L.~G. and {Ramachandran}, V. and {Sabhahit}, G.~N. and {Vink}, J.~S.},
        title = "{The temperature dependency of Wolf-Rayet-type mass loss. An exploratory study for winds launched by the hot iron bump}",
      journal = {\aap},
         year = 2023,
        month = feb,
       volume = {670},
          eid = {A83},
        pages = {A83},
          doi = {10.1051/0004-6361/202245110},
archivePrefix = {arXiv},
       eprint = {2301.01785},
 primaryClass = {astro-ph.SR},
       adsurl = {https://ui.adsabs.harvard.edu/abs/2023A&A...670A..83S}
}

@ARTICLE{2019A&A...629A.134G,
       author = {{G{\"o}tberg}, Y. and {de Mink}, S.~E. and {Groh}, J.~H. and {Leitherer}, C. and {Norman}, C.},
        title = "{The impact of stars stripped in binaries on the integrated spectra of stellar populations}",
      journal = {\aap},
         year = 2019,
        month = sep,
       volume = {629},
          eid = {A134},
        pages = {A134},
          doi = {10.1051/0004-6361/201834525},
archivePrefix = {arXiv},
       eprint = {1908.06102},
 primaryClass = {astro-ph.GA},
       adsurl = {https://ui.adsabs.harvard.edu/abs/2019A&A...629A.134G}
}

@ARTICLE{1967ZA.....65..251K,
       author = {{Kippenhahn}, R. and {Weigert}, A.},
        title = "{Entwicklung in engen Doppelsternsystemen I. Massenaustausch vor und nach Beendigung des zentralen Wasserstoff-Brennens}",
      journal = {\zap},
         year = 1967,
        month = jan,
       volume = {65},
        pages = {251},
       adsurl = {https://ui.adsabs.harvard.edu/abs/1967ZA.....65..251K}
}

@ARTICLE{1967AcA....17..355P,
       author = {{Paczy{\'n}ski}, B.},
        title = "{Evolution of Close Binaries. V. The Evolution of Massive Binaries and the Formation of the Wolf-Rayet Stars}",
      journal = {\actaa},
         year = 1967,
        month = jan,
       volume = {17},
        pages = {355},
       adsurl = {https://ui.adsabs.harvard.edu/abs/1967AcA....17..355P}
}

@ARTICLE{2002ApJ...573..283P,
       author = {{Pfahl}, Eric and {Rappaport}, Saul and {Podsiadlowski}, Philipp},
        title = "{A Comprehensive Study of Neutron Star Retention in Globular Clusters}",
      journal = {\apj},
         year = 2002,
        month = jul,
       volume = {573},
       number = {1},
        pages = {283-305},
          doi = {10.1086/340494},
archivePrefix = {arXiv},
       eprint = {astro-ph/0106141},
 primaryClass = {astro-ph},
       adsurl = {https://ui.adsabs.harvard.edu/abs/2002ApJ...573..283P}
}

@INPROCEEDINGS{2008AIPC..990..230D,
       author = {{de Mink}, S.~E. and {Pols}, O.~R. and {Yoon}, S.-C.},
        title = "{Binaries at Low Metallicity: Ranges For Case A, B and C Mass Transfer}",
    booktitle = {First Stars III},
         year = 2008,
       editor = {{O'Shea}, Brian W. and {Heger}, Alexander},
       series = {American Institute of Physics Conference Series},
       volume = {990},
        month = mar,
    publisher = {AIP},
        pages = {230-232},
          doi = {10.1063/1.2905549},
archivePrefix = {arXiv},
       eprint = {0710.1010},
 primaryClass = {astro-ph},
       adsurl = {https://ui.adsabs.harvard.edu/abs/2008AIPC..990..230D}
}

@ARTICLE{2012MNRAS.419..479E,
       author = {{Eldridge}, John J. and {Stanway}, Elizabeth R.},
        title = "{The effect of stellar evolution uncertainties on the rest-frame ultraviolet stellar lines of C IV and He II in high-redshift Lyman-break galaxies}",
      journal = {\mnras},
         year = 2012,
        month = jan,
       volume = {419},
       number = {1},
        pages = {479-489},
          doi = {10.1111/j.1365-2966.2011.19713.x},
archivePrefix = {arXiv},
       eprint = {1109.0288},
 primaryClass = {astro-ph.CO},
       adsurl = {https://ui.adsabs.harvard.edu/abs/2012MNRAS.419..479E}
}

@ARTICLE{2016MNRAS.459.2992K,
       author = {{Kehrig}, C. and {V{\'\i}lchez}, J.~M. and {P{\'e}rez-Montero}, E. and {Iglesias-P{\'a}ramo}, J. and {Hern{\'a}ndez-Fern{\'a}ndez}, J.~D. and {Duarte Puertas}, S. and {Brinchmann}, J. and {Durret}, F. and {Kunth}, D.},
        title = "{Spatially resolved integral field spectroscopy of the ionized gas in IZw18}",
      journal = {\mnras},
         year = 2016,
        month = jul,
       volume = {459},
       number = {3},
        pages = {2992-3004},
          doi = {10.1093/mnras/stw806},
archivePrefix = {arXiv},
       eprint = {1604.08555},
 primaryClass = {astro-ph.GA},
       adsurl = {https://ui.adsabs.harvard.edu/abs/2016MNRAS.459.2992K}
}

@ARTICLE{2012MNRAS.421.1043S,
       author = {{Shirazi}, Maryam and {Brinchmann}, Jarle},
        title = "{Strongly star forming galaxies in the local Universe with nebular He II{\ensuremath{\lambda}}4686 emission}",
      journal = {\mnras},
         year = 2012,
        month = apr,
       volume = {421},
       number = {2},
        pages = {1043-1063},
          doi = {10.1111/j.1365-2966.2012.20439.x},
archivePrefix = {arXiv},
       eprint = {1201.1290},
 primaryClass = {astro-ph.CO},
       adsurl = {https://ui.adsabs.harvard.edu/abs/2012MNRAS.421.1043S}
}

@ARTICLE{2022ApJS..262...37J,
       author = {{James}, Bethan L. and {Berg}, Danielle A. and {King}, Teagan and {Sahnow}, David J. and {Mingozzi}, Matilde and {Chisholm}, John and {Heckman}, Timothy and {Martin}, Crystal L. and {Stark}, Dan P. and {Aloisi}, Alessandra and {Amor{\'\i}n}, Ricardo O. and {Arellano-C{\'o}rdova}, Karla Z. and {Bayliss}, Matthew and {Bordoloi}, Rongmon and {Brinchmann}, Jarle and {Charlot}, St{\'e}phane and {Chen}, Zuyi and {Chevallard}, Jacopo and {Clark}, Ilyse and {Erb}, Dawn K. and {Feltre}, Anna and {Hayes}, Matthew and {Henry}, Alaina and {Hernandez}, Svea and {Jaskot}, Anne and {Kewley}, Lisa J. and {Kumari}, Nimisha and {Leitherer}, Claus and {Llerena}, Mario and {Maseda}, Michael and {Nanayakkara}, Themiya and {Ouchi}, Masami and {Plat}, Adele and {Pogge}, Richard W. and {Ravindranath}, Swara and {Rigby}, Jane R. and {Scarlata}, Claudia and {Senchyna}, Peter and {Skillman}, Evan D. and {Steidel}, Charles C. and {Strom}, Allison L. and {Sugahara}, Yuma and {Wilkins}, Stephen M. and {Wofford}, Aida and {Xu}, Xinfeng and {Classy Team}},
        title = "{CLASSY. II. A Technical Overview of the COS Legacy Archive Spectroscopic Survey}",
      journal = {\apjs},
         year = 2022,
        month = oct,
       volume = {262},
       number = {2},
          eid = {37},
        pages = {37},
          doi = {10.3847/1538-4365/ac8008},
archivePrefix = {arXiv},
       eprint = {2206.01224},
 primaryClass = {astro-ph.GA},
       adsurl = {https://ui.adsabs.harvard.edu/abs/2022ApJS..262...37J}
}

@ARTICLE{2019A&A...622L..10S,
       author = {{Schaerer}, D. and {Fragos}, T. and {Izotov}, Y.~I.},
        title = "{X-ray binaries as the origin of nebular He II emission in low-metallicity star-forming galaxies}",
      journal = {\aap},
         year = 2019,
        month = feb,
       volume = {622},
          eid = {L10},
        pages = {L10},
          doi = {10.1051/0004-6361/201935005},
archivePrefix = {arXiv},
       eprint = {1902.10496},
 primaryClass = {astro-ph.GA},
       adsurl = {https://ui.adsabs.harvard.edu/abs/2019A&A...622L..10S}
}

@ARTICLE{2020MNRAS.496.3796S,
       author = {{Saxena}, A. and {Pentericci}, L. and {Schaerer}, D. and {Schneider}, R. and {Amorin}, R. and {Bongiorno}, A. and {Calabr{\`o}}, A. and {Castellano}, M. and {Cimatti}, A. and {Cullen}, F. and {Fontana}, A. and {Fynbo}, J.~P.~U. and {Hathi}, N. and {McLeod}, D.~J. and {Talia}, M. and {Zamorani}, G.},
        title = "{X-ray properties of He II {\ensuremath{\lambda}} 1640 emitting galaxies in VANDELS}",
      journal = {\mnras},
         year = 2020,
        month = aug,
       volume = {496},
       number = {3},
        pages = {3796-3807},
          doi = {10.1093/mnras/staa1805},
archivePrefix = {arXiv},
       eprint = {2003.13800},
 primaryClass = {astro-ph.GA},
       adsurl = {https://ui.adsabs.harvard.edu/abs/2020MNRAS.496.3796S}
}

@ARTICLE{2005ApJS..161..240T,
       author = {{Thuan}, Trinh X. and {Izotov}, Yuri I.},
        title = "{High-Ionization Emission in Metal-deficient Blue Compact Dwarf Galaxies}",
      journal = {\apjs},
         year = 2005,
        month = dec,
       volume = {161},
       number = {2},
        pages = {240-270},
          doi = {10.1086/491657},
archivePrefix = {arXiv},
       eprint = {astro-ph/0507209},
 primaryClass = {astro-ph},
       adsurl = {https://ui.adsabs.harvard.edu/abs/2005ApJS..161..240T}
}

@article{Allen_2008,
doi = {10.1086/589652},
url = {https://doi.org/10.1086/589652},
year = {2008},
month = {sep},
publisher = {},
volume = {178},
number = {1},
pages = {20},
author = {Allen, Mark G. and Groves, Brent A. and Dopita, Michael A. and Sutherland, Ralph S. and Kewley, Lisa J.},
title = {The MAPPINGS III Library of Fast Radiative Shock Models},
journal = {The Astrophysical Journal Supplement Series},
}

@ARTICLE{1996ApJS..102..161D,
       author = {{Dopita}, Michael A. and {Sutherland}, Ralph S.},
        title = "{Spectral Signatures of Fast Shocks. I. Low-Density Model Grid}",
      journal = {\apjs},
         year = 1996,
        month = jan,
       volume = {102},
        pages = {161},
          doi = {10.1086/192255},
       adsurl = {https://ui.adsabs.harvard.edu/abs/1996ApJS..102..161D}
}

@ARTICLE{2012MNRAS.427.1229I,
       author = {{Izotov}, Y.~I. and {Thuan}, T.~X. and {Privon}, G.},
        title = "{The detection of [Ne V] emission in five blue compact dwarf galaxies}",
      journal = {\mnras},
         year = 2012,
        month = dec,
       volume = {427},
       number = {2},
        pages = {1229-1237},
          doi = {10.1111/j.1365-2966.2012.22051.x},
archivePrefix = {arXiv},
       eprint = {1209.5265},
 primaryClass = {astro-ph.CO},
       adsurl = {https://ui.adsabs.harvard.edu/abs/2012MNRAS.427.1229I}
}

@ARTICLE{1902RSPTA.199....1J,
       author = {{Jeans}, J.~H.},
        title = "{The Stability of a Spherical Nebula}",
      journal = {Philosophical Transactions of the Royal Society of London Series A},
         year = 1902,
        month = jan,
       volume = {199},
        pages = {1-53},
          doi = {10.1098/rsta.1902.0012},
       adsurl = {https://ui.adsabs.harvard.edu/abs/1902RSPTA.199....1J}
}

@ARTICLE{2004ARA&A..42...79B,
       author = {{Bromm}, Volker and {Larson}, Richard B.},
        title = "{The First Stars}",
      journal = {\araa},
         year = 2004,
        month = sep,
       volume = {42},
       number = {1},
        pages = {79-118},
          doi = {10.1146/annurev.astro.42.053102.134034},
archivePrefix = {arXiv},
       eprint = {astro-ph/0311019},
 primaryClass = {astro-ph},
       adsurl = {https://ui.adsabs.harvard.edu/abs/2004ARA&A..42...79B}
}

@ARTICLE{2015A&A...578A...7V,
       author = {{Verhamme}, Anne and {Orlitov{\'a}}, Ivana and {Schaerer}, Daniel and {Hayes}, Matthew},
        title = "{Using Lyman-{\ensuremath{\alpha}} to detect galaxies that leak Lyman continuum}",
      journal = {\aap},
         year = 2015,
        month = jun,
       volume = {578},
          eid = {A7},
        pages = {A7},
          doi = {10.1051/0004-6361/201423978},
archivePrefix = {arXiv},
       eprint = {1404.2958},
 primaryClass = {astro-ph.GA},
       adsurl = {https://ui.adsabs.harvard.edu/abs/2015A&A...578A...7V}
}

@ARTICLE{2018MNRAS.478L..60V,
       author = {{Verhamme}, A. and {Garel}, T. and {Ventou}, E. and {Contini}, T. and {Bouch{\'e}}, N. and {Herenz}, EC and {Richard}, J. and {Bacon}, R. and {Schmidt}, KB and {Maseda}, M. and {Marino}, RA and {Brinchmann}, J. and {Cantalupo}, S. and {Caruana}, J. and {Cl{\'e}ment}, B. and {Diener}, C. and {Drake}, AB and {Hashimoto}, T. and {Inami}, H. and {Kerutt}, J. and {Kollatschny}, W. and {Leclercq}, F. and {Patr{\'\i}cio}, V. and {Schaye}, J. and {Wisotzki}, L. and {Zabl}, J.},
        title = "{Recovering the systemic redshift of galaxies from their Lyman alpha line profile}",
      journal = {\mnras},
         year = 2018,
        month = jul,
       volume = {478},
       number = {1},
        pages = {L60-L65},
          doi = {10.1093/mnrasl/sly058},
archivePrefix = {arXiv},
       eprint = {1804.01883},
 primaryClass = {astro-ph.GA},
       adsurl = {https://ui.adsabs.harvard.edu/abs/2018MNRAS.478L..60V}
}

@ARTICLE{2006ApJ...649...14D,
       author = {{Dijkstra}, Mark and {Haiman}, Zolt{\'a}n and {Spaans}, Marco},
        title = "{Ly{\ensuremath{\alpha}} Radiation from Collapsing Protogalaxies. I. Characteristics of the Emergent Spectrum}",
      journal = {\apj},
         year = 2006,
        month = sep,
       volume = {649},
       number = {1},
        pages = {14-36},
          doi = {10.1086/506243},
archivePrefix = {arXiv},
       eprint = {astro-ph/0510407},
 primaryClass = {astro-ph},
       adsurl = {https://ui.adsabs.harvard.edu/abs/2006ApJ...649...14D}
}

@ARTICLE{2023A&A...680L...5M,
       author = {{Mukherjee}, T. and {Zafar}, T. and {Nanayakkara}, T. and {Wisnioski}, E. and {Battisti}, A. and {Gupta}, A. and {Lagos}, C.~D.~P. and {Harborne}, K.~E. and {Foster}, C. and {Mendel}, T. and {Croom}, S.~M. and {Mailvaganam}, A. and {Prathap}, J.},
        title = "{Compact to extended Lyman-{\ensuremath{\alpha}} emitters in MAGPI: Strong blue-peak emission at z {\ensuremath{\gtrsim}} 3}",
      journal = {\aap},
         year = 2023,
        month = dec,
       volume = {680},
          eid = {L5},
        pages = {L5},
          doi = {10.1051/0004-6361/202348202},
archivePrefix = {arXiv},
       eprint = {2311.07019},
 primaryClass = {astro-ph.GA},
       adsurl = {https://ui.adsabs.harvard.edu/abs/2023A&A...680L...5M}
}

@ARTICLE{2011ApJ...730....5H,
       author = {{Heckman}, Timothy M. and {Borthakur}, Sanchayeeta and {Overzier}, Roderik and {Kauffmann}, Guinevere and {Basu-Zych}, Antara and {Leitherer}, Claus and {Sembach}, Ken and {Martin}, D. Chris and {Rich}, R. Michael and {Schiminovich}, David and {Seibert}, Mark},
        title = "{Extreme Feedback and the Epoch of Reionization: Clues in the Local Universe}",
      journal = {\apj},
         year = 2011,
        month = mar,
       volume = {730},
       number = {1},
          eid = {5},
        pages = {5},
          doi = {10.1088/0004-637X/730/1/5},
archivePrefix = {arXiv},
       eprint = {1101.4219},
 primaryClass = {astro-ph.CO},
       adsurl = {https://ui.adsabs.harvard.edu/abs/2011ApJ...730....5H}
}

@ARTICLE{2006A&A...447..863N,
       author = {{Nagao}, T. and {Maiolino}, R. and {Marconi}, A.},
        title = "{Gas metallicity in the narrow-line regions of high-redshift active galactic nuclei}",
      journal = {\aap},
         year = 2006,
        month = mar,
       volume = {447},
       number = {3},
        pages = {863-876},
          doi = {10.1051/0004-6361:20054127},
archivePrefix = {arXiv},
       eprint = {astro-ph/0508652},
 primaryClass = {astro-ph},
       adsurl = {https://ui.adsabs.harvard.edu/abs/2006A&A...447..863N}
}

@ARTICLE{2020ApJ...893....1B,
       author = {{Byler}, Nell and {Kewley}, Lisa J. and {Rigby}, Jane R. and {Acharyya}, Ayan and {Berg}, Danielle A. and {Bayliss}, Matthew and {Sharon}, Keren},
        title = "{A Comparison of UV and Optical Metallicities in Star-forming Galaxies}",
      journal = {\apj},
         year = 2020,
        month = apr,
       volume = {893},
       number = {1},
          eid = {1},
        pages = {1},
          doi = {10.3847/1538-4357/ab7ea9},
archivePrefix = {arXiv},
       eprint = {2003.06925},
 primaryClass = {astro-ph.GA},
       adsurl = {https://ui.adsabs.harvard.edu/abs/2020ApJ...893....1B}
}

@ARTICLE{2017MNRAS.470.3532V,
       author = {{Vidal-Garc{\'\i}a}, A. and {Charlot}, S. and {Bruzual}, G. and {Hubeny}, I.},
        title = "{Modelling ultraviolet-line diagnostics of stars, the ionized and the neutral interstellar medium in star-forming galaxies}",
      journal = {\mnras},
         year = 2017,
        month = sep,
       volume = {470},
       number = {3},
        pages = {3532-3556},
          doi = {10.1093/mnras/stx1324},
archivePrefix = {arXiv},
       eprint = {1705.10320},
 primaryClass = {astro-ph.GA},
       adsurl = {https://ui.adsabs.harvard.edu/abs/2017MNRAS.470.3532V}
}

@ARTICLE{2016MNRAS.462.1415C,
       author = {{Chevallard}, Jacopo and {Charlot}, St{\'e}phane},
        title = "{Modelling and interpreting spectral energy distributions of galaxies with BEAGLE}",
      journal = {\mnras},
         year = 2016,
        month = oct,
       volume = {462},
       number = {2},
        pages = {1415-1443},
          doi = {10.1093/mnras/stw1756},
archivePrefix = {arXiv},
       eprint = {1603.03037},
 primaryClass = {astro-ph.GA},
       adsurl = {https://ui.adsabs.harvard.edu/abs/2016MNRAS.462.1415C}
}

@ARTICLE{2017ApJ...837..170L,
       author = {{Leja}, Joel and {Johnson}, Benjamin D. and {Conroy}, Charlie and {van Dokkum}, Pieter G. and {Byler}, Nell},
        title = "{Deriving Physical Properties from Broadband Photometry with Prospector: Description of the Model and a Demonstration of its Accuracy Using 129 Galaxies in the Local Universe}",
      journal = {\apj},
         year = 2017,
        month = mar,
       volume = {837},
       number = {2},
          eid = {170},
        pages = {170},
          doi = {10.3847/1538-4357/aa5ffe},
archivePrefix = {arXiv},
       eprint = {1609.09073},
 primaryClass = {astro-ph.GA},
       adsurl = {https://ui.adsabs.harvard.edu/abs/2017ApJ...837..170L}
}

@ARTICLE{2023MNRAS.520.5903H,
       author = {{Hayes}, Matthew J. and {Runnholm}, Axel and {Scarlata}, Claudia and {Gronke}, Max and {Rivera-Thorsen}, T. Emil},
        title = "{Spectral shapes of the Ly {\ensuremath{\alpha}} emission from galaxies - II. The influence of stellar properties and nebular conditions on the emergent Ly {\ensuremath{\alpha}} profiles}",
      journal = {\mnras},
         year = 2023,
        month = apr,
       volume = {520},
       number = {4},
        pages = {5903-5927},
          doi = {10.1093/mnras/stad477},
archivePrefix = {arXiv},
       eprint = {2302.04875},
 primaryClass = {astro-ph.GA},
       adsurl = {https://ui.adsabs.harvard.edu/abs/2023MNRAS.520.5903H}
}

@ARTICLE{2015ApJ...805...14R,
       author = {{Rivera-Thorsen}, Th{\o}ger E. and {Hayes}, Matthew and {{\"O}stlin}, G{\"o}ran and {Duval}, Florent and {Orlitov{\'a}}, Ivana and {Verhamme}, Anne and {Mas-Hesse}, J. Miguel and {Schaerer}, Daniel and {Cannon}, John M. and {Ot{\'\i}-Floranes}, H{\'e}ctor and {Sandberg}, Andreas and {Guaita}, Lucia and {Adamo}, Angela and {Atek}, Hakim and {Herenz}, E. Christian and {Kunth}, Daniel and {Laursen}, Peter and {Melinder}, Jens},
        title = "{The Lyman Alpha Reference Sample. V. The Impact of Neutral ISM Kinematics and Geometry on Ly{\ensuremath{\alpha}} Escape}",
      journal = {\apj},
         year = 2015,
        month = may,
       volume = {805},
       number = {1},
          eid = {14},
        pages = {14},
          doi = {10.1088/0004-637X/805/1/14},
archivePrefix = {arXiv},
       eprint = {1503.01157},
 primaryClass = {astro-ph.GA},
       adsurl = {https://ui.adsabs.harvard.edu/abs/2015ApJ...805...14R}
}

@article{Reddy_2022,
doi = {10.3847/1538-4357/ac3b4c},
url = {https://doi.org/10.3847/1538-4357/ac3b4c},
year = {2022},
month = {feb},
publisher = {The American Astronomical Society},
volume = {926},
number = {1},
pages = {31},
author = {Reddy, Naveen A. and Topping, Michael W. and Shapley, Alice E. and Steidel, Charles C. and Sanders, Ryan L. and Du, Xinnan and Coil, Alison L. and Mobasher, Bahram and Price, Sedona H. and Shivaei, Irene},
title = {The Effects of Stellar Population and Gas Covering Fraction on the Emergent LyŒ± Emission of High-redshift Galaxies*},
journal = {The Astrophysical Journal}
}

@ARTICLE{2020ApJ...892...48R,
       author = {{Runnholm}, Axel and {Hayes}, Matthew and {Melinder}, Jens and {Rivera-Thorsen}, Emil and {{\"O}stlin}, G{\"o}ran and {Cannon}, John and {Kunth}, Daniel},
        title = "{The Lyman Alpha Reference Sample. X. Predicting Ly{\ensuremath{\alpha}} Output from Star-forming Galaxies Using Multivariate Regression}",
      journal = {\apj},
         year = 2020,
        month = mar,
       volume = {892},
       number = {1},
          eid = {48},
        pages = {48},
          doi = {10.3847/1538-4357/ab7a91},
archivePrefix = {arXiv},
       eprint = {2002.12378},
 primaryClass = {astro-ph.GA},
       adsurl = {https://ui.adsabs.harvard.edu/abs/2020ApJ...892...48R}
}

@ARTICLE{2019ApJ...887...85T,
       author = {{Trainor}, Ryan F. and {Strom}, Allison L. and {Steidel}, Charles C. and {Rudie}, Gwen C. and {Chen}, Yuguang and {Theios}, Rachel L.},
        title = "{Predicting Ly{\ensuremath{\alpha}} Emission from Galaxies via Empirical Markers of Production and Escape in the KBSS}",
      journal = {\apj},
         year = 2019,
        month = dec,
       volume = {887},
       number = {1},
          eid = {85},
        pages = {85},
          doi = {10.3847/1538-4357/ab4993},
archivePrefix = {arXiv},
       eprint = {1908.04794},
 primaryClass = {astro-ph.GA},
       adsurl = {https://ui.adsabs.harvard.edu/abs/2019ApJ...887...85T}
}

@ARTICLE{2018MNRAS.477..904X,
       author = {{Xiao}, Lin and {Stanway}, Elizabeth R. and {Eldridge}, J.~J.},
        title = "{Emission-line diagnostics of nearby H II regions including interacting binary populations}",
      journal = {\mnras},
         year = 2018,
        month = jun,
       volume = {477},
       number = {1},
        pages = {904-934},
          doi = {10.1093/mnras/sty646},
archivePrefix = {arXiv},
       eprint = {1801.07068},
 primaryClass = {astro-ph.GA},
       adsurl = {https://ui.adsabs.harvard.edu/abs/2018MNRAS.477..904X}
}

@ARTICLE{2025A&A...703L..11G,
       author = {{Gonz{\'a}lez-Tor{\`a}}, G. and {Sander}, A.~A.~C. and {Egorova}, E. and {Lefever}, R.~R. and {Ramachandran}, V. and {Egorov}, O.~V. and {Josiek}, J. and {Sch{\"o}sser}, E.~C. and {Bernini-Peron}, M. and {Kreckel}, K. and {Wofford}, A. and {Telford}, O.~G. and {Senchyna}, P. and {Leitherer}, C. and {Liang}, F.-H. and {Blanc}, G.~A. and {Drory}, N. and {Fern{\'a}ndez-Trincado}, J.~G. and {Johnston}, E.~J. and {Mej{\'\i}a-Narv{\'a}ez}, A.~J. and {Sanchez}, S.~F.},
        title = "{SDSS-V LVM: Detectability of Wolf-Rayet stars and their He II ionizing flux in low-metallicity environments: I. The weak-lined, early-type WN3 stars in the SMC}",
      journal = {\aap},
         year = 2025,
        month = nov,
       volume = {703},
          eid = {L11},
        pages = {L11},
          doi = {10.1051/0004-6361/202557041},
archivePrefix = {arXiv},
       eprint = {2509.04569},
 primaryClass = {astro-ph.GA},
       adsurl = {https://ui.adsabs.harvard.edu/abs/2025A&A...703L..11G}
}

@ARTICLE{2015A&A...578L...2G,
       author = {{Gr{\"a}fener}, G. and {Vink}, J.~S.},
        title = "{Narrow He II emission in star-forming galaxies at low metallicity. Stellar wind emission from a population of very massive stars}",
      journal = {\aap},
         year = 2015,
        month = jun,
       volume = {578},
          eid = {L2},
        pages = {L2},
          doi = {10.1051/0004-6361/201425287},
archivePrefix = {arXiv},
       eprint = {1505.02994},
 primaryClass = {astro-ph.GA},
       adsurl = {https://ui.adsabs.harvard.edu/abs/2015A&A...578L...2G}
}

@ARTICLE{2019ApJ...885...65F,
       author = {{Fornasini}, Francesca M. and {Kriek}, Mariska and {Sanders}, Ryan L. and {Shivaei}, Irene and {Civano}, Francesca and {Reddy}, Naveen A. and {Shapley}, Alice E. and {Coil}, Alison L. and {Mobasher}, Bahram and {Siana}, Brian and {Aird}, James and {Azadi}, Mojegan and {Freeman}, William R. and {Leung}, Gene C.~K. and {Price}, Sedona H. and {Fetherolf}, Tara and {Zick}, Tom and {Barro}, Guillermo},
        title = "{The MOSDEF Survey: The Metallicity Dependence of X-Ray Binary Populations at z {\ensuremath{\sim}} 2}",
      journal = {\apj},
         year = 2019,
        month = nov,
       volume = {885},
       number = {1},
          eid = {65},
        pages = {65},
          doi = {10.3847/1538-4357/ab4653},
archivePrefix = {arXiv},
       eprint = {1909.08635},
 primaryClass = {astro-ph.HE},
       adsurl = {https://ui.adsabs.harvard.edu/abs/2019ApJ...885...65F}
}

@ARTICLE{2015A&A...579A..44D,
       author = {{Douna}, V.~M. and {Pellizza}, L.~J. and {Mirabel}, I.~F. and {Pedrosa}, S.~E.},
        title = "{Metallicity dependence of high-mass X-ray binary populations}",
      journal = {\aap},
         year = 2015,
        month = jul,
       volume = {579},
          eid = {A44},
        pages = {A44},
          doi = {10.1051/0004-6361/201525617},
archivePrefix = {arXiv},
       eprint = {1505.05483},
 primaryClass = {astro-ph.GA},
       adsurl = {https://ui.adsabs.harvard.edu/abs/2015A&A...579A..44D}
}

@ARTICLE{2016MNRAS.457.4081B,
       author = {{Brorby}, M. and {Kaaret}, P. and {Prestwich}, A. and {Mirabel}, I.~F.},
        title = "{Enhanced X-ray emission from Lyman break analogues and a possible L$_{X}$-SFR-metallicity plane}",
      journal = {\mnras},
         year = 2016,
        month = apr,
       volume = {457},
       number = {4},
        pages = {4081-4088},
          doi = {10.1093/mnras/stw284},
archivePrefix = {arXiv},
       eprint = {1602.01091},
 primaryClass = {astro-ph.HE},
       adsurl = {https://ui.adsabs.harvard.edu/abs/2016MNRAS.457.4081B}
}

@ARTICLE{2025arXiv251116752H,
       author = {{Hovis-Afflerbach}, Beryl and {Strom}, Allison L. and {Saldana-Lopez}, Alberto and {Flury}, Sophia R.},
        title = "{Finding He II: Testing Novel Models of Binary Populations across Cosmic Time}",
      journal = {arXiv e-prints},
         year = 2025,
        month = nov,
          eid = {arXiv:2511.16752},
        pages = {arXiv:2511.16752},
          doi = {10.48550/arXiv.2511.16752},
archivePrefix = {arXiv},
       eprint = {2511.16752},
 primaryClass = {astro-ph.GA},
       adsurl = {https://ui.adsabs.harvard.edu/abs/2025arXiv251116752H}
}

@ARTICLE{2017PASA...34...58E,
       author = {{Eldridge}, J.~J. and {Stanway}, E.~R. and {Xiao}, L. and {McClelland}, L.~A.~S. and {Taylor}, G. and {Ng}, M. and {Greis}, S.~M.~L. and {Bray}, J.~C.},
        title = "{Binary Population and Spectral Synthesis Version 2.1: Construction, Observational Verification, and New Results}",
      journal = {\pasa},
         year = 2017,
        month = nov,
       volume = {34},
          eid = {e058},
        pages = {e058},
          doi = {10.1017/pasa.2017.51},
archivePrefix = {arXiv},
       eprint = {1710.02154},
 primaryClass = {astro-ph.SR},
       adsurl = {https://ui.adsabs.harvard.edu/abs/2017PASA...34...58E}
}

@ARTICLE{1991ApJ...373..458G,
       author = {{Garnett}, Donald R. and {Kennicutt}, Jr., Robert C. and {Chu}, You-Hua and {Skillman}, Evan D.},
        title = "{He II Emission in Extragalactic H II Regions}",
      journal = {\apj},
         year = 1991,
        month = jun,
       volume = {373},
        pages = {458},
          doi = {10.1086/170065},
       adsurl = {https://ui.adsabs.harvard.edu/abs/1991ApJ...373..458G}
}

@ARTICLE{2025ApJ...992...91R,
       author = {{Roy}, Namrata and {Heckman}, Timothy and {Henry}, Alaina and {Chisholm}, John and {Flury}, Sophia and {Leitherer}, Claus and {Hayes}, Matthew J. and {Jaskot}, Anne and {Ji}, Zhiyuan and {Schaerer}, Daniel and {Wang}, Bingjie and {Borthakur}, Sanchayeeta and {Xu}, Xinfeng and {{\"O}stlin}, G{\"o}ran},
        title = "{Lyman Continuum Leakage from Massive Leaky Starbursts: A Different Class of Emitters?}",
      journal = {\apj},
         year = 2025,
        month = oct,
       volume = {992},
       number = {1},
          eid = {91},
        pages = {91},
          doi = {10.3847/1538-4357/adff5e},
archivePrefix = {arXiv},
       eprint = {2410.13254},
 primaryClass = {astro-ph.GA},
       adsurl = {https://ui.adsabs.harvard.edu/abs/2025ApJ...992...91R}
}

@ARTICLE{2015MNRAS.447.3402B,
       author = {{Becker}, George D. and {Bolton}, James S. and {Madau}, Piero and {Pettini}, Max and {Ryan-Weber}, Emma V. and {Venemans}, Bram P.},
        title = "{Evidence of patchy hydrogen reionization from an extreme Ly{\ensuremath{\alpha}} trough below redshift six}",
      journal = {\mnras},
         year = 2015,
        month = mar,
       volume = {447},
       number = {4},
        pages = {3402-3419},
          doi = {10.1093/mnras/stu2646},
archivePrefix = {arXiv},
       eprint = {1407.4850},
 primaryClass = {astro-ph.CO},
       adsurl = {https://ui.adsabs.harvard.edu/abs/2015MNRAS.447.3402B}
}

@article{10.1093/mnras/stv2661,
    author = {Stanway, Elizabeth R. and Eldridge, J. J. and Becker, George D.},
    title = {Stellar population effects on the inferred photon density at reionization},
    journal = {Monthly Notices of the Royal Astronomical Society},
    volume = {456},
    number = {1},
    pages = {485-499},
    year = {2016},
    month = {02},
    issn = {0035-8711},
    doi = {10.1093/mnras/stv2661},
    url = {https://doi.org/10.1093/mnras/stv2661},
    eprint = {https://academic.oup.com/mnras/article-pdf/456/1/485/3678348/stv2661.pdf},
}

@ARTICLE{2019ApJ...882..178K,
       author = {{Kulkarni}, Mihir and {Visbal}, Eli and {Bryan}, Greg L.},
        title = "{Fragmentation in Population III Galaxies Formed through Ionizing Radiation}",
      journal = {\apj},
         year = 2019,
        month = sep,
       volume = {882},
       number = {2},
          eid = {178},
        pages = {178},
          doi = {10.3847/1538-4357/ab35e2},
archivePrefix = {arXiv},
       eprint = {1907.11724},
 primaryClass = {astro-ph.GA},
       adsurl = {https://ui.adsabs.harvard.edu/abs/2019ApJ...882..178K}
}

@ARTICLE{2021ApJ...917L..37C,
       author = {{Cain}, Christopher and {D'Aloisio}, Anson and {Gangolli}, Nakul and {Becker}, George D.},
        title = "{A Short Mean Free Path at z = 6 Favors Late and Rapid Reionization by Faint Galaxies}",
      journal = {\apjl},
         year = 2021,
        month = aug,
       volume = {917},
       number = {2},
          eid = {L37},
        pages = {L37},
          doi = {10.3847/2041-8213/ac1ace},
archivePrefix = {arXiv},
       eprint = {2105.10511},
 primaryClass = {astro-ph.CO},
       adsurl = {https://ui.adsabs.harvard.edu/abs/2021ApJ...917L..37C}
}

@ARTICLE{2022MNRAS.514...55B,
       author = {{Bosman}, Sarah E.~I. and {Davies}, Frederick B. and {Becker}, George D. and {Keating}, Laura C. and {Davies}, Rebecca L. and {Zhu}, Yongda and {Eilers}, Anna-Christina and {D'Odorico}, Valentina and {Bian}, Fuyan and {Bischetti}, Manuela and {Cristiani}, Stefano V. and {Fan}, Xiaohui and {Farina}, Emanuele P. and {Haehnelt}, Martin G. and {Hennawi}, Joseph F. and {Kulkarni}, Girish and {Mesinger}, Andrei and {Meyer}, Romain A. and {Onoue}, Masafusa and {Pallottini}, Andrea and {Qin}, Yuxiang and {Ryan-Weber}, Emma and {Schindler}, Jan-Torge and {Walter}, Fabian and {Wang}, Feige and {Yang}, Jinyi},
        title = "{Hydrogen reionization ends by z = 5.3: Lyman-{\ensuremath{\alpha}} optical depth measured by the XQR-30 sample}",
      journal = {\mnras},
         year = 2022,
        month = jul,
       volume = {514},
       number = {1},
        pages = {55-76},
          doi = {10.1093/mnras/stac1046},
archivePrefix = {arXiv},
       eprint = {2108.03699},
 primaryClass = {astro-ph.CO},
       adsurl = {https://ui.adsabs.harvard.edu/abs/2022MNRAS.514...55B}
}

@ARTICLE{2026PASA...43...21M,
       author = {{Mukherjee}, Tamal and {Zafar}, Tayyaba and {Nanayakkara}, Themiya and {Gurung-L{\'o}pez}, Siddhartha and {Gupta}, Anshu and {Croom}, Scott and {Battisti}, Andrew and {Glazebrook}, Karl and {Papaderos}, Polychronis and {Riggs}, Melissa and {Wisnioski}, Emily and {Foster}, Caroline and {Harborne}, Katherine and {Lagos}, Claudia and {Mendel}, Jon Trevor and {Prathap}, Jahang and {Barsanti}, Stefania and {Sweet}, Sarah and {Valenzuela}, Lucas and {Mailvaganam}, Anilkumar},
        title = "{A census of double-peaked Lyman-{\ensuremath{\alpha}} emitters in MAGPI: Classification, global characteristics, and spatially resolved properties}",
      journal = {\pasa},
         year = 2026,
        month = jan,
       volume = {43},
          eid = {e021},
        pages = {e021},
          doi = {10.1017/pasa.2026.10145},
archivePrefix = {arXiv},
       eprint = {2510.18398},
 primaryClass = {astro-ph.GA},
       adsurl = {https://ui.adsabs.harvard.edu/abs/2026PASA...43...21M}
}

@ARTICLE{2006A&A...460..397V,
       author = {{Verhamme}, A. and {Schaerer}, D. and {Maselli}, A.},
        title = "{3D Ly{\ensuremath{\alpha}} radiation transfer. I. Understanding Ly{\ensuremath{\alpha}} line profile morphologies}",
      journal = {\aap},
         year = 2006,
        month = dec,
       volume = {460},
       number = {2},
        pages = {397-413},
          doi = {10.1051/0004-6361:20065554},
archivePrefix = {arXiv},
       eprint = {astro-ph/0608075},
 primaryClass = {astro-ph},
       adsurl = {https://ui.adsabs.harvard.edu/abs/2006A&A...460..397V}
}

@ARTICLE{2025ApJ...988..171M,
       author = {{Mondal}, Chayan and {Saha}, Kanak and {Borgohain}, Anshuman and {Smith}, Brent M. and {Windhorst}, Rogier A. and {Reddy}, Naveen and {Chen}, Chian-Chou and {Umetsu}, Keiichi and {Jansen}, Rolf A.},
        title = "{GNHeII J1236+6215: A He II {\ensuremath{\lambda}}1640 Emitting and Potentially LyC Leaking Galaxy at z = 2.9803 Unveiled through JWST and Keck Observations}",
      journal = {\apj},
         year = 2025,
        month = aug,
       volume = {988},
       number = {2},
          eid = {171},
        pages = {171},
          doi = {10.3847/1538-4357/ade2cd},
archivePrefix = {arXiv},
       eprint = {2506.06831},
 primaryClass = {astro-ph.GA},
       adsurl = {https://ui.adsabs.harvard.edu/abs/2025ApJ...988..171M}
}

@INPROCEEDINGS{1993oee..conf...15G,
       author = {{Grevesse}, N. and {Noels}, A.},
        title = "{Cosmic abundances of the elements.}",
    booktitle = {Origin and Evolution of the Elements},
         year = 1993,
       editor = {{Prantzos}, N. and {Vangioni-Flam}, E. and {Casse}, M.},
        month = jan,
        pages = {15-25},
       adsurl = {https://ui.adsabs.harvard.edu/abs/1993oee..conf...15G}
}

@ARTICLE{2024ApJ...967L..42W,
       author = {{Wang}, Xin and {Cheng}, Cheng and {Ge}, Junqiang and {Meng}, Xiao-Lei and {Daddi}, Emanuele and {Yan}, Haojing and {Ji}, Zhiyuan and {Jin}, Yifei and {Jones}, Tucker and {Malkan}, Matthew A. and {Arrabal Haro}, Pablo and {Brammer}, Gabriel and {Oguri}, Masamune and {Hou}, Meicun and {Zhang}, Shiwu},
        title = "{A Strong He II {\ensuremath{\lambda}}1640 Emitter with an Extremely Blue UV Spectral Slope at z = 8.16: Presence of Population III Stars?}",
      journal = {\apjl},
         year = 2024,
        month = jun,
       volume = {967},
       number = {2},
          eid = {L42},
        pages = {L42},
          doi = {10.3847/2041-8213/ad4ced},
archivePrefix = {arXiv},
       eprint = {2212.04476},
 primaryClass = {astro-ph.GA},
       adsurl = {https://ui.adsabs.harvard.edu/abs/2024ApJ...967L..42W}
}

@ARTICLE{2016ARA&A..54..761S,
       author = {{Stark}, Daniel P.},
        title = "{Galaxies in the First Billion Years After the Big Bang}",
      journal = {\araa},
         year = 2016,
        month = sep,
       volume = {54},
        pages = {761-803},
          doi = {10.1146/annurev-astro-081915-023417},
       adsurl = {https://ui.adsabs.harvard.edu/abs/2016ARA&A..54..761S}
}

@ARTICLE{2019ApJ...887..168G,
       author = {{Gburek}, Timothy and {Siana}, Brian and {Alavi}, Anahita and {Emami}, Najmeh and {Richard}, Johan and {Freeman}, William R. and {Stark}, Daniel P. and {Snapp-Kolas}, Christopher and {Lucero}, Breanna},
        title = "{The Detection of [O III] {\ensuremath{\lambda}}4363 in a Lensed, Dwarf Galaxy at z = 2.59: Testing Metallicity Indicators and Scaling Relations at High Redshift and Low Mass}",
      journal = {\apj},
         year = 2019,
        month = dec,
       volume = {887},
       number = {2},
          eid = {168},
        pages = {168},
          doi = {10.3847/1538-4357/ab5713},
archivePrefix = {arXiv},
       eprint = {1906.11849},
 primaryClass = {astro-ph.GA},
       adsurl = {https://ui.adsabs.harvard.edu/abs/2019ApJ...887..168G}
}

@misc{maiolino2026searchpopulationiiiconfirmation,
      title={The search for Population III: Confirmation of a HeII emitter with no metal lines at z=10.6}, 
      author={Roberto Maiolino and Hannah Übler and Michele Perna and Joris Witstok and Gareth C. Jones and Pablo G. Perez-Gonzalez and Kimihiko Nakajima and Elka Rusta and Stefania Salvadori and Sandro Tacchella and Piero Madau and James A. A. Trussler and Francesco D'Eugenio and Xihan Ji and Jan Scholtz and Stefano Carniani and Yuki Isobe and Harley Katz and Santiago Arribas and William M. Baker and Torsten Böker and Volker Bromm and Andrew J. Bunker and Stephane Charlot and Jacopo Chevallard and Mirko Curti and Emma Curtis-Lake and Daniel Eisenstein and Eiichi Egami and Andrea Ferrara and Luca Graziani and Kevin Hainline and Jakob M. Helton and Lucy Ivey and Benjamin Jonson and Maria Koller and Nimisha Kumari and Alessandro Marconi and Giovanni Mazzolari and Nicolas Laporte and Eleonora Parlanti and Robert Pascalau and Laura Pentericci and Pierluigi Rinaldi and Brant Robertson and Bruno Rodríguez Del Pino and Raffaella Schneider and Alessandra Venditti and Giacomo Venturi and Christopher N. A. Willmer and Callum Witten and Sandra Zamora},
      year={2026},
      eprint={2603.20362},
      archivePrefix={arXiv},
      primaryClass={astro-ph.GA},
      url={https://arxiv.org/abs/2603.20362}, 
}

@ARTICLE{2017ApJ...840...44B,
       author = {{Byler}, Nell and {Dalcanton}, Julianne J. and {Conroy}, Charlie and {Johnson}, Benjamin D.},
        title = "{Nebular Continuum and Line Emission in Stellar Population Synthesis Models}",
      journal = {\apj},
         year = 2017,
        month = may,
       volume = {840},
       number = {1},
          eid = {44},
        pages = {44},
          doi = {10.3847/1538-4357/aa6c66},
archivePrefix = {arXiv},
       eprint = {1611.08305},
 primaryClass = {astro-ph.GA},
       adsurl = {https://ui.adsabs.harvard.edu/abs/2017ApJ...840...44B}
}

@ARTICLE{2018ApJ...863...14B,
       author = {{Byler}, Nell and {Dalcanton}, Julianne J. and {Conroy}, Charlie and {Johnson}, Benjamin D. and {Levesque}, Emily M. and {Berg}, Danielle A.},
        title = "{Stellar and Nebular Diagnostics in the Ultraviolet for Star-forming Galaxies}",
      journal = {\apj},
         year = 2018,
        month = aug,
       volume = {863},
       number = {1},
          eid = {14},
        pages = {14},
          doi = {10.3847/1538-4357/aacd50},
archivePrefix = {arXiv},
       eprint = {1803.04425},
 primaryClass = {astro-ph.GA},
       adsurl = {https://ui.adsabs.harvard.edu/abs/2018ApJ...863...14B}
}

@ARTICLE{2003ApJ...588...65S,
       author = {{Shapley}, Alice E. and {Steidel}, Charles C. and {Pettini}, Max and {Adelberger}, Kurt L.},
        title = "{Rest-Frame Ultraviolet Spectra of z\raisebox{-0.5ex}\textasciitilde3 Lyman Break Galaxies}",
      journal = {\apj},
         year = 2003,
        month = may,
       volume = {588},
       number = {1},
        pages = {65-89},
          doi = {10.1086/373922},
archivePrefix = {arXiv},
       eprint = {astro-ph/0301230},
 primaryClass = {astro-ph},
       adsurl = {https://ui.adsabs.harvard.edu/abs/2003ApJ...588...65S}
}

@ARTICLE{1972ApJ...173...25S,
       author = {{Searle}, Leonard and {Sargent}, Wallace L.~W.},
        title = "{Inferences from the Composition of Two Dwarf Blue Galaxies}",
      journal = {\apj},
         year = 1972,
        month = apr,
       volume = {173},
        pages = {25},
          doi = {10.1086/151398},
       adsurl = {https://ui.adsabs.harvard.edu/abs/1972ApJ...173...25S}
}

@ARTICLE{1999ApJ...527..757I,
       author = {{Izotov}, Yuri I. and {Chaffee}, Frederic H. and {Foltz}, Craig B. and {Green}, Richard F. and {Guseva}, Natalia G. and {Thuan}, Trinh X.},
        title = "{Helium Abundance in the Most Metal-deficient Blue Compact Galaxies: I ZW 18 and SBS 0335-052}",
      journal = {\apj},
         year = 1999,
        month = dec,
       volume = {527},
       number = {2},
        pages = {757-777},
          doi = {10.1086/308119},
archivePrefix = {arXiv},
       eprint = {astro-ph/9907228},
 primaryClass = {astro-ph},
       adsurl = {https://ui.adsabs.harvard.edu/abs/1999ApJ...527..757I}
}

@ARTICLE{2004ApJ...606..213T,
       author = {{Thuan}, Trinh X. and {Bauer}, Franz E. and {Papaderos}, Polichronis and {Izotov}, Yuri I.},
        title = "{Chandra Observations of the Three Most Metal Deficient Blue Compact Dwarf Galaxies Known in the Local Universe, SBS 0335-052, SBS 0335-052W, and I Zw 18}",
      journal = {\apj},
         year = 2004,
        month = may,
       volume = {606},
       number = {1},
        pages = {213-220},
          doi = {10.1086/382949},
archivePrefix = {arXiv},
       eprint = {astro-ph/0401349},
 primaryClass = {astro-ph},
       adsurl = {https://ui.adsabs.harvard.edu/abs/2004ApJ...606..213T}
}

@ARTICLE{2005ApJ...632..210I,
       author = {{Izotov}, Yuri I. and {Thuan}, Trinh X. and {Guseva}, Natalia G.},
        title = "{SBS 0335-052W: The Lowest Metallicity Star-forming Galaxy Known}",
      journal = {\apj},
         year = 2005,
        month = oct,
       volume = {632},
       number = {1},
        pages = {210-216},
          doi = {10.1086/432874},
archivePrefix = {arXiv},
       eprint = {astro-ph/0506498},
 primaryClass = {astro-ph},
       adsurl = {https://ui.adsabs.harvard.edu/abs/2005ApJ...632..210I}
}

@ARTICLE{2009A&A...503...61I,
       author = {{Izotov}, Y.~I. and {Guseva}, N.~G. and {Fricke}, K.~J. and {Papaderos}, P.},
        title = "{SBS 0335-052E+W: deep VLT/FORS+UVES spectroscopy of the pair of the lowest-metallicity blue compact dwarf galaxies}",
      journal = {\aap},
         year = 2009,
        month = aug,
       volume = {503},
       number = {1},
        pages = {61-72},
          doi = {10.1051/0004-6361/200911965},
archivePrefix = {arXiv},
       eprint = {0907.2116},
 primaryClass = {astro-ph.CO},
       adsurl = {https://ui.adsabs.harvard.edu/abs/2009A&A...503...61I}
}

@ARTICLE{2006A&A...454..119P,
       author = {{Papaderos}, P. and {Izotov}, Y.~I. and {Guseva}, N.~G. and {Thuan}, T.~X. and {Fricke}, K.~J.},
        title = "{Oxygen abundance variations in the system of the two blue compact dwarf galaxies SBS 0335-052E and SBS 0335-052W}",
      journal = {\aap},
         year = 2006,
        month = jul,
       volume = {454},
       number = {1},
        pages = {119-123},
          doi = {10.1051/0004-6361:20065110},
archivePrefix = {arXiv},
       eprint = {astro-ph/0604270},
 primaryClass = {astro-ph},
       adsurl = {https://ui.adsabs.harvard.edu/abs/2006A&A...454..119P}
}

@BOOK{2006agna.book.....O,
       author = {{Osterbrock}, Donald E. and {Ferland}, Gary J.},
        title = "{Astrophysics of gaseous nebulae and active galactic nuclei}",
         year = 2006,
       adsurl = {https://ui.adsabs.harvard.edu/abs/2006agna.book.....O}
}

@ARTICLE{2024MNRAS.527.6139S,
       author = {{Simmonds}, C. and {Tacchella}, S. and {Hainline}, K. and {Johnson}, B.~D. and {McClymont}, W. and {Robertson}, B. and {Saxena}, A. and {Sun}, F. and {Witten}, C. and {Baker}, W.~M. and {Bhatawdekar}, R. and {Boyett}, K. and {Bunker}, A.~J. and {Charlot}, S. and {Curtis-Lake}, E. and {Egami}, E. and {Eisenstein}, D.~J. and {Hausen}, R. and {Maiolino}, R. and {Maseda}, M.~V. and {Scholtz}, J. and {Williams}, C.~C. and {Willott}, C. and {Witstok}, J.},
        title = "{Low-mass bursty galaxies in JADES efficiently produce ionizing photons and could represent the main drivers of reionization}",
      journal = {\mnras},
         year = 2024,
        month = jan,
       volume = {527},
       number = {3},
        pages = {6139-6157},
          doi = {10.1093/mnras/stad3605},
archivePrefix = {arXiv},
       eprint = {2310.01112},
 primaryClass = {astro-ph.GA},
       adsurl = {https://ui.adsabs.harvard.edu/abs/2024MNRAS.527.6139S}
}

@ARTICLE{2006ApJ...646L..33Y,
       author = {{Yang}, Yujin and {Tremonti}, Christy A. and {Zabludoff}, Ann I. and {Zaritsky}, Dennis},
        title = "{E+A Galaxies with Blue Cores: Active Galaxies in Transition}",
      journal = {\apjl},
         year = 2006,
        month = jul,
       volume = {646},
       number = {1},
        pages = {L33-L36},
          doi = {10.1086/506909},
archivePrefix = {arXiv},
       eprint = {astro-ph/0606368},
 primaryClass = {astro-ph},
       adsurl = {https://ui.adsabs.harvard.edu/abs/2006ApJ...646L..33Y}
}

@ARTICLE{2025arXiv251111314V,
       author = {{van Veenen}, Lisanne and {Sharda}, Piyush and {Viti}, Serena and {Menon}, Shyam H.},
        title = "{Radiation magnetohydrodynamics simulations of Population III star formation during the Epoch of Reionization}",
      journal = {arXiv e-prints},
         year = 2025,
        month = nov,
          eid = {arXiv:2511.11314},
        pages = {arXiv:2511.11314},
          doi = {10.48550/arXiv.2511.11314},
archivePrefix = {arXiv},
       eprint = {2511.11314},
 primaryClass = {astro-ph.GA},
       adsurl = {https://ui.adsabs.harvard.edu/abs/2025arXiv251111314V}
}

@ARTICLE{2020MNRAS.494..941S,
       author = {{Senchyna}, Peter and {Stark}, Daniel P. and {Mirocha}, Jordan and {Reines}, Amy E. and {Charlot}, St{\'e}phane and {Jones}, Tucker and {Mulchaey}, John S.},
        title = "{High-mass X-ray binaries in nearby metal-poor galaxies: on the contribution to nebular He II emission}",
      journal = {\mnras},
         year = 2020,
        month = may,
       volume = {494},
       number = {1},
        pages = {941-957},
          doi = {10.1093/mnras/staa586},
archivePrefix = {arXiv},
       eprint = {1909.10574},
 primaryClass = {astro-ph.GA},
       adsurl = {https://ui.adsabs.harvard.edu/abs/2020MNRAS.494..941S}
}

@ARTICLE{2010ApJ...724..341H,
       author = {{Heger}, Alexander and {Woosley}, S.~E.},
        title = "{Nucleosynthesis and Evolution of Massive Metal-free Stars}",
      journal = {\apj},
         year = 2010,
        month = nov,
       volume = {724},
       number = {1},
        pages = {341-373},
          doi = {10.1088/0004-637X/724/1/341},
archivePrefix = {arXiv},
       eprint = {0803.3161},
 primaryClass = {astro-ph},
       adsurl = {https://ui.adsabs.harvard.edu/abs/2010ApJ...724..341H}
}

@ARTICLE{2022MNRAS.517....1S,
       author = {{Smith}, Aaron and {Kannan}, Rahul and {Tacchella}, Sandro and {Vogelsberger}, Mark and {Hernquist}, Lars and {Marinacci}, Federico and {Sales}, Laura V. and {Torrey}, Paul and {Li}, Hui and {Yeh}, Jessica Y.-C. and {Qi}, Jia},
        title = "{The physics of Lyman-{\ensuremath{\alpha}} escape from disc-like galaxies}",
      journal = {\mnras},
         year = 2022,
        month = nov,
       volume = {517},
       number = {1},
        pages = {1-27},
          doi = {10.1093/mnras/stac2641},
archivePrefix = {arXiv},
       eprint = {2111.13721},
 primaryClass = {astro-ph.GA},
       adsurl = {https://ui.adsabs.harvard.edu/abs/2022MNRAS.517....1S}
}

@ARTICLE{2017MNRAS.466.4403N,
       author = {{Nicholls}, David C. and {Sutherland}, Ralph S. and {Dopita}, Michael A. and {Kewley}, Lisa J. and {Groves}, Brent A.},
        title = "{Abundance scaling in stars, nebulae and galaxies}",
      journal = {\mnras},
         year = 2017,
        month = apr,
       volume = {466},
       number = {4},
        pages = {4403-4422},
          doi = {10.1093/mnras/stw3235},
archivePrefix = {arXiv},
       eprint = {1612.03546},
 primaryClass = {astro-ph.GA},
       adsurl = {https://ui.adsabs.harvard.edu/abs/2017MNRAS.466.4403N}
}

@ARTICLE{2023MNRAS.518.3985S,
       author = {{Sharda}, Piyush and {Amarsi}, Anish M. and {Grasha}, Kathryn and {Krumholz}, Mark R. and {Yong}, David and {Chiaki}, Gen and {Roy}, Arpita and {Nordlander}, Thomas},
        title = "{The impact of carbon and oxygen abundances on the metal-poor initial mass function}",
      journal = {\mnras},
         year = 2023,
        month = jan,
       volume = {518},
       number = {3},
        pages = {3985-3998},
          doi = {10.1093/mnras/stac3315},
archivePrefix = {arXiv},
       eprint = {2211.05505},
 primaryClass = {astro-ph.GA},
       adsurl = {https://ui.adsabs.harvard.edu/abs/2023MNRAS.518.3985S}
}

@ARTICLE{2025ApJ...987...12M,
       author = {{Menon}, Shyam H. and {Burkhart}, Blakesley and {Somerville}, Rachel S. and {Thompson}, Todd A. and {Sternberg}, Amiel},
        title = "{Bursts of Star Formation and Radiation-driven Outflows Produce Efficient LyC Leakage from Dense Compact Star Clusters}",
      journal = {\apj},
         year = 2025,
        month = jul,
       volume = {987},
       number = {1},
          eid = {12},
        pages = {12},
          doi = {10.3847/1538-4357/add2f9},
archivePrefix = {arXiv},
       eprint = {2408.14591},
 primaryClass = {astro-ph.GA},
       adsurl = {https://ui.adsabs.harvard.edu/abs/2025ApJ...987...12M}
}

@ARTICLE{2015MNRAS.454.2691M,
       author = {{Muratov}, Alexander L. and {Kere{\v{s}}}, Du{\v{s}}an and {Faucher-Gigu{\`e}re}, Claude-Andr{\'e} and {Hopkins}, Philip F. and {Quataert}, Eliot and {Murray}, Norman},
        title = "{Gusty, gaseous flows of FIRE: galactic winds in cosmological simulations with explicit stellar feedback}",
      journal = {\mnras},
         year = 2015,
        month = dec,
       volume = {454},
       number = {3},
        pages = {2691-2713},
          doi = {10.1093/mnras/stv2126},
archivePrefix = {arXiv},
       eprint = {1501.03155},
 primaryClass = {astro-ph.GA},
       adsurl = {https://ui.adsabs.harvard.edu/abs/2015MNRAS.454.2691M}
}

@ARTICLE{2020A&A...639A..85G,
       author = {{Gazagnes}, S. and {Chisholm}, J. and {Schaerer}, D. and {Verhamme}, A. and {Izotov}, Y.},
        title = "{The origin of the escape of Lyman {\ensuremath{\alpha}} and ionizing photons in Lyman continuum emitters}",
      journal = {\aap},
         year = 2020,
        month = jul,
       volume = {639},
          eid = {A85},
        pages = {A85},
          doi = {10.1051/0004-6361/202038096},
archivePrefix = {arXiv},
       eprint = {2005.07215},
 primaryClass = {astro-ph.GA},
       adsurl = {https://scixplorer.org/abs/2020A&A...639A..85G}
}

@ARTICLE{2013ApJ...777...39Z,
       author = {{Zackrisson}, Erik and {Inoue}, Akio K. and {Jensen}, Hannes},
        title = "{The Spectral Evolution of the First Galaxies. II. Spectral Signatures of Lyman Continuum Leakage from Galaxies in the Reionization Epoch}",
      journal = {\apj},
         year = 2013,
        month = nov,
       volume = {777},
       number = {1},
          eid = {39},
        pages = {39},
          doi = {10.1088/0004-637X/777/1/39},
archivePrefix = {arXiv},
       eprint = {1304.6404},
 primaryClass = {astro-ph.CO},
       adsurl = {https://ui.adsabs.harvard.edu/abs/2013ApJ...777...39Z}
}

@ARTICLE{2022ApJS..259...21K,
       author = {{Kimm}, Taysun and {Bieri}, Rebekka and {Geen}, Sam and {Rosdahl}, Joakim and {Blaizot}, J{\'e}r{\'e}my and {Michel-Dansac}, L{\'e}o and {Garel}, Thibault},
        title = "{A Systematic Study of the Escape of LyC and Ly{\ensuremath{\alpha}} Photons from Star-forming, Magnetized Turbulent Clouds}",
      journal = {\apjs},
         year = 2022,
        month = mar,
       volume = {259},
       number = {1},
          eid = {21},
        pages = {21},
          doi = {10.3847/1538-4365/ac426d},
archivePrefix = {arXiv},
       eprint = {2110.02975},
 primaryClass = {astro-ph.GA},
       adsurl = {https://ui.adsabs.harvard.edu/abs/2022ApJS..259...21K}
}

@article{Bressan_2012,
   title={<scp>parsec</scp>: stellar tracks and isochrones with the PAdova and TRieste Stellar Evolution Code: <scp>parsec</scp>:tracks and isochrones},
   volume={427},
   ISSN={1365-2966},
   url={http://dx.doi.org/10.1111/j.1365-2966.2012.21948.x},
   DOI={10.1111/j.1365-2966.2012.21948.x},
   number={1},
   journal={Monthly Notices of the Royal Astronomical Society},
   publisher={Oxford University Press (OUP)},
   author={Bressan, Alessandro and Marigo, Paola and Girardi, Léo and Salasnich, Bernardo and Dal Cero, Claudia and Rubele, Stefano and Nanni, Ambra},
   year={2012},
   month=oct, pages={127–145} }

@misc{dsilva2025dissectingreionisationcosmicstar,
      title={Dissecting Reionisation with the Cosmic Star Formation and Active Galactic Nuclei Luminosity History}, 
      author={Jordan C. J. D'Silva and Simon P. Driver and Claudia D. P. Lagos and Aaron S. G. Robotham and Nathan J. Adams and Christopher J. Conselice and Brenda Frye and Nimish P. Hathi and Thomas Harvey and Anton M. Koekemoer and Rafael Ortiz III and Massimo Ricotti and Clayton Robertson and Ross M. Silver and Stephen M. Wilkins and Christopher N. A. Willmer and Rogier A. Windhorst and Seth H. Cohen and Rolf A. Jansen and Jake Summers and Dan Coe and Norman A. Grogin and Madeline A. Marshall and Nor Pirzkal and Russell E. Ryan Jr. and Haojing Yan},
      year={2025},
      eprint={2507.16112},
      archivePrefix={arXiv},
      primaryClass={astro-ph.GA},
      url={https://arxiv.org/abs/2507.16112}, 
}

@article{Sobral_2015,
doi = {10.1088/0004-637X/808/2/139},
url = {https://doi.org/10.1088/0004-637X/808/2/139},
year = {2015},
month = {jul},
publisher = {The American Astronomical Society},
volume = {808},
number = {2},
pages = {139},
author = {Sobral, David and Matthee, Jorryt and Darvish, Behnam and Schaerer, Daniel and Mobasher, Bahram and R√∂ttgering, Huub J. A. and Santos, S√©rgio and Hemmati, Shoubaneh},
title = {EVIDENCE FOR PopIII-LIKE STELLAR POPULATIONS IN THE MOST LUMINOUS LyŒ± EMITTERS AT THE EPOCH OF REIONIZATION: SPECTROSCOPIC CONFIRMATION*},
journal = {The Astrophysical Journal}
}

@article{Stark_2016,
   title={Lyα and C iii] emission inz= 7–9 Galaxies: accelerated reionization around luminous star-forming systems?},
   volume={464},
   ISSN={1365-2966},
   url={http://dx.doi.org/10.1093/mnras/stw2233},
   DOI={10.1093/mnras/stw2233},
   number={1},
   journal={Monthly Notices of the Royal Astronomical Society},
   publisher={Oxford University Press (OUP)},
   author={Stark, Daniel P. and Ellis, Richard S. and Charlot, Stéphane and Chevallard, Jacopo and Tang, Mengtao and Belli, Sirio and Zitrin, Adi and Mainali, Ramesh and Gutkin, Julia and Vidal-García, Alba and Bouwens, Rychard and Oesch, Pascal},
   year={2016},
   month=sep, pages={469–479} }

@ARTICLE{2025MNRAS.543.3172G,
       author = {{Gunawardhana}, M.~L.~P. and {Brinchmann}, J. and {Croom}, S. and {Bunker}, A.~J. and {Bryant}, J. and {Oh}, S.},
        title = "{JADES NIRSpec spectroscopy of GN-z11: evidence for Wolf─Rayet contribution to stellar populations at 430 Myr after big bang?}",
      journal = {\mnras},
         year = 2025,
        month = nov,
       volume = {543},
       number = {4},
        pages = {3172-3195},
          doi = {10.1093/mnras/staf1620},
archivePrefix = {arXiv},
       eprint = {2504.12584},
 primaryClass = {astro-ph.GA},
       adsurl = {https://ui.adsabs.harvard.edu/abs/2025MNRAS.543.3172G}
}

@ARTICLE{2013MNRAS.432.1709C,
       author = {{Cappellari}, Michele and {Scott}, Nicholas and {Alatalo}, Katherine and {Blitz}, Leo and {Bois}, Maxime and {Bournaud}, Fr{\'e}d{\'e}ric and {Bureau}, M. and {Crocker}, Alison F. and {Davies}, Roger L. and {Davis}, Timothy A. and {de Zeeuw}, P.~T. and {Duc}, Pierre-Alain and {Emsellem}, Eric and {Khochfar}, Sadegh and {Krajnovi{\'c}}, Davor and {Kuntschner}, Harald and {McDermid}, Richard M. and {Morganti}, Raffaella and {Naab}, Thorsten and {Oosterloo}, Tom and {Sarzi}, Marc and {Serra}, Paolo and {Weijmans}, Anne-Marie and {Young}, Lisa M.},
        title = "{The ATLAS$^{3D}$ project - XV. Benchmark for early-type galaxies scaling relations from 260 dynamical models: mass-to-light ratio, dark matter, Fundamental Plane and Mass Plane}",
      journal = {\mnras},
         year = 2013,
        month = jul,
       volume = {432},
       number = {3},
        pages = {1709-1741},
          doi = {10.1093/mnras/stt562},
archivePrefix = {arXiv},
       eprint = {1208.3522},
 primaryClass = {astro-ph.CO},
       adsurl = {https://ui.adsabs.harvard.edu/abs/2013MNRAS.432.1709C}
}

@article{Atek_2024,
   title={Most of the photons that reionized the Universe came from dwarf galaxies},
   volume={626},
   ISSN={1476-4687},
   url={http://dx.doi.org/10.1038/s41586-024-07043-6},
   DOI={10.1038/s41586-024-07043-6},
   number={8001},
   journal={Nature},
   publisher={Springer Science and Business Media LLC},
   author={Atek, Hakim and Labbé, Ivo and Furtak, Lukas J. and Chemerynska, Iryna and Fujimoto, Seiji and Setton, David J. and Miller, Tim B. and Oesch, Pascal and Bezanson, Rachel and Price, Sedona H. and Dayal, Pratika and Zitrin, Adi and Kokorev, Vasily and Weaver, John R. and Brammer, Gabriel and Dokkum, Pieter van and Williams, Christina C. and Cutler, Sam E. and Feldmann, Robert and Fudamoto, Yoshinobu and Greene, Jenny E. and Leja, Joel and Maseda, Michael V. and Muzzin, Adam and Pan, Richard and Papovich, Casey and Nelson, Erica J. and Nanayakkara, Themiya and Stark, Daniel P. and Stefanon, Mauro and Suess, Katherine A. and Wang, Bingjie and Whitaker, Katherine E.},
   year={2024},
   month=feb, pages={975–978} }

@ARTICLE{2010Ap&SS.328..179G,
       author = {{Grevesse}, N. and {Asplund}, M. and {Sauval}, A.~J. and {Scott}, P.},
        title = "{The chemical composition of the Sun}",
      journal = {\apss},
         year = 2010,
        month = jul,
       volume = {328},
       number = {1-2},
        pages = {179-183},
          doi = {10.1007/s10509-010-0288-z},
       adsurl = {https://ui.adsabs.harvard.edu/abs/2010Ap&SS.328..179G}
}

@ARTICLE{2013ApJS..208...10D,
       author = {{Dopita}, Michael A. and {Sutherland}, Ralph S. and {Nicholls}, David C. and {Kewley}, Lisa J. and {Vogt}, Fr{\'e}d{\'e}ric P.~A.},
        title = "{New Strong-line Abundance Diagnostics for H II Regions: Effects of {\ensuremath{\kappa}}-distributed Electron Energies and New Atomic Data}",
      journal = {\apjs},
         year = 2013,
        month = sep,
       volume = {208},
       number = {1},
          eid = {10},
        pages = {10},
          doi = {10.1088/0067-0049/208/1/10},
archivePrefix = {arXiv},
       eprint = {1307.5950},
 primaryClass = {astro-ph.CO},
       adsurl = {https://ui.adsabs.harvard.edu/abs/2013ApJS..208...10D}
}

@ARTICLE{1965ApJ...142.1633G,
       author = {{Gunn}, James E. and {Peterson}, Bruce A.},
        title = "{On the Density of Neutral Hydrogen in Intergalactic Space.}",
      journal = {\apj},
         year = 1965,
        month = nov,
       volume = {142},
        pages = {1633-1636},
          doi = {10.1086/148444},
       adsurl = {https://ui.adsabs.harvard.edu/abs/1965ApJ...142.1633G}
}

@article{Johnson_2009,
   title={The first galaxies: signatures of the initial starburst},
   volume={399},
   ISSN={1365-2966},
   url={http://dx.doi.org/10.1111/j.1365-2966.2009.15158.x},
   DOI={10.1111/j.1365-2966.2009.15158.x},
   number={1},
   journal={Monthly Notices of the Royal Astronomical Society},
   publisher={Oxford University Press (OUP)},
   author={Johnson, Jarrett L. and Greif, Thomas H. and Bromm, Volker and Klessen, Ralf S. and Ippolito, Joseph},
   year={2009},
   month=oct, pages={37–47} }

@article{Sana_2012,
   title={Binary Interaction Dominates the Evolution of Massive Stars},
   volume={337},
   ISSN={1095-9203},
   url={http://dx.doi.org/10.1126/science.1223344},
   DOI={10.1126/science.1223344},
   number={6093},
   journal={Science},
   publisher={American Association for the Advancement of Science (AAAS)},
   author={Sana, H. and de Mink, S. E. and de Koter, A. and Langer, N. and Evans, C. J. and Gieles, M. and Gosset, E. and Izzard, R. G. and Le Bouquin, J.-B. and Schneider, F. R. N.},
   year={2012},
   month=july, pages={444–446} }

@article{Mason_2009,
   title={THE HIGH ANGULAR RESOLUTION MULTIPLICITY OF MASSIVE STARS},
   volume={137},
   ISSN={1538-3881},
   url={http://dx.doi.org/10.1088/0004-6256/137/2/3358},
   DOI={10.1088/0004-6256/137/2/3358},
   number={2},
   journal={The Astronomical Journal},
   publisher={American Astronomical Society},
   author={Mason, Brian D. and Hartkopf, William I. and Gies, Douglas R. and Henry, Todd J. and Helsel, John W.},
   year={2009},
   month=jan, pages={3358–3377} }

@article{Sana_2025,
   title={A high fraction of close massive binary stars at low metallicity},
   volume={9},
   ISSN={2397-3366},
   url={http://dx.doi.org/10.1038/s41550-025-02610-x},
   DOI={10.1038/s41550-025-02610-x},
   number={9},
   journal={Nature Astronomy},
   publisher={Springer Science and Business Media LLC},
   author={Sana, H. and Shenar, T. and Bodensteiner, J. and Britavskiy, N. and Langer, N. and Lennon, D. J. and Mahy, L. and Mandel, I. and de Mink, S. E. and Patrick, L. R. and Villaseñor, J. I. and Dirickx, M. and Abdul-Masih, M. and Almeida, L. A. and Backs, F. and Berlanas, S. R. and Bernini-Peron, M. and Bowman, D. M. and Bronner, V. A. and Crowther, P. A. and Deshmukh, K. and Evans, C. J. and Fabry, M. and Gieles, M. and Gilkis, A. and González-Torà, G. and Gräfener, G. and Götberg, Y. and Hawcroft, C. and Hénault-Brunet, V. and Herrero, A. and Holgado, G. and Izzard, R. G. and de Koter, A. and Janssens, S. and Johnston, C. and Josiek, J. and Justham, S. and Kalari, V. M. and Klencki, J. and Kubát, J. and Kubátová, B. and Lefever, R. R. and van Loon, J. Th. and Ludwig, B. and Mackey, J. and Maíz Apellániz, J. and Maravelias, G. and Marchant, P. and Mazeh, T. and Menon, A. and Moe, M. and Najarro, F. and Oskinova, L. M. and Ovadia, R. and Pauli, D. and Pawlak, M. and Ramachandran, V. and Renzo, M. and Rocha, D. F. and Sander, A. A. C. and Schneider, F. R. N. and Schootemeijer, A. and Schösser, E. C. and Schürmann, C. and Sen, K. and Shahaf, S. and Simón-Díaz, S. and van Son, L. A. C. and Stoop, M. and Toonen, S. and Tramper, F. and Valli, R. and Vigna-Gómez, A. and Vink, J. S. and Wang, C. and Willcox, R.},
   year={2025},
   month=sept, pages={1337–1346} }

@article{Stanway_2018,
   title={Re-evaluating old stellar populations},
   volume={479},
   ISSN={1365-2966},
   url={http://dx.doi.org/10.1093/mnras/sty1353},
   DOI={10.1093/mnras/sty1353},
   number={1},
   journal={Monthly Notices of the Royal Astronomical Society},
   publisher={Oxford University Press (OUP)},
   author={Stanway, E R and Eldridge, J J},
   year={2018},
   month=may, pages={75–93} }

\appendix
\section{Spectra of \heii\ emitting galaxies in our sample}
\label{appendix}

Here, we present the spectra of all
our \heii\ emitting galaxies in our sample.

\begin{figure*}[ht!]
    \centering
\includegraphics[width=0.98\linewidth]
{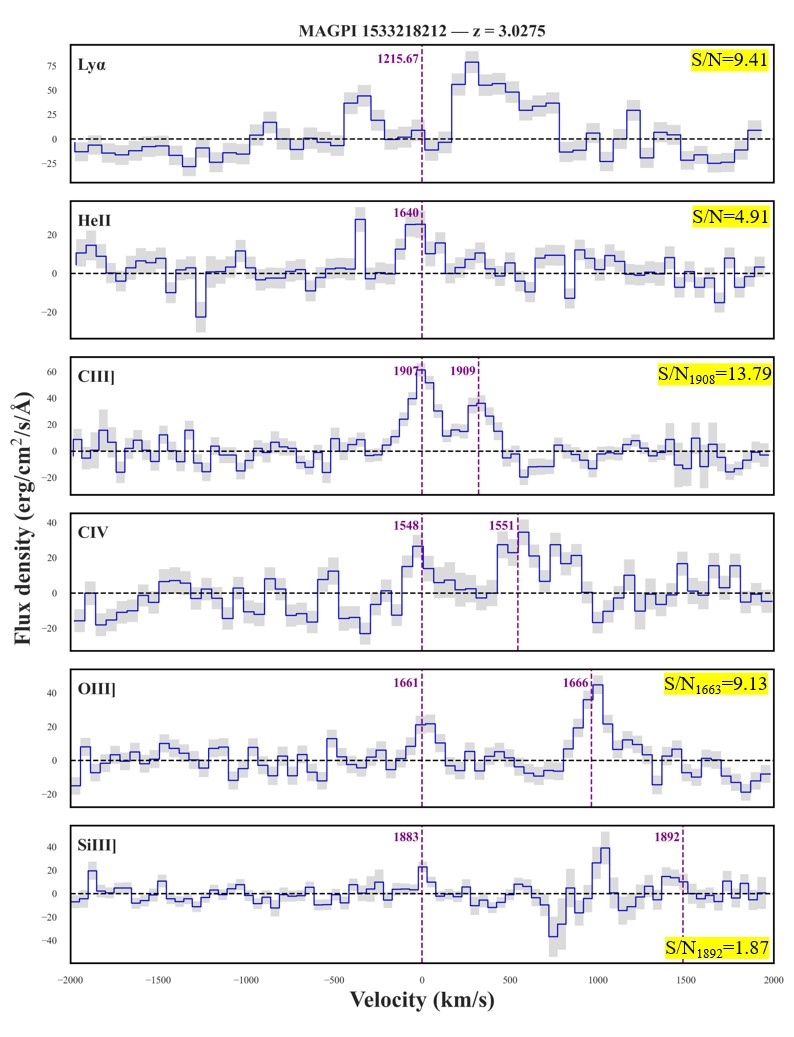}
    \caption{Multipanel figure presenting continuum-subtracted spectra of rest-UV emission lines. From top to bottom: \lya\ $\lambda1215.67$, \heii\ $\lambda1640$, \ciii\ $\lambda\lambda1907,09$, \civ\ $\lambda\lambda1548,51$, \oiii\ $\lambda\lambda1661,66$, and \siiii\ $\lambda\lambda1883,92$. The shaded region in grey represents the MUSE $1\sigma$ uncertainties. The $S/N$ is indicated for each emission line. For \ciii, \oiii, and \siiii, $S/N_{1908}$, $S/N_{1663}$, and $S/N_{1888}$ denote the combined $S/N$ of the two doublet components, respectively. When only one component is detected, the wavelength of that component is used in the $S/N$ label (e.g. $S/N_{1907}$, $S/N_{1666}$, or $S/N_{1883}$). The \lya\ $S/N$ values are adopted from \citep{2026PASA...43...21M}.}
    \label{fig:1}
\end{figure*}

\begin{figure*}[ht!]
    \centering
\includegraphics[width=0.98\linewidth]
{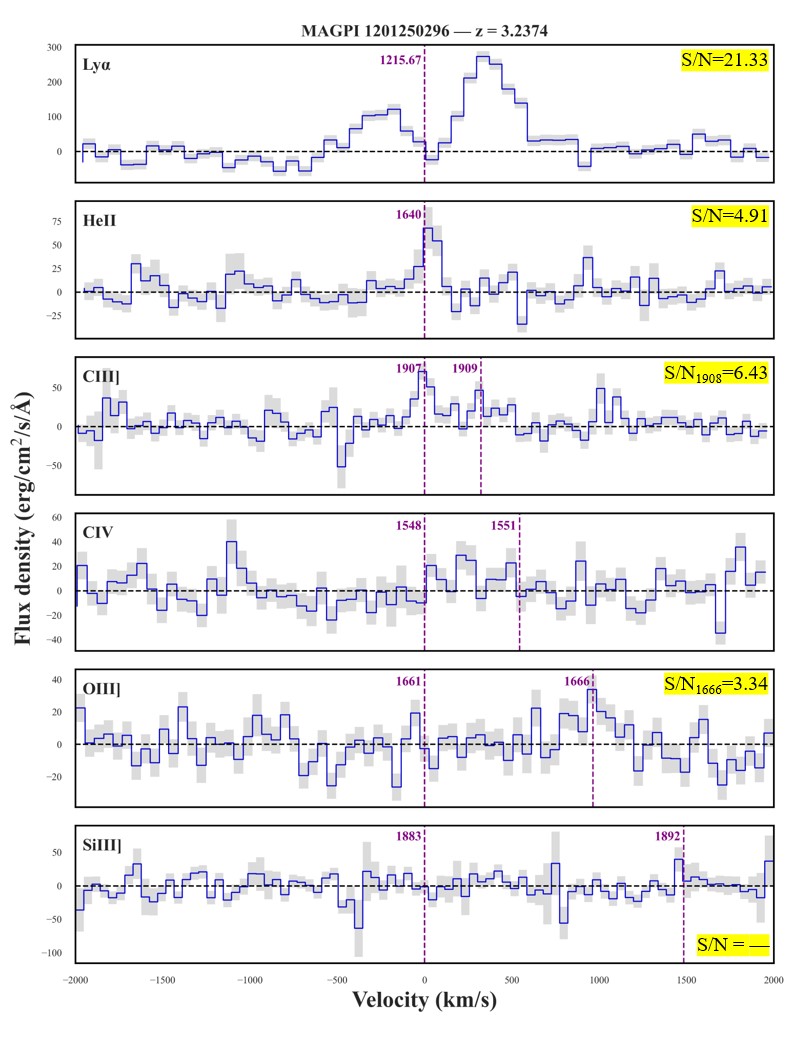}
    \caption{continued}
    \label{fig:2}
\end{figure*}

\begin{figure*}[ht!]
    \centering
\includegraphics[width=0.98\linewidth]
{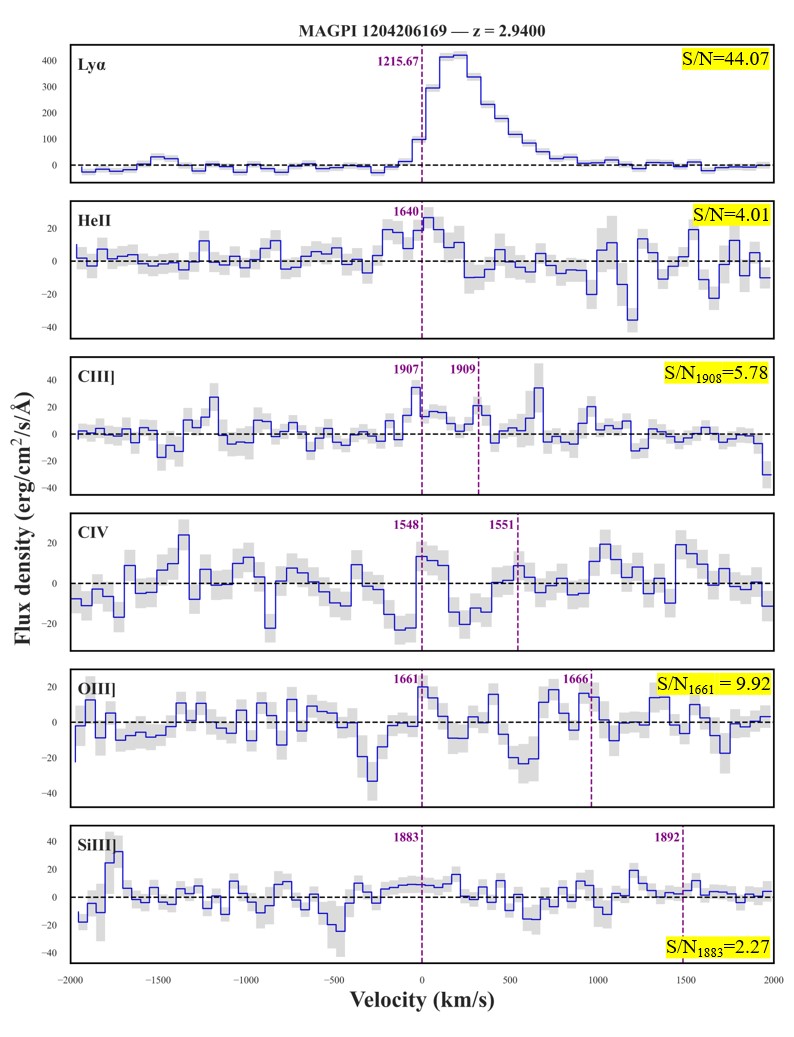}
    \caption{continued}
    \label{fig:3}
\end{figure*}

\begin{figure*}[ht!]
    \centering
\includegraphics[width=0.98\linewidth]
{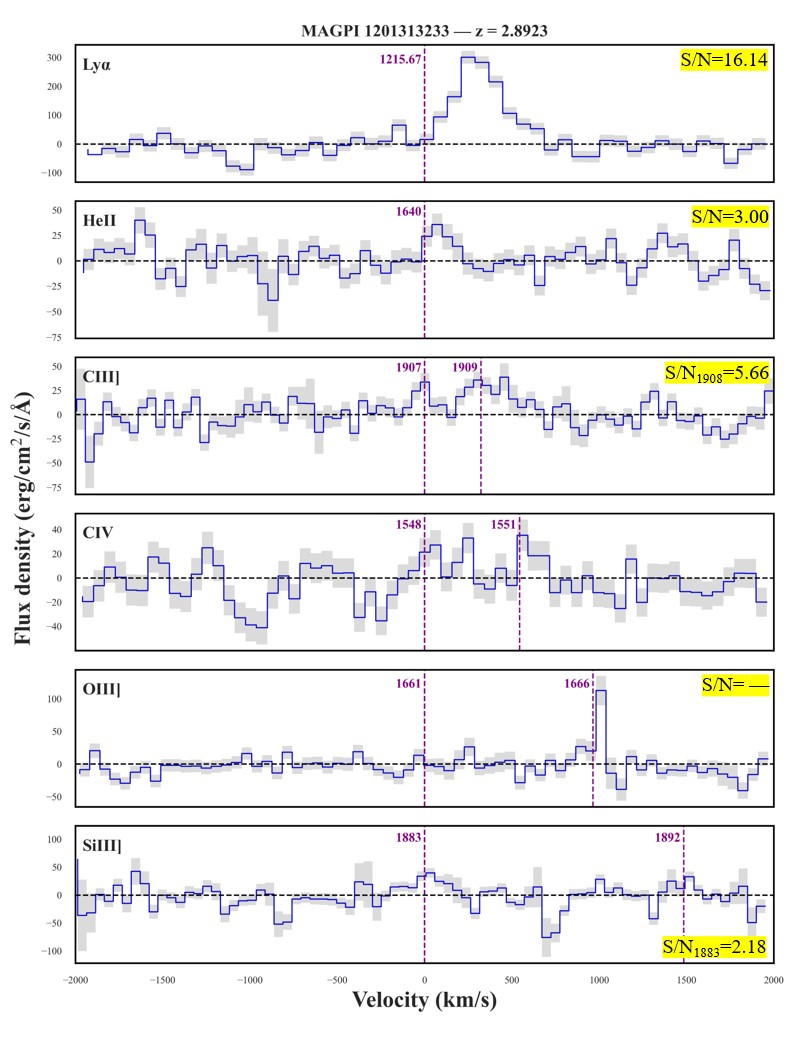}
    \caption{continued}
    \label{fig:4}
\end{figure*}

\begin{figure*}[ht!]
    \centering
\includegraphics[width=0.98\linewidth]
{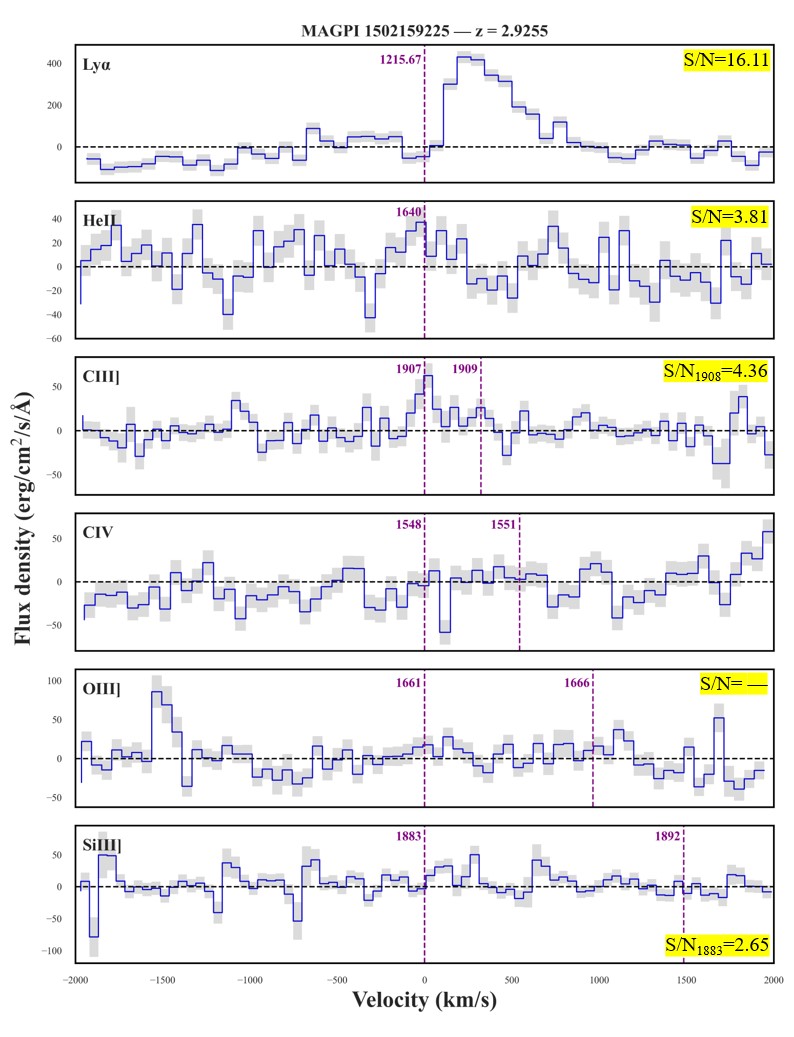}
    \caption{continued}
    \label{fig:5}
\end{figure*}

\begin{figure*}[ht!]
    \centering
\includegraphics[width=0.98\linewidth]
{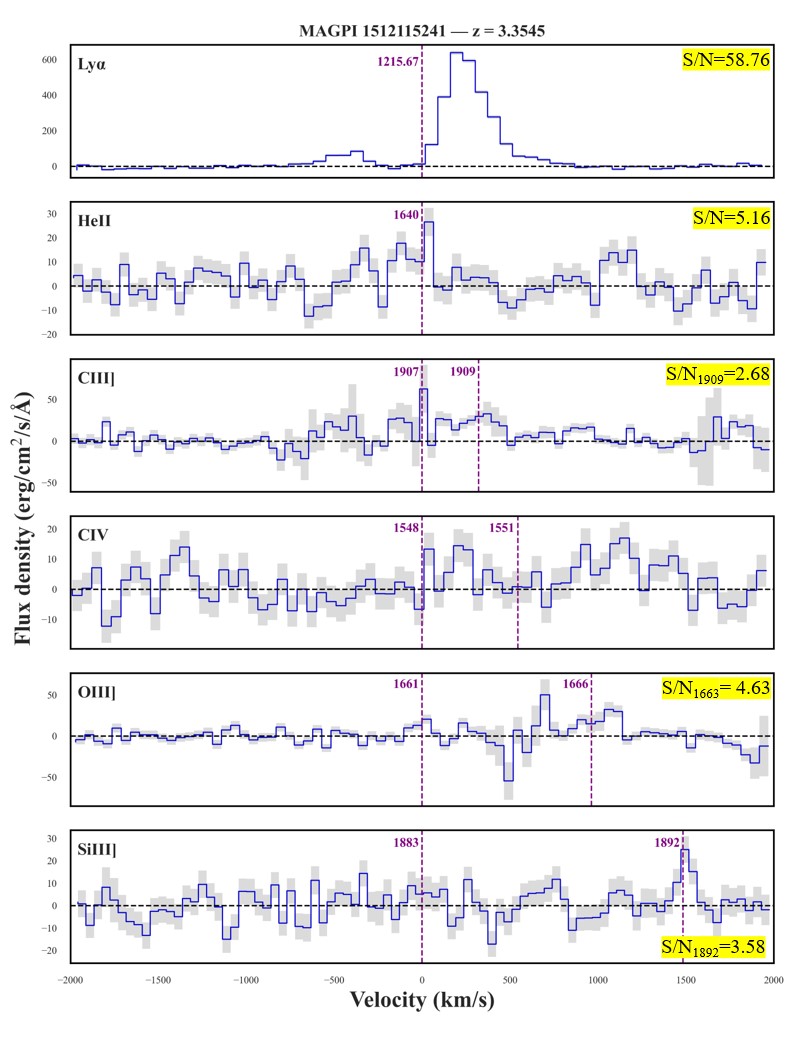}
    \caption{continued}
    \label{fig:6}
\end{figure*}

\begin{figure*}[ht!]
    \centering
\includegraphics[width=0.98\linewidth]
{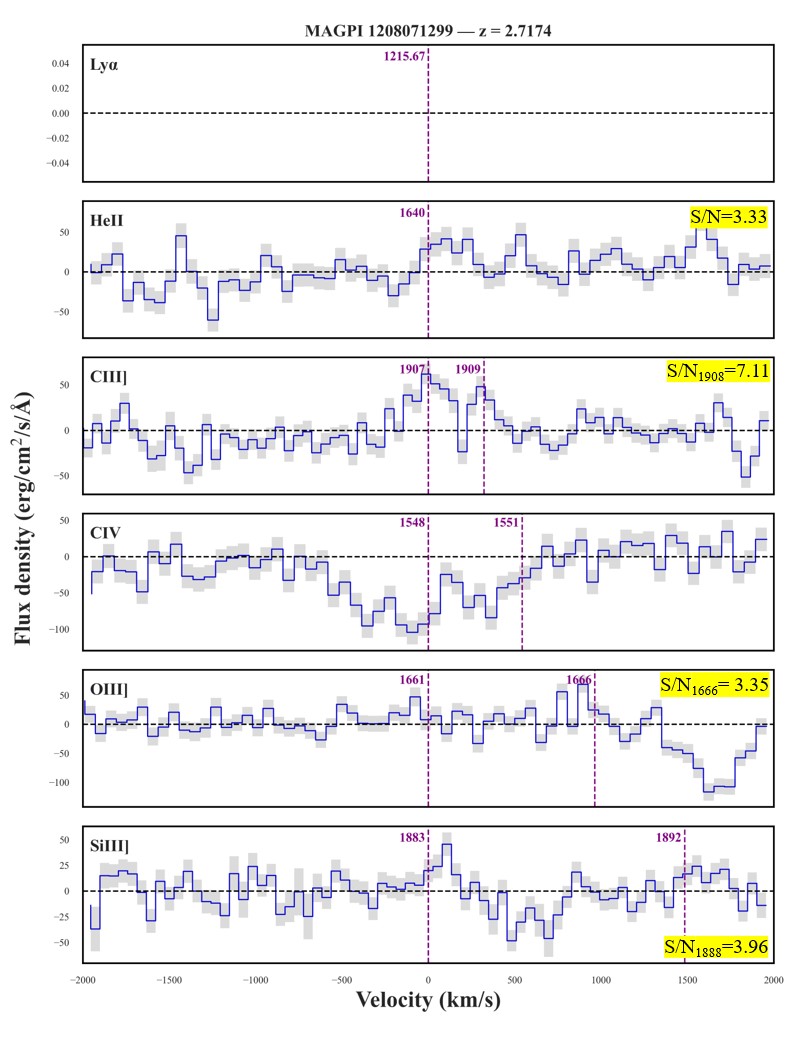}
    \caption{continued}
    \label{fig:7}
\end{figure*}

\begin{figure*}[ht!]
    \centering
\includegraphics[width=0.98\linewidth]
{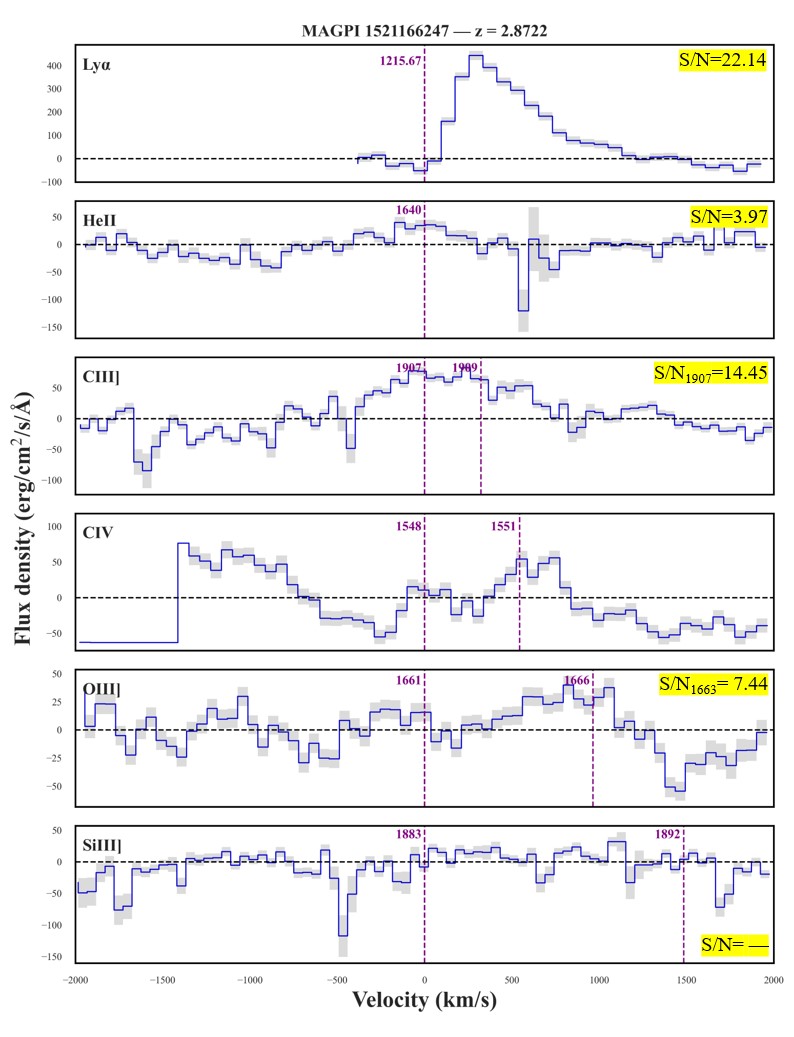}
    \caption{continued}
    \label{fig:8}
\end{figure*}

\begin{figure*}[ht!]
    \centering
\includegraphics[width=0.98\linewidth]
{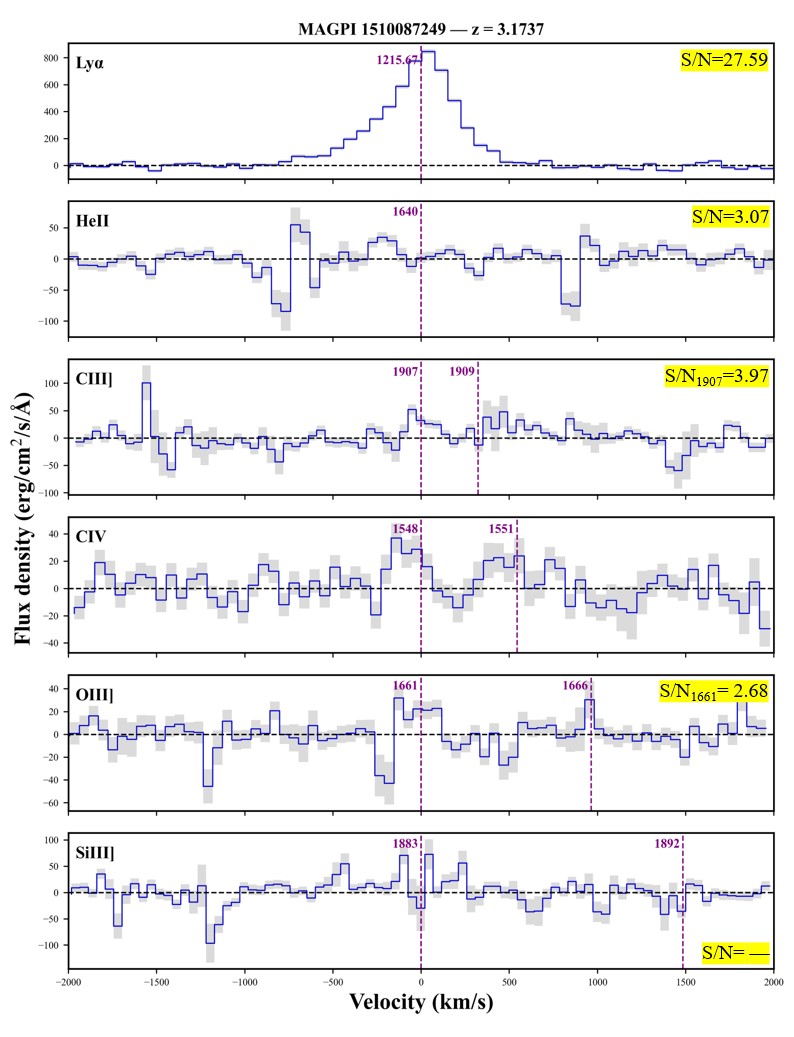}
    \caption{continued}
    \label{fig:8}
\end{figure*}

\begin{figure*}[ht!]
    \centering
\includegraphics[width=0.98\linewidth]
{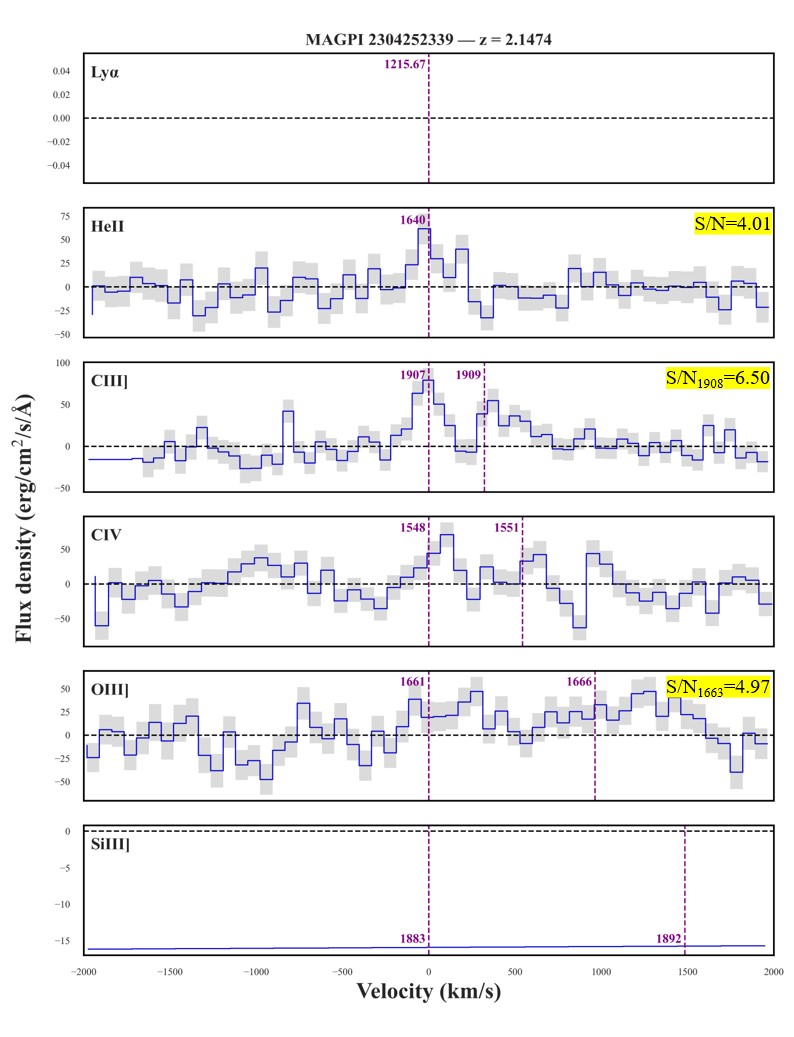}
    \caption{continued}
    \label{fig:9}
\end{figure*}

\begin{figure*}[ht!]
    \centering
\includegraphics[width=0.98\linewidth]
    {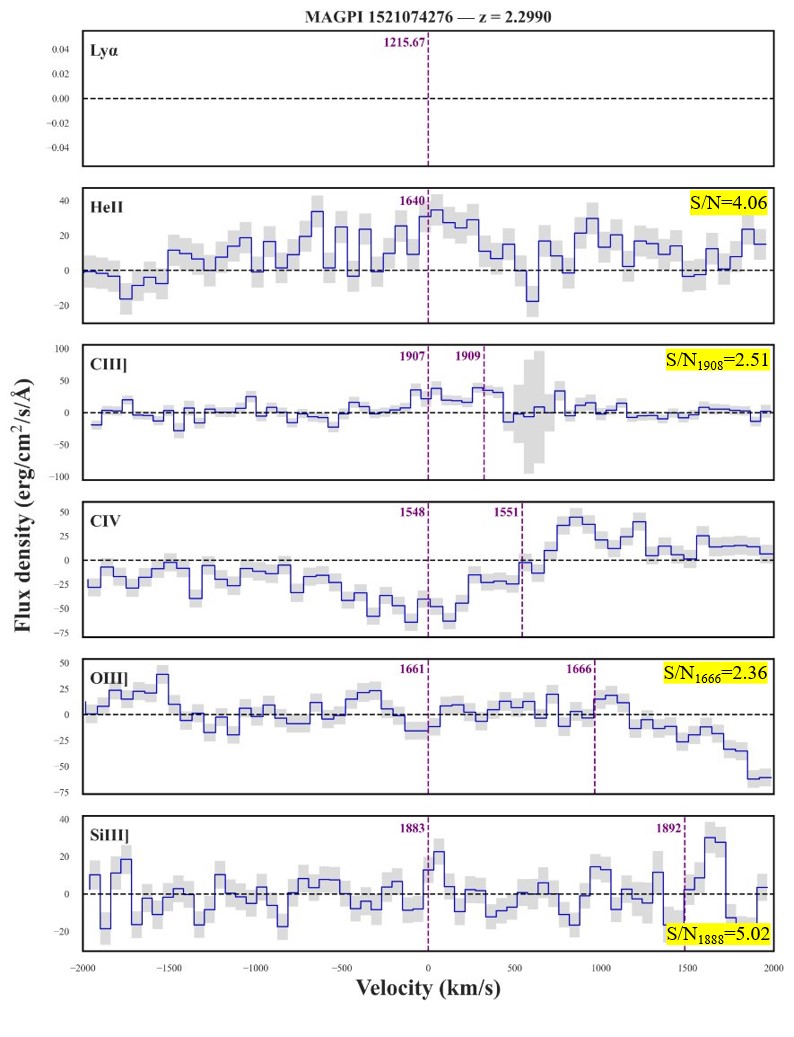}
    \caption{continued}
    \label{fig:10}
\end{figure*}

\begin{figure*}[ht!]
    \centering
\includegraphics[width=0.98\linewidth]
{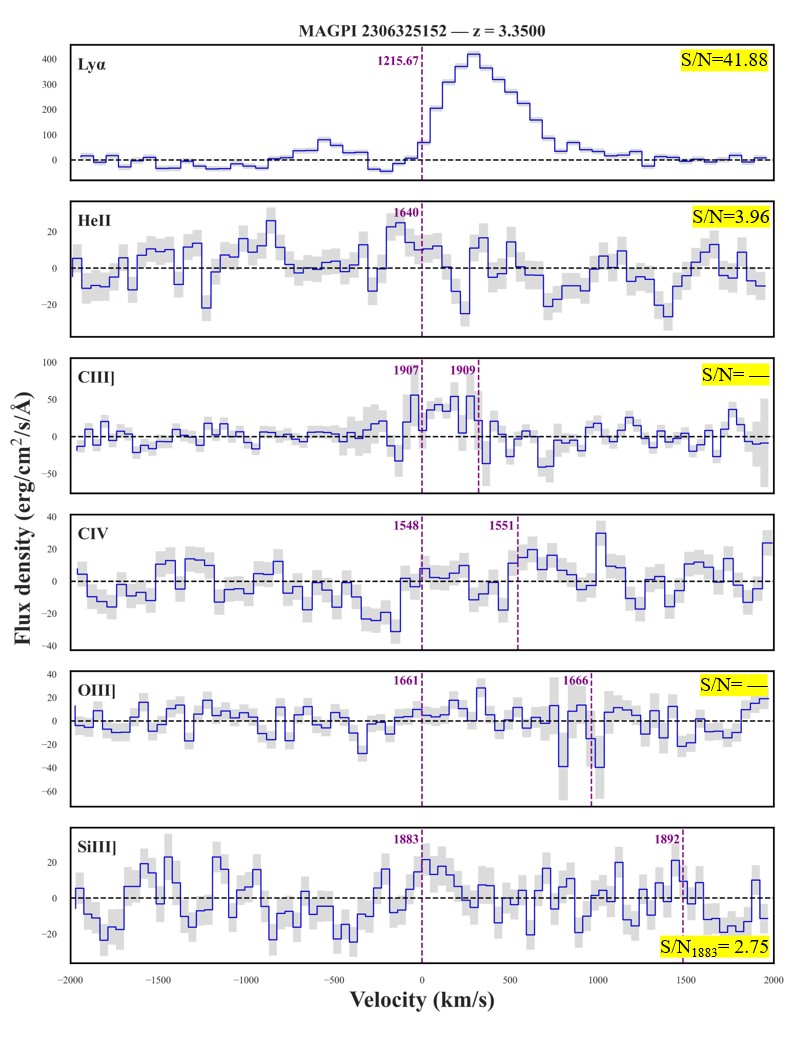}
    \caption{continued}
    \label{fig:11}
\end{figure*}

\begin{figure*}[ht!]
    \centering
\includegraphics[width=0.98\linewidth]
{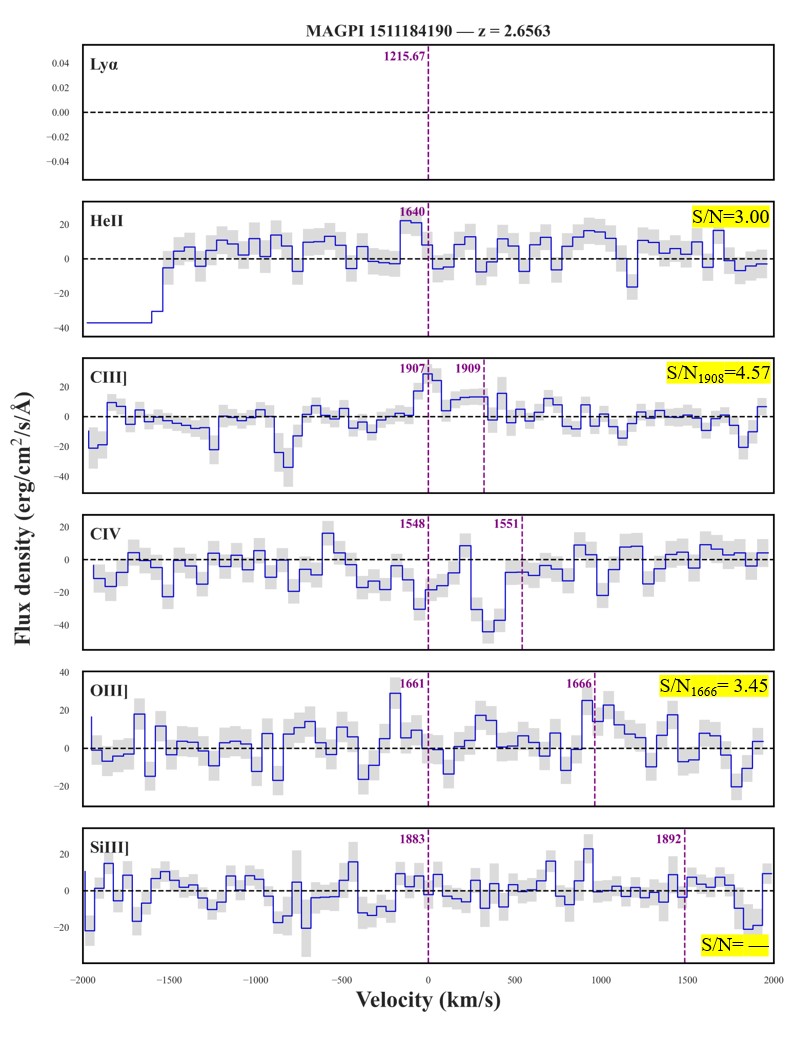}
    \caption{continued}
    \label{fig:12}
\end{figure*}

\begin{figure*}[ht!]
    \centering
\includegraphics[width=0.98\linewidth]
{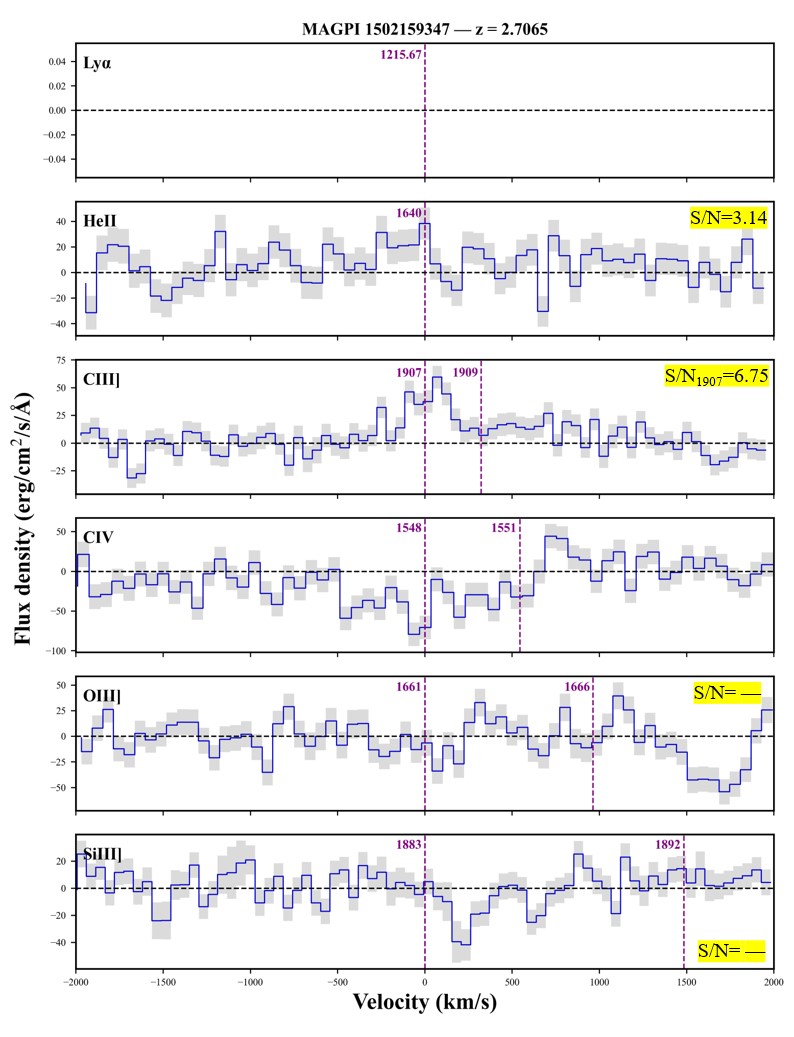}
    \caption{continued}
    \label{fig:13}
\end{figure*}

\begin{figure*}[ht!]
    \centering
\includegraphics[width=0.98\linewidth]
{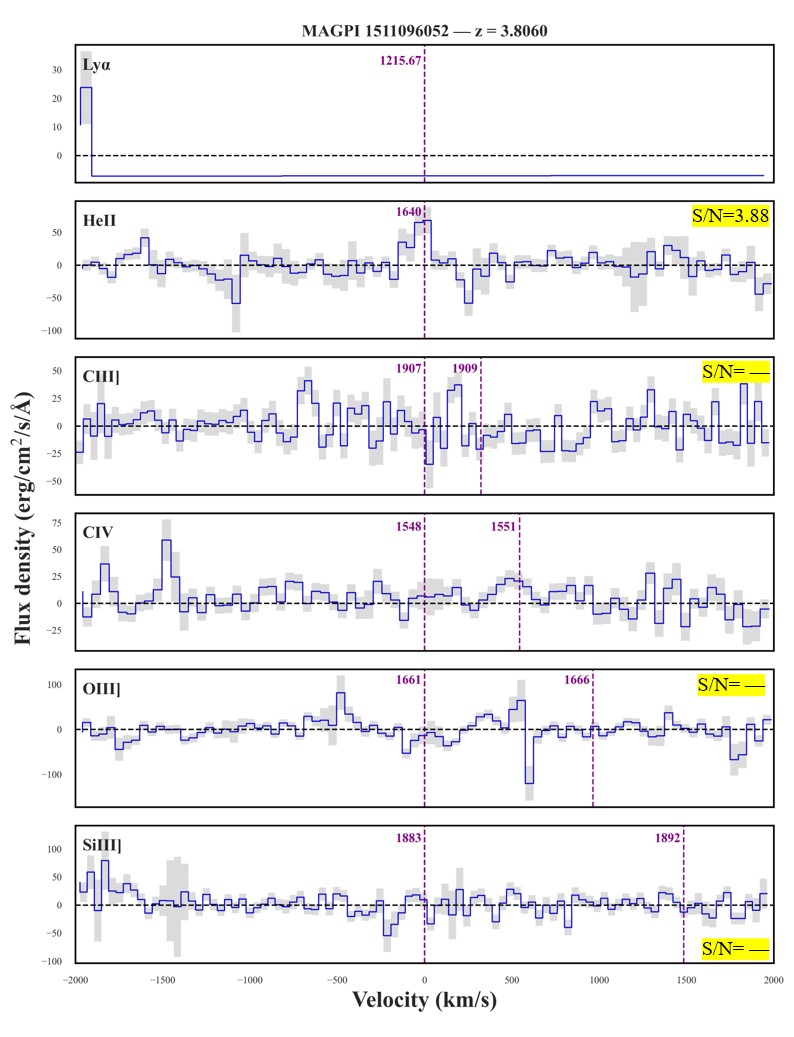}
    \caption{continued}
    \label{fig:14}
\end{figure*}

\begin{figure*}[ht!]
    \centering
\includegraphics[width=0.98\linewidth]
{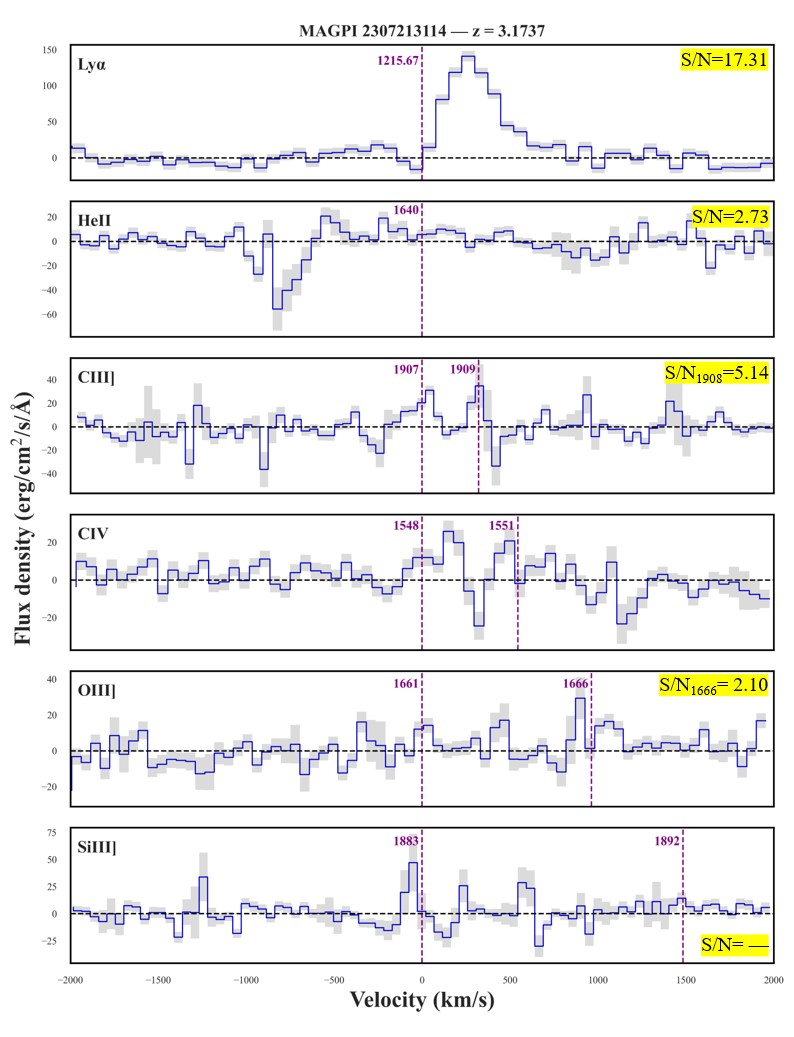}
    \caption{continued}
    \label{fig:15}
\end{figure*}

\begin{figure*}[ht!]
    \centering
\includegraphics[width=0.98\linewidth]
{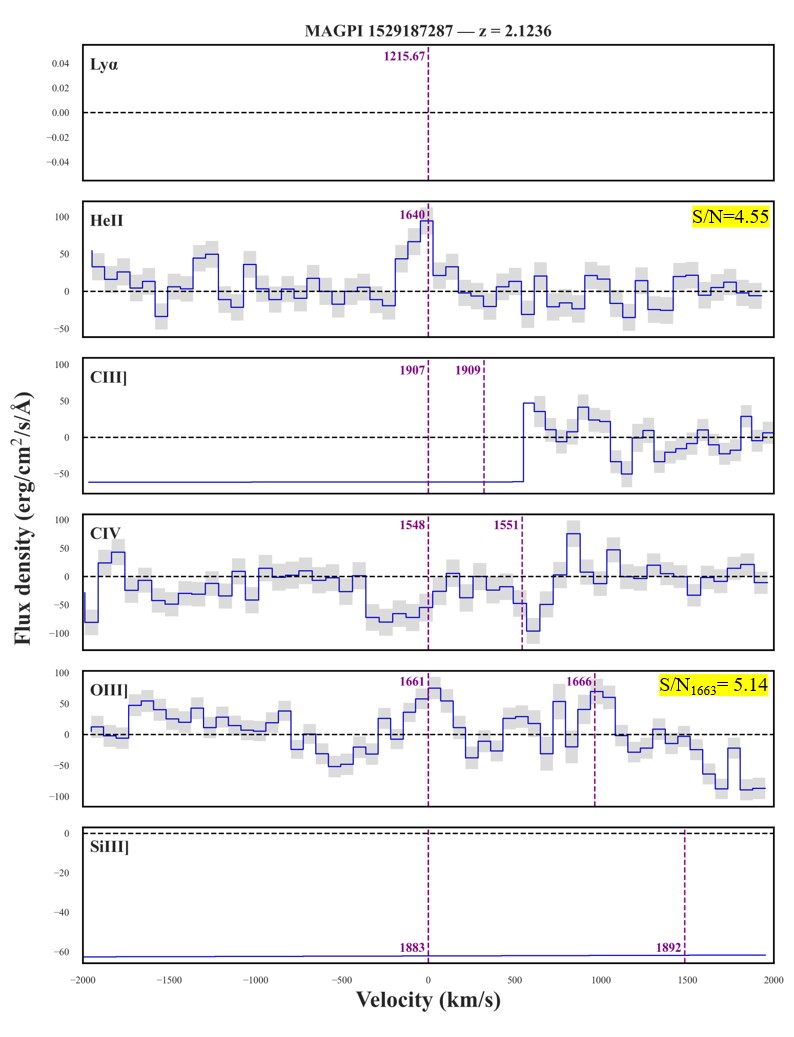}
    \caption{continued}
    \label{fig:16}
\end{figure*}

\begin{figure*}[ht!]
    \centering
\includegraphics[width=0.98\linewidth]
{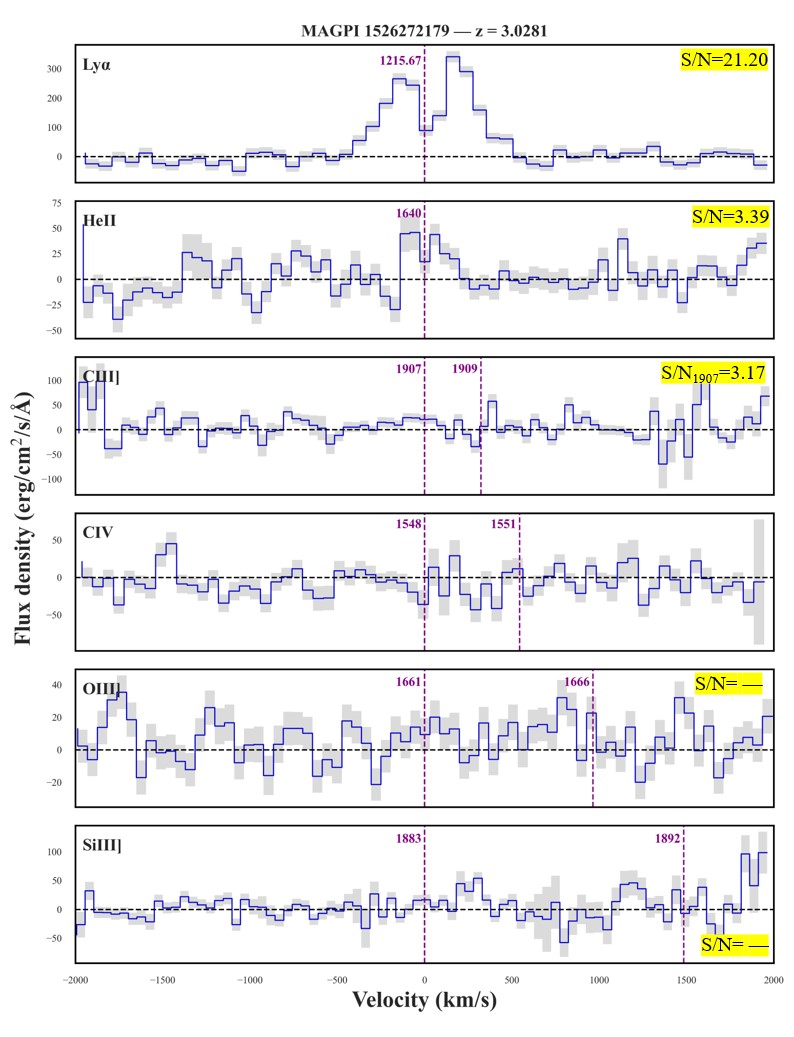}
    \caption{continued}
    \label{fig:18}
\end{figure*}

\begin{figure*}[ht!]
    \centering
\includegraphics[width=0.98\linewidth]
{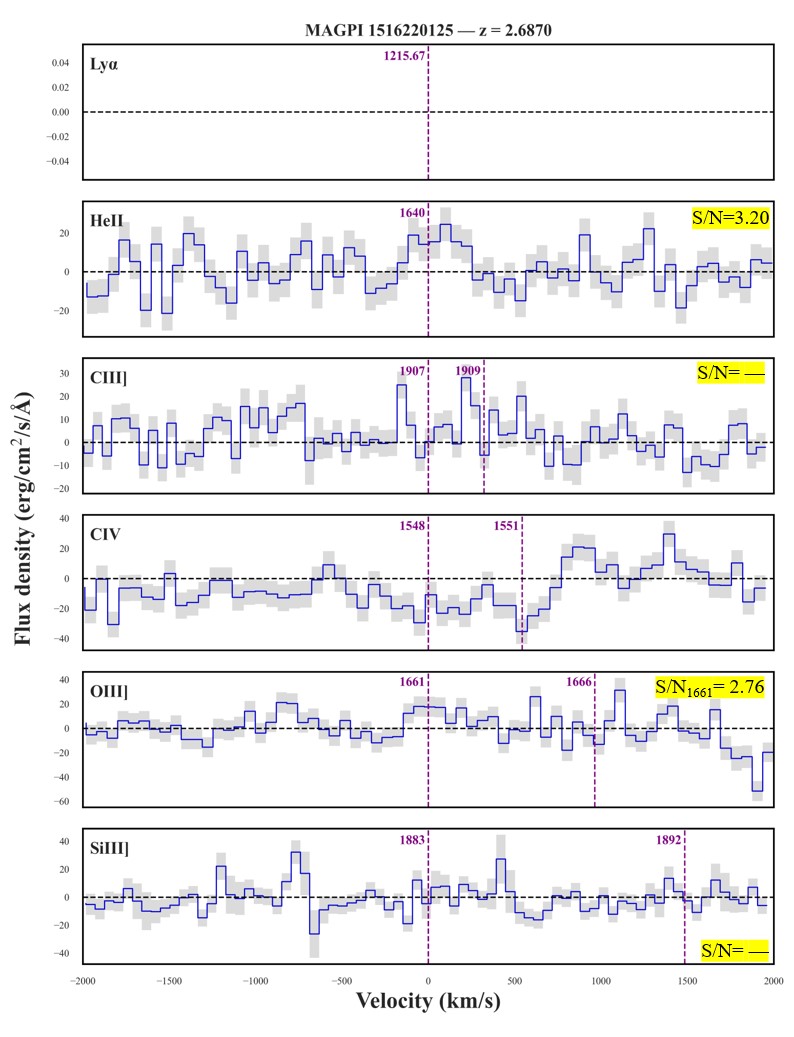}
    \caption{continued}
    \label{fig:19}
\end{figure*}

\begin{figure*}[ht!]
    \centering
\includegraphics[width=0.98\linewidth]
{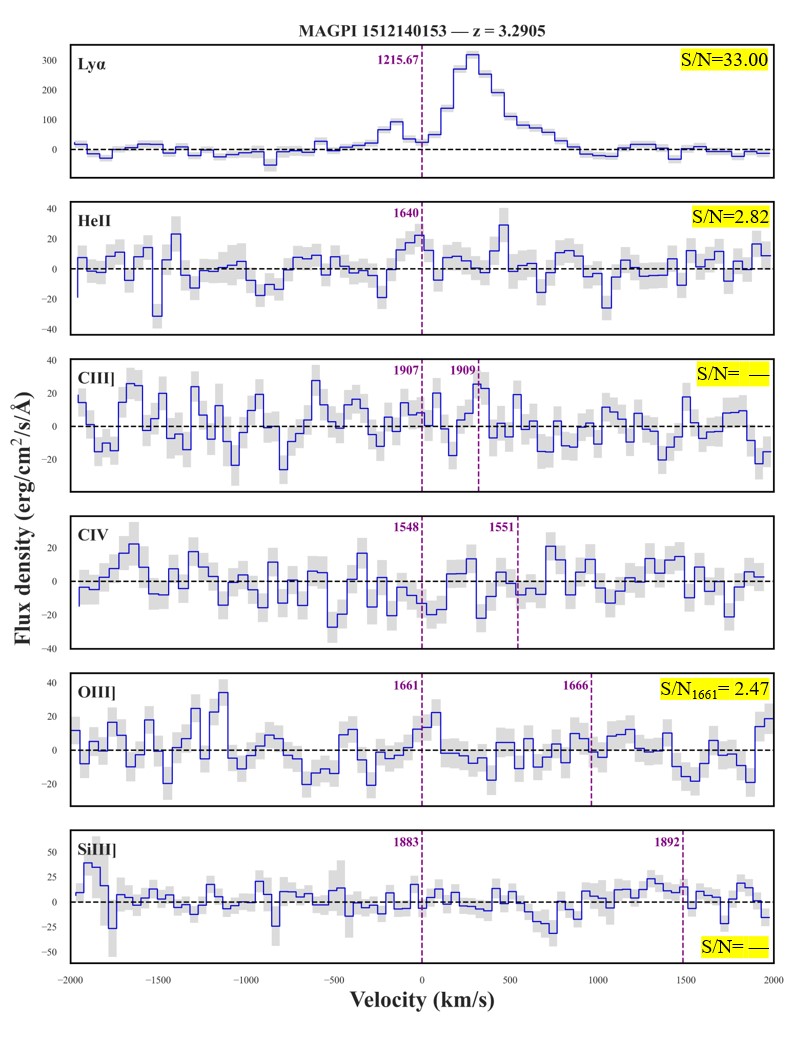}
    \caption{continued}
    \label{fig:20}
\end{figure*}

\begin{figure*}[ht!]
    \centering
\includegraphics[width=0.98\linewidth]
{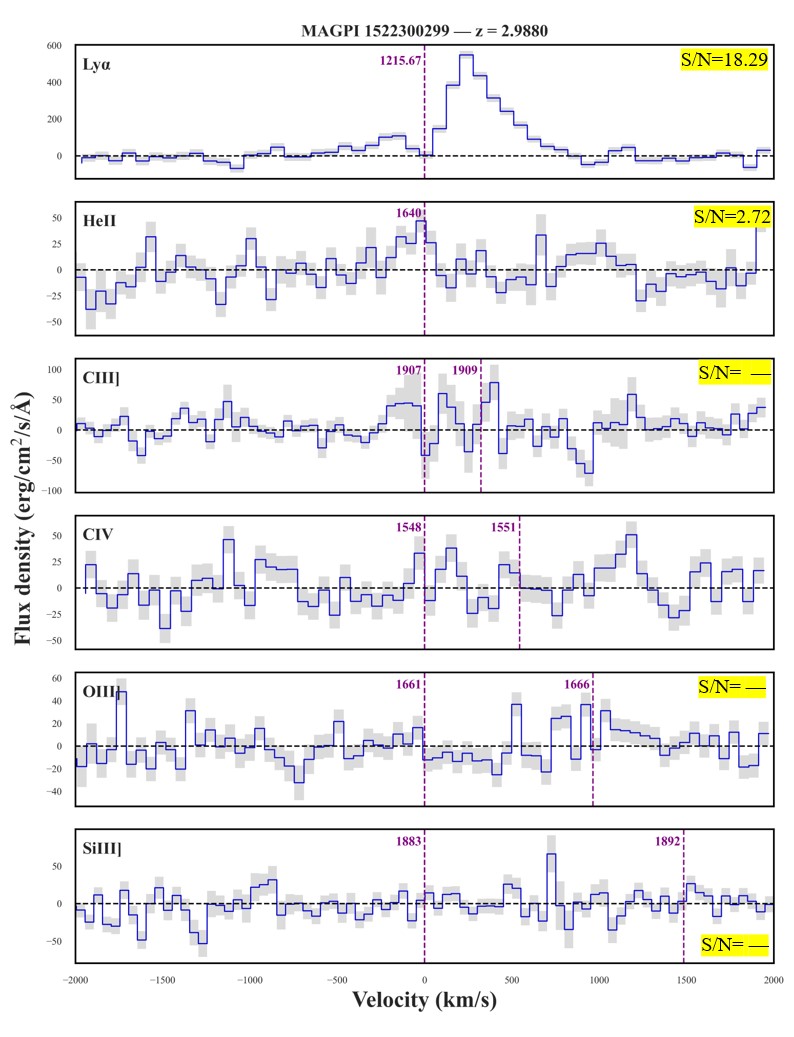}
\caption{continued}
    \label{fig:21}
\end{figure*}

\begin{figure*}[ht!]
    \centering
\includegraphics[width=0.98\linewidth]
{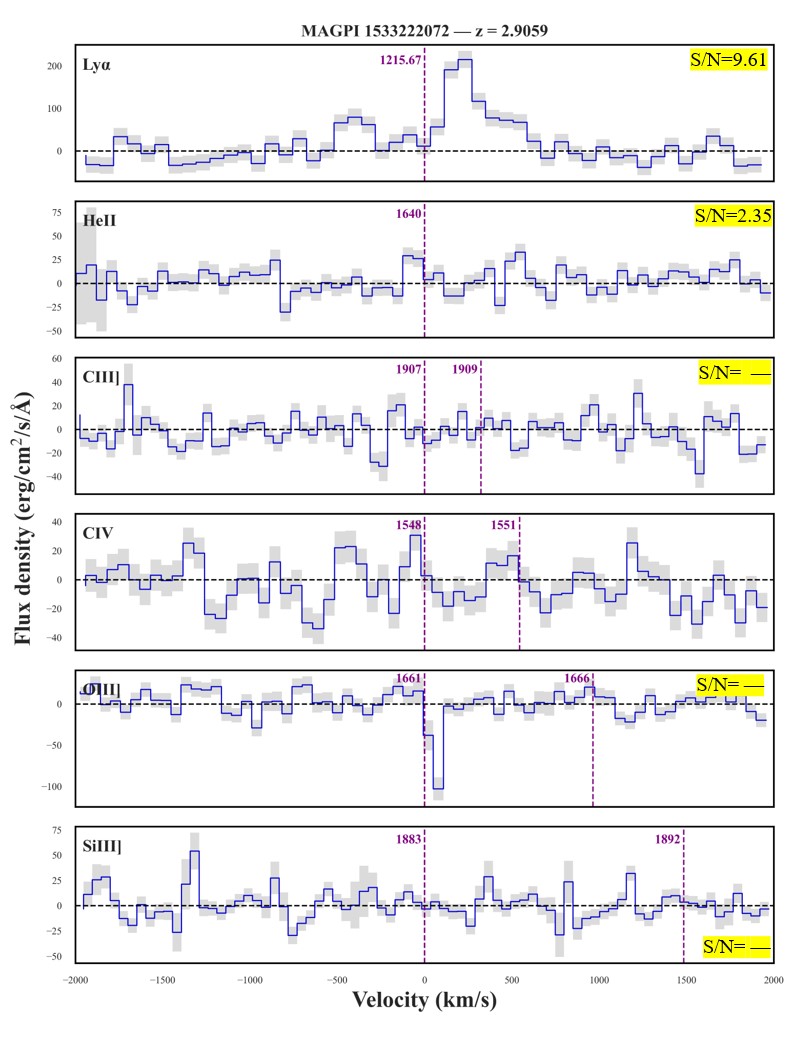}
    \caption{continued}
    \label{fig:22}
\end{figure*}

\begin{figure*}[ht!]
    \centering
\includegraphics[width=0.98\linewidth]
{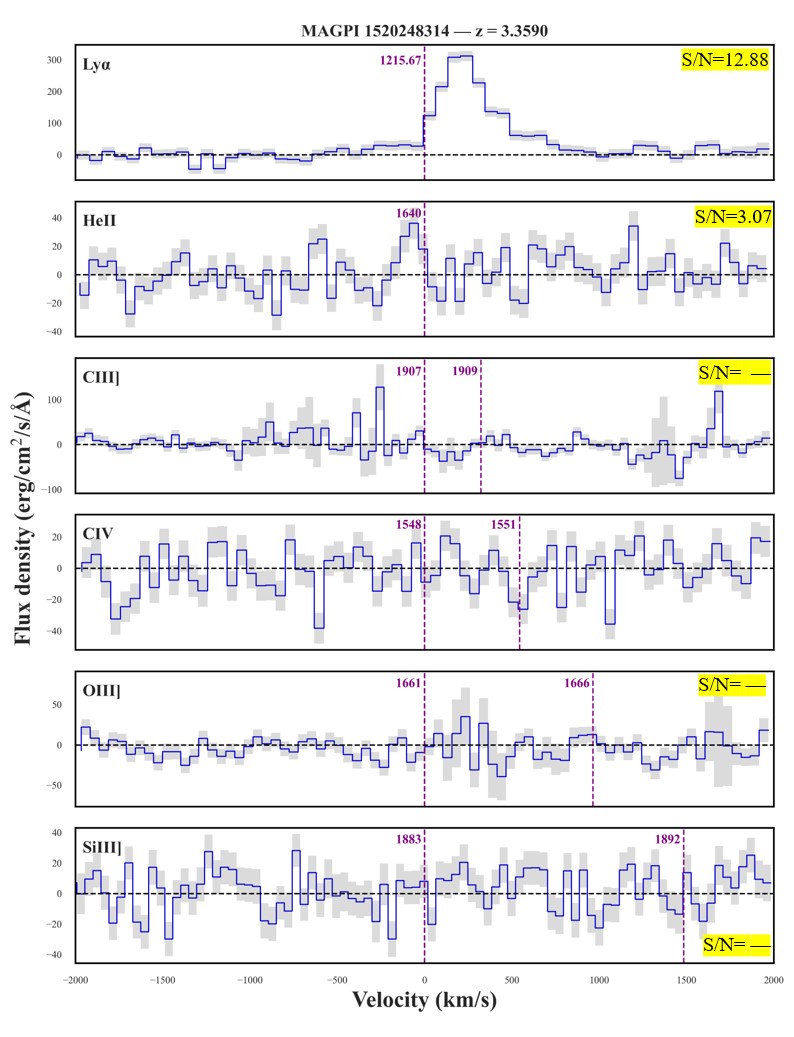}
    \caption{continued}
    \label{fig:23}
\end{figure*}

\begin{figure*}[ht!]
    \centering
\includegraphics[width=0.98\linewidth]
{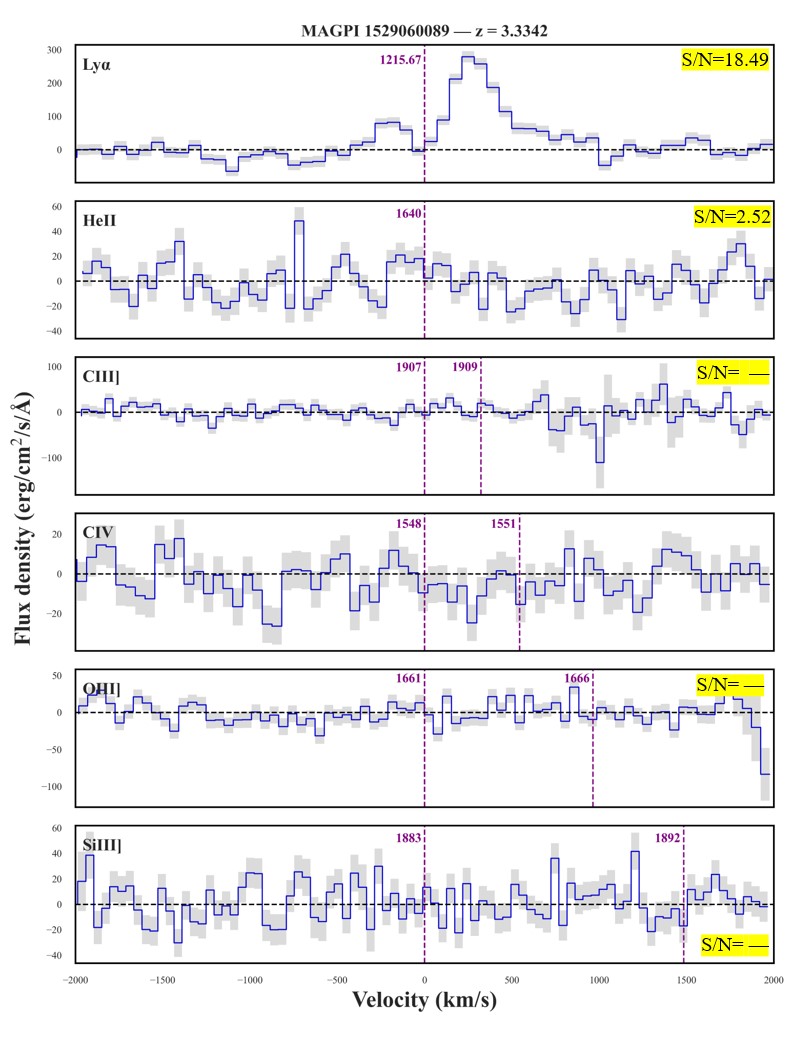}
    \caption{continued}
    \label{fig:24}
\end{figure*}

\begin{figure*}[ht!]
    \centering
\includegraphics[width=0.98\linewidth]
{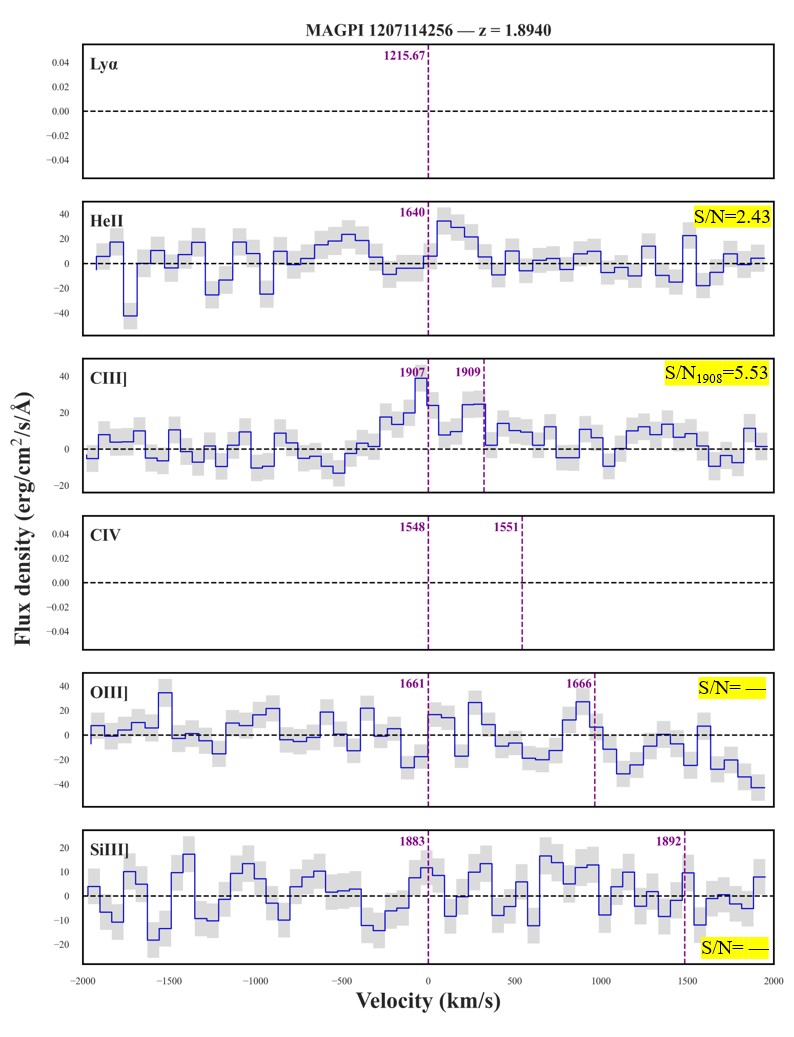}
    \caption{continued}
    \label{fig:25}
\end{figure*}

\begin{figure*}[ht!]
    \centering
\includegraphics[width=0.98\linewidth]
{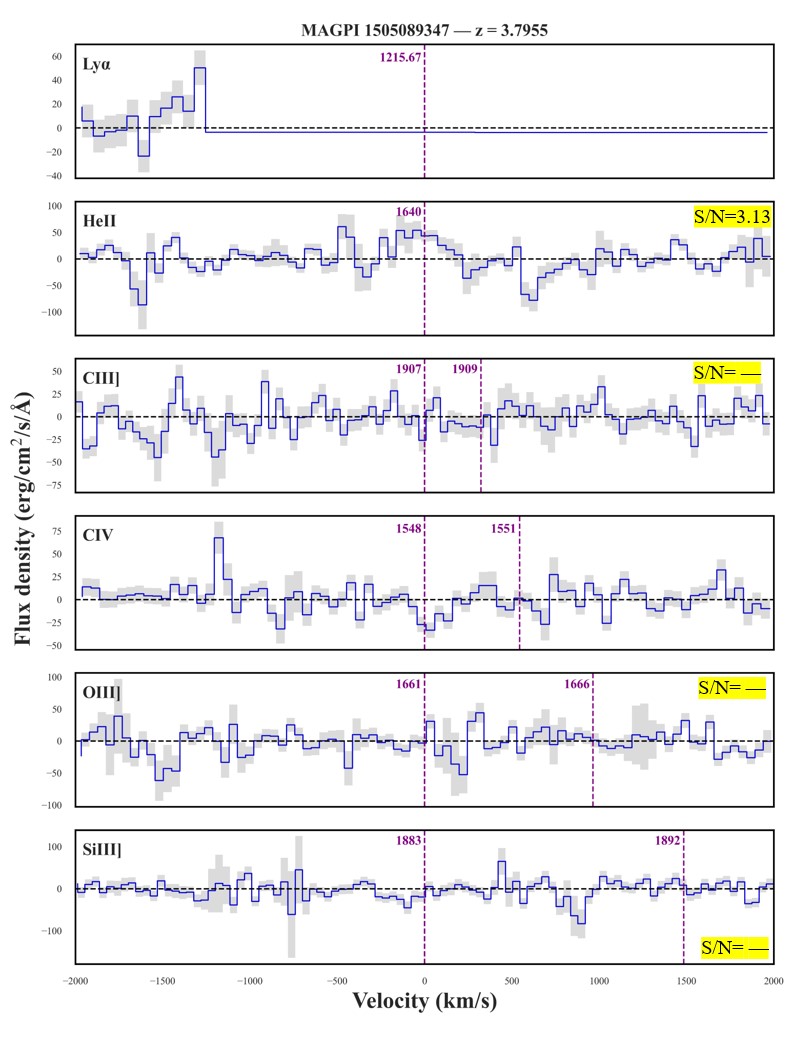}
    \caption{continued}
    \label{fig:26}
    
\end{figure*}
\begin{figure*}[ht!]
    \centering
\includegraphics[width=0.98\linewidth]
{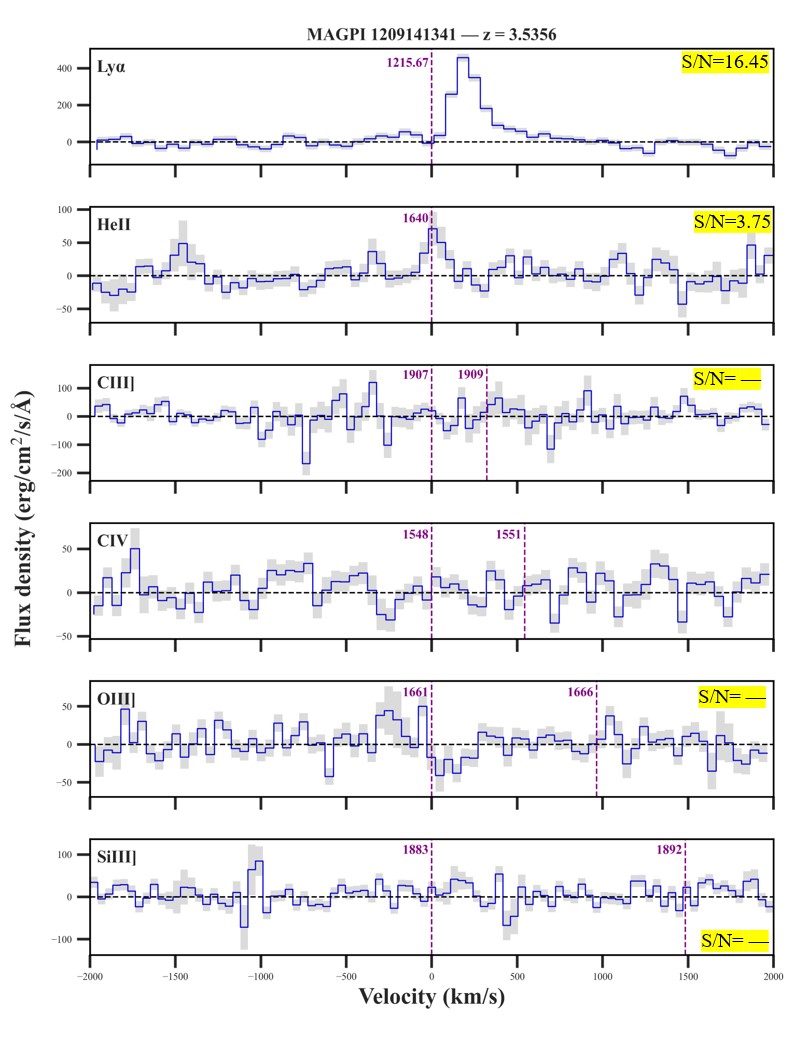}
    \caption{continued}
    \label{fig:27}
\end{figure*}

\begin{figure*}[ht!]
    \centering
\includegraphics[width=0.98\linewidth]
{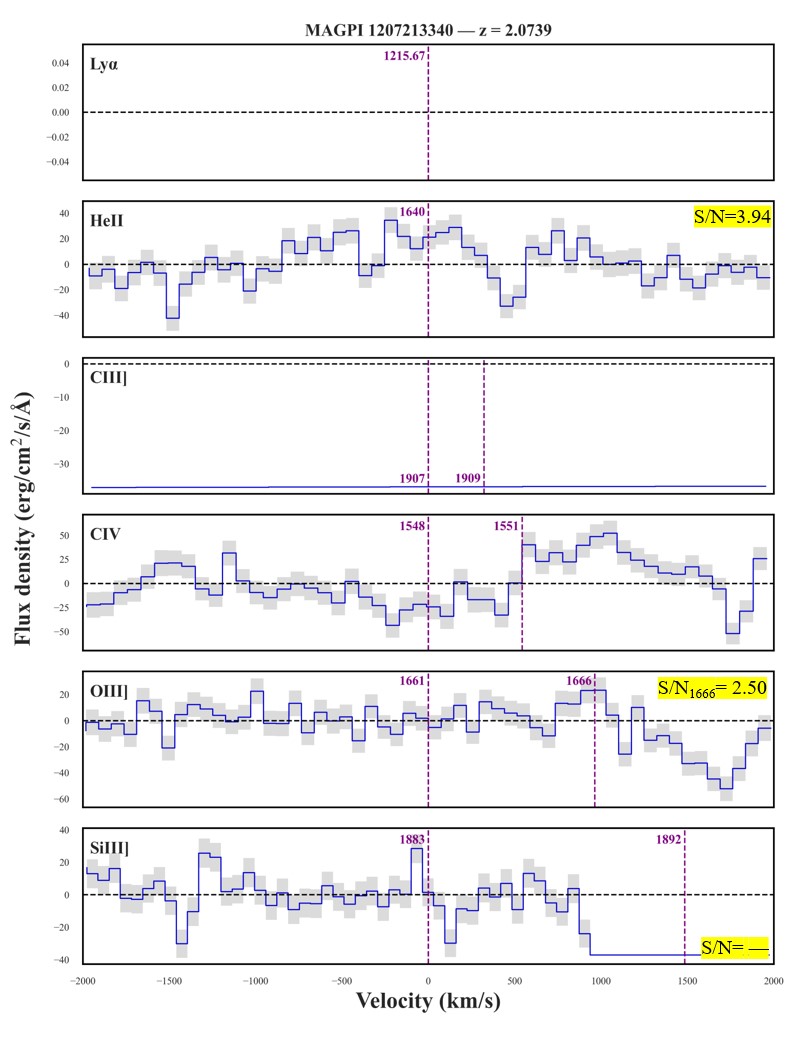 }
    \caption{continued}
    \label{fig:27}
\end{figure*}

% --- PAGE 1: First two spectra (a and b) ---
\begin{figure*}[p]
    \centering
    \begin{subfigure}{\textwidth}
        \centering
        \includegraphics[width=1.0\linewidth]{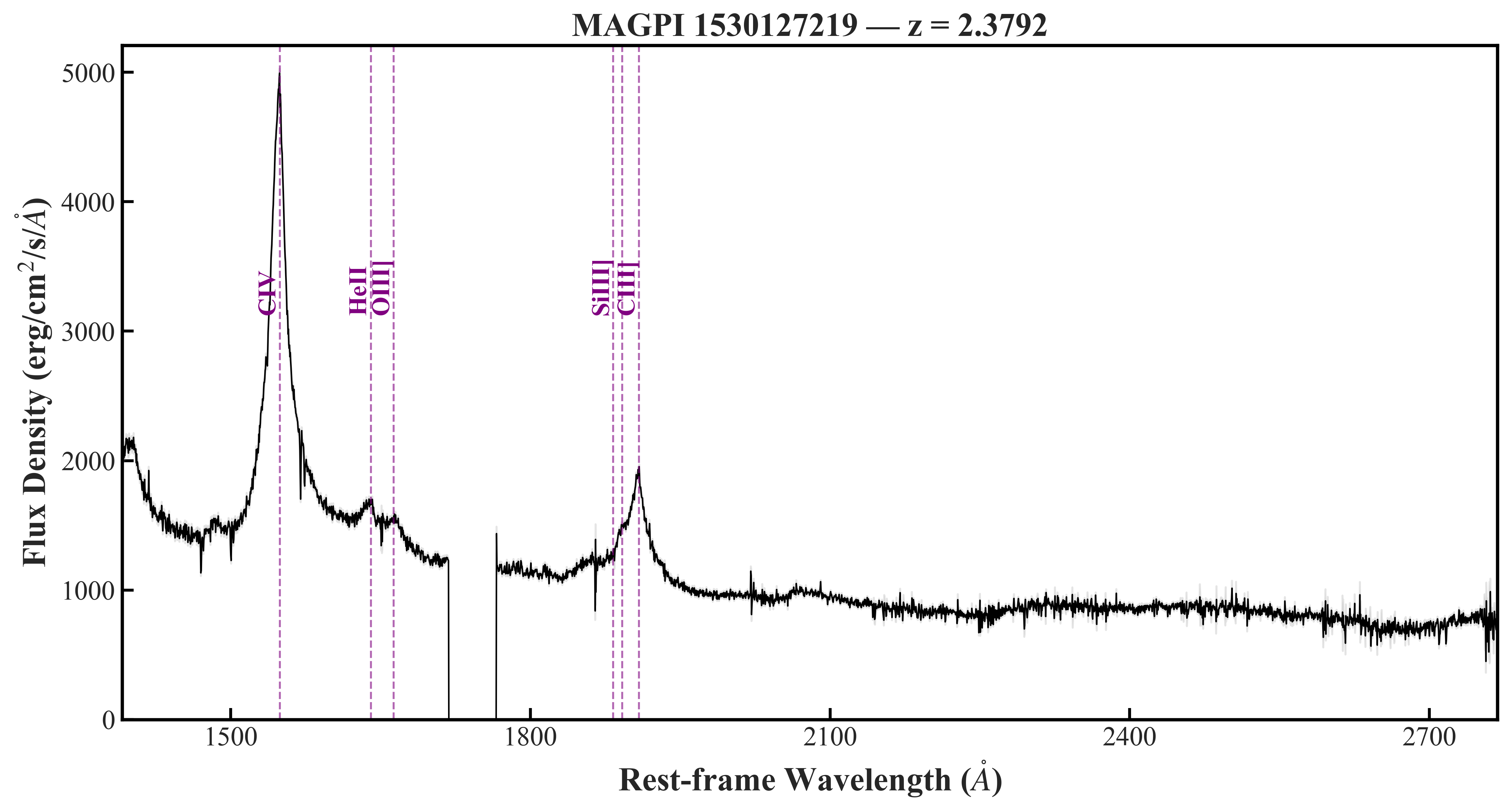}
        \caption{}
        \label{fig:QS1}
    \end{subfigure}
    
    \vspace{0.6cm} 

    \begin{subfigure}{\textwidth}
        \centering
        \includegraphics[width=1.0\linewidth]{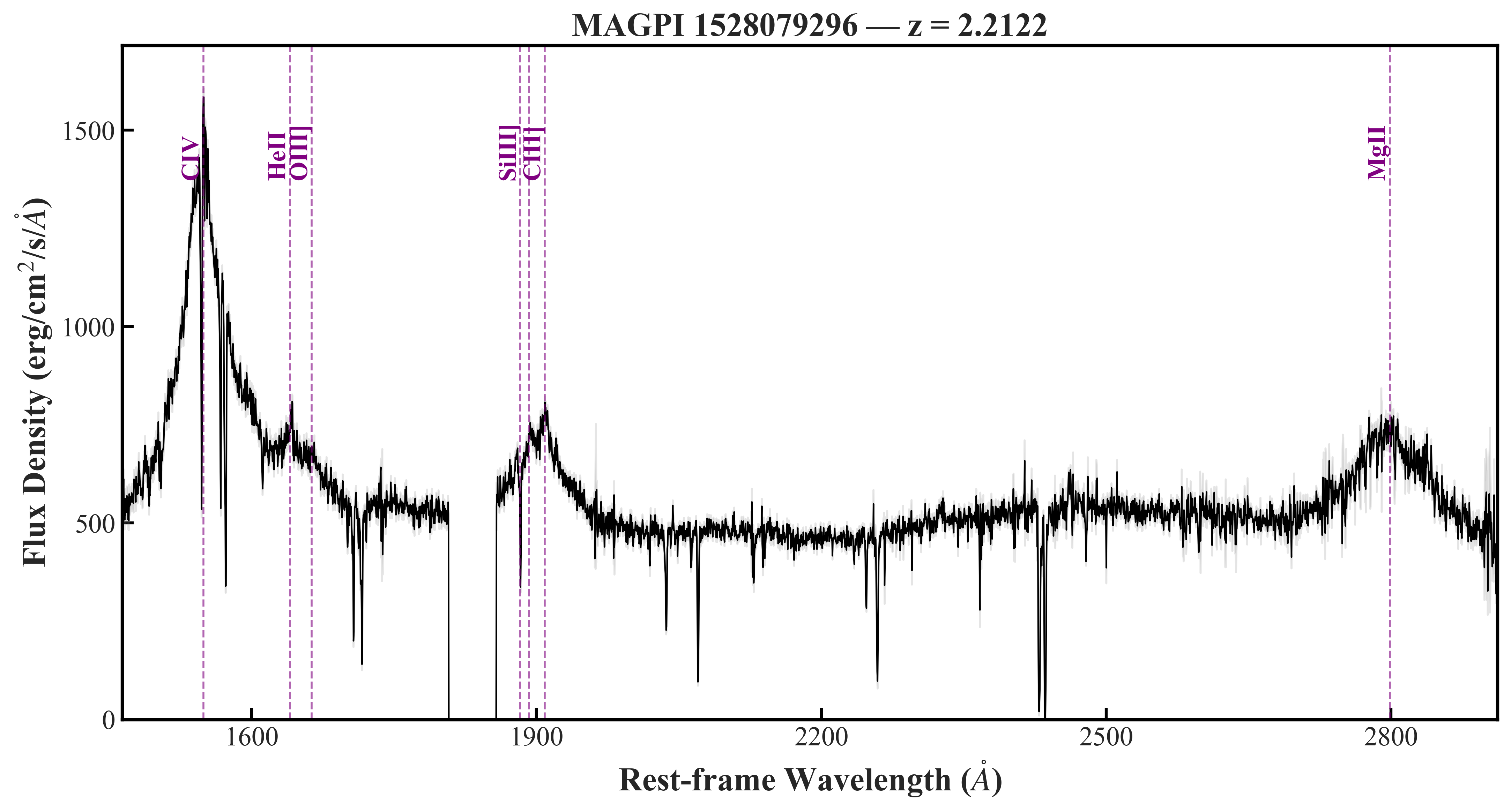}
        \caption{}
        \label{fig:QS2}
    \end{subfigure}
    
    \caption{Full spectra of excluded quasar-like sources. (\textit{Continued on next page})}
    \label{fig:QS_Combined} % <-- MOVED HERE
\end{figure*}

% --- PAGE 2: Third spectrum (c) ---
\begin{figure*}[ht!]
    \centering
    \ContinuedFloat % Keeps the same Figure number and continues subfigure labeling to (c)
    \begin{subfigure}{\textwidth}
        \centering
        \includegraphics[width=1.0\linewidth]{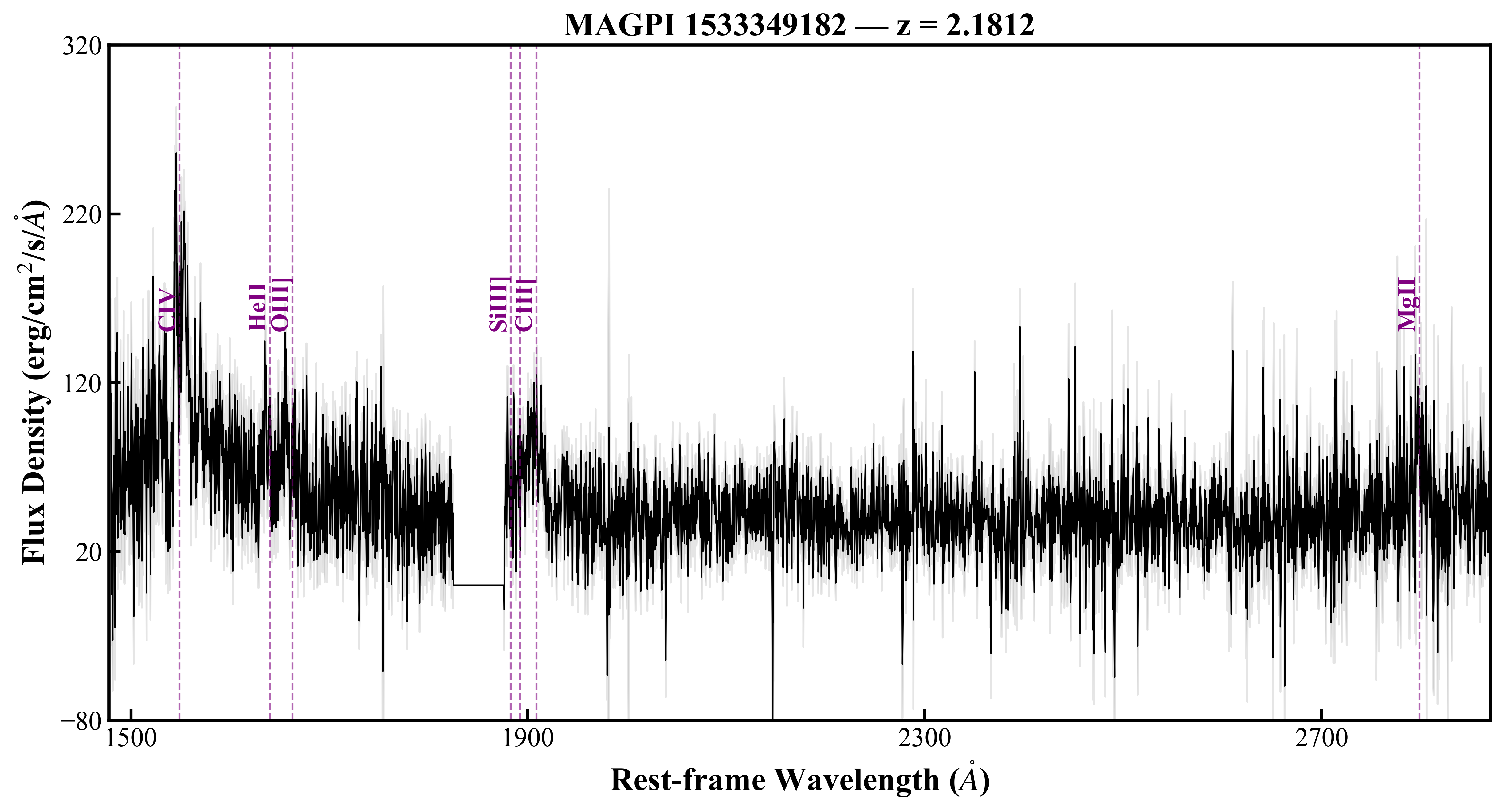}
        \caption{}
        \label{fig:QS3}
    \end{subfigure}
    
    \caption{Full spectra of excluded quasar-like sources (continued)}
\end{figure*}

\end{document}